\documentclass[12pt,a4paper]{article} 
\pdfoutput=1
\usepackage{jheppub} 
\usepackage{graphicx} 
\usepackage{amsmath} 
\usepackage{amssymb} 
\usepackage{slashed} 
\usepackage{bigstrut} 
\usepackage{multirow} 
\usepackage{url}

\newcommand{\beq}{\begin{equation}} 
\newcommand{\eeq}{\end{equation}} 
\def\bsp#1\esp{\begin{split}#1\end{split}} 
\def\bal#1\eal{\begin{align}#1\end{align}} 
\newcommand{\beeq}{\begin{eqnarray}} 
\newcommand{\eeeq}{\end{eqnarray}}

\newcommand{\msbar}{$\overline{\text{MS}}\, $}

\title{Prompt neutrinos from atmospheric charm in the general-mass variable-flavor-number scheme} 
\author{M.~Benzke,} 
\author{M.~V.~Garzelli,} 
\author{B.~A.~Kniehl,} 
\author{G.~Kramer,} 
\author{S.~Moch,} 
\author{G.~Sigl} 
\affiliation{II.~Institute for Theoretical Physics, Hamburg University
\\
Luruper Chaussee 149, D--22761 Hamburg, Germany}

\emailAdd{michael.benzke@desy.de}
\emailAdd{maria.vittoria.garzelli@desy.de} 
\emailAdd{bernd.kniehl@desy.de} 
\emailAdd{gustav.kramer@desy.de} 
\emailAdd{sven-olaf.moch@desy.de} 
\emailAdd{guenter.sigl@desy.de}  
  
\abstract{
We present predictions for the prompt-neutrino flux arising from the decay of charmed mesons and baryons produced by the interactions of high-energy cosmic rays in the Earth's atmosphere, making use of a QCD approach on the basis of the general-mass variable-flavor-number scheme for the description of charm hadroproduction at NLO, complemented by a consistent set of fragmentation functions. We compare the theoretical results to those already obtained by our and other groups with different theoretical approaches. We provide comparisons with the experimental results obtained by the IceCube Collaboration in two different analyses and we discuss the implications for parton distribution functions. 
}  
 

\preprint{DESY 17-073}
\begin{document} 
\maketitle 

\section{Introduction} 
\label{sec:intro} 
Prompt neutrinos produced in the atmosphere are expected to contribute to the total leptonic signal observed at Very Large Volume Neutrino Telescopes (VLV$\nu$Ts). Although, at present, there are no experiments which separately measure their contribution~\cite{Aartsen:2016xlq}, their existence is predicted theoretically by many different approaches. As a consequence, a series of dedicated searches is planned, which will benefit from the increasing statistics accumulated over the years and from the extension of the fiducial volume of some present experimental apparata, as foreseen for the near future~\cite{Adrian-Martinez:2016fdl,Aartsen:2014njl}. 
 
In principle, Cosmic Rays (CR) impinging on the upper layers of the Earth's atmosphere interact with the air nuclei, fragmenting into many different hadrons. The heaviest ones, i.e. those containing heavy quarks as valence quarks in their composition, are characterized by decay lengths shorter than their interaction lengths. Thus they decay promptly, emitting, in case of semi-leptonic decays, prompt neutrinos. 

On the other hand, the lighter abundant mesons, i.e. charged pions and kaons, whose leptonic decays are sources of the so-called conventional neutrino flux, are cha\-racterized by larger decay lengths, suppressing their decays at large enough energies. As a consequence, the prompt-neutrino flux is supposed to become dominant with respect to the conventional one for those energies. Although several uncertainties characterize the exact position of the transition point between the two domains, different available estimates suggest that it should be well within the energy interval presently explored by VLV$\nu$Ts.  
The big uncertainties in the transition energy reflect the big uncertainties affecting present predictions of prompt-neutrino fluxes, arising both from some poorly constrained astrophysical inputs and from the still not precise enough description of charm hadroproduction, the core process at the basis of the production of prompt neutrinos. 

Charm hadroproduction in the astrophysical context has been estimated along the years making use of many different approaches, ranging from phenomenological models to QCD theory. In the QCD framework, tree-level computations as available in Shower Monte Carlo (SMC) event generators were used for this purpose already more than ten years ago \cite{Gondolo:1995fq}, whereas, more recently, calculations including NLO QCD corrections matched to parton showers have been adopted. In particular, in our previous papers~\cite{Garzelli:2015psa,Garzelli:2016xmx}, we considered NLO QCD corrections to charm hadroproduction in an implementation with matrix elements in the fixed-flavor-number scheme, as available in the {\texttt{POWHEGBOX}} approach, matched with parton shower and hadronization, as available in the {\texttt{PYTHIA}} event generator \cite{Sjostrand:2014zea}. 
In the present paper, we follow a different QCD approach, utilizing the general-mass variable-flavor-number scheme (GM-VFNS), which allows for a transition between different numbers of flavors (from 3 to 4, in the case of charm hadroproduction) according to the region of phase space under study. Matrix elements for the hadroproduction of light and heavy partons are combined with a consistent set of fragmentation functions (FFs), which describe the transition from these partons to charmed hadrons. The validity and flexibility of this approach has been studied and cross-checked by means of comparisons with experimental data obtained at the LHC.

This paper is organized as follows: in Section~\ref{sec:fns}, we summarize the basic features of the variable-flavor-number schemes and we briefly sketch the differences with respect to other flavor number schemes; in Section~\ref{sec:gmvfns}, we give details of the specific implementation used in this work and compare theoretical predictions on charm meson hadroproduction from our approach to LHCb experimental data, also for small values of transverse momentum ($p_T$); in Section~\ref{sec:astro}, we summarize the methodology adopted for computing prompt-neutrino fluxes, listing the astrophysical aspects in Subsection~\ref{sec:equa} and focusing on the QCD input in Subsection~\ref{sec:spectra}; in Section~\ref{sec:fluxes}, we present our predictions for prompt-neutrino fluxes, together with the associated uncertainties, and compare them with other recent theoretical predictions, in particular with those we obtained in the {\texttt{POWHEGBOX~+~PYTHIA}} approach; in Section~\ref{sec:implications}, we summarize the implications for searches at VLV$\nu$Ts and related PDF fit constraints; finally, we present our conclusions in Section~\ref{sec:conclusions}. 

\section{Flavor Number Schemes: basic features} 
\label{sec:fns} 

When calculating cross sections of inclusive heavy-quark production, the quark mass $m_Q$ appears as a relevant scale. Depending on the kinematic region, different calculation schemes are appropriate. In the center-of-mass frame, one may introduce the produced-quark transverse momentum $p_T$ relative to the collision axis. When considering the kinematic region where $p_T$ is of the same order as $m_Q$ or even lower, one uses a massive or fixed-flavor-number scheme (FFNS) \cite{Nason:1987xz,Nason:1989zy,Beenakker:1988bq,Beenakker:1990maa}. In that scheme, one calculates the cross section assuming only the heavy quark to be massive while all the others are massless and may appear as active flavors in the initial state. Due to the mass, there are no collinear singularities associated with the heavy quark and, consequently, no requirement to absorb them into the components of a factorized expression. Explicitly, there is no need for a FF, except to model non-perturbative effects of hadronization. However, instead of collinear singularities, logarithms of the ratio of the relevant scales $\ln (m_Q/p_T)$ appear in the calculation at every order in the perturbative expansion. If one considers a kinematic region where these scales are very different from each other, the logarithms become large and may invalidate the truncation of the perturbative series at fixed order.
In the context of charm production through cosmic rays, the whole $p_T$ range is of interest in principle, and energies can become very large.
While the differential cross section in $p_T$ is dominated by the low-$p_T$ region (see e.g. figure \ref{fig:lhcbcompare13}), at high energies, the high-$p_T$ region is still probed and may yield a noticeable contribution.

In order to make the perturbative series converge in the whole kinematic range, the potentially large logarithms can be resummed by properly factorizing the cross section and running the components to their appropriate scales. A suitable framework for this is the zero-mass variable-flavor number scheme (ZM-VFNS) \cite{Cacciari:1993mq, Kniehl:1995em, Cacciari:1995ej, Binnewies:1997gz, Kniehl:1996we, Binnewies:1997xq, Binnewies:1998vm, Kniehl:1999vf, Kniehl:2002xn, Kniehl:2005de, Kniehl:2006mw, Cacciari:2012ny}. Here, also the heavy quark is considered massless and may appear in the initial state. The collinear singularities of the massless calculation are absorbed into the initial-state parton distribution functions (PDFs) and the final-state FFs. Using the Dokshitzer-Gribov-Lipatov-Altarelli-Parisi (DGLAP) evolution equations the corresponding logarithms may be resummed. However, the assumption of the heavy quark being massless is, of course, inappropriate in the low-$p_T$ region. Specifically, the calculation misses contributions proportional to $m_Q^2/p_T^2$, which are present in the FFNS approach. In summary, the differential cross section at low and intermediate $p_T$ is well described by the FFNS, while, at large $p_T$, it becomes necessary to use the ZM-VFNS.

Both approaches may be combined using a GM-VFNS \cite{Kramer:2001gd, Kramer:2003cw, Kramer:2003jw, Kniehl:2004fy, Kniehl:2005mk, Kniehl:2005st, Kniehl:2005ej, Kniehl:2008zza}. Here, the terms proportional to $m_Q^2/p_T^2$ are kept in the hard-scattering cross sections, while, at the same time, the large logarithms are resummed using DGLAP evolution. The running of the PDFs and FFs is determined using the appropriate number of active flavors at each scale and performing a matching at the transition points. Here, we will use a specific implementation of the GM-VFNS, described in the next section, to compute the charm production cross sections needed to determine the prompt-neutrino fluxes.

\section{General-Mass Variable-Flavor-Number Scheme: details of our NLO implementation and comparison with LHCb experimental data}
\label{sec:gmvfns}

In this work, we will use the GM-VFNS as it was introduced in Ref.~\cite{Kniehl:2004fy}. The basis is formed by the factorized expression for the differential cross section of the inclusive production of a hadron $h$ in $pp$ collisions,
\begin{align}
d\sigma_{pp\to hX}(P,S) 
= F_{i/p}(x_1,\mu_i)F_{j/p}(x_2,\mu_i)\,\otimes\, d{\hat\sigma}_{ij\to kX}(p,s,\mu_r,\mu_i,\mu_f)\,\otimes\, D_{h/k}(z,\mu_f)\,,
\label{eqn:fact}
\end{align}
where $F_{i/p}$ are the PDFs, $D_{h/k}$ are the FFs, the $\otimes$ symbol denotes convolutions with respect to the scaling variables $x_1$, $x_2$, $z$, and a sum over all possible partons $i$, $j$ and $k$ is implied. 
The partonic quantities $p$ and $s$ depend on the final-state-hadron momentum $P$ and the hadronic center-of-mass energy $\sqrt{S}$ via a suitable definition of the scaling variables.
In the conventional parton model approach, the partonic cross section $d{\hat\sigma}$ is calculated assuming all partons to be massless. It will be denoted by $d{\hat\sigma}^\mathrm{ZM}$. In this case, the hadronic momenta $P_i$ are simply proportional to the partonic ones $p_i$, with the scaling variables being the corresponding factors
\begin{equation}
  p_1=x_1 P_1\,,\quad p_2=x_2 P_2\,,\quad p=P/z\,,
  \label{eqn:zdef}
\end{equation}
where $P_1$ and $P_2$ are the proton momenta, which also implies $s=x_1x_2S$.
Considering the partonic Mandelstam variables $s$, $t$, $u$ and introducing the commonly used kinematic invariants $v=1+t/s$, $w=-u/(s+t)$ and their hadronic (capital) equivalents leads to the explicit form of the factorization formula,
\begin{align}
\frac{1}{p_T}\frac{d\sigma_{pp\to hX}}{dp_T dy}(S,p_T,y) &=
\,\frac{2}{S}\sum_{i,j,k}\int_{1-V+VW}^1\frac{dz}{z^2}\int_{\frac{VW}{z}}^{1-\frac{1-V}{z}}\frac{dv}{1-v}\int_{\frac{VW}{vz}}^1\frac{dw}{w}\,\nonumber\\
&F_{i/p}(x_1,\mu_i)F_{j/p}(x_2,\mu_i)\, \frac{1}{v}\frac{d{\hat\sigma}_{ij\to kX}^\mathrm{ZM}}{dvdw}(s,v,w,\mu_r,\mu_i,\mu_f)\, D_{h/k}(z,\mu_f)\,,
\label{eqn:explFact}
\end{align}
where $p_T$ and $y$ denote the transverse momentum and the rapidity of the produced hadron. The next-to-leading order (NLO) results were derived in Ref.~\cite{Aversa:1988vb}. The large logarithms were subsequently resummed in the next-to-leading-log (NLL) approximation in Ref.~\cite{Cacciari:1993mq}.

The factorization formula still holds true in the case of non-vanishing quark masses~\cite{Collins:1998rz}. The partonic cross section $d\hat\sigma$ in eq.~(\ref{eqn:fact}) is replaced by the corresponding massive version $d\hat\sigma(m_c)$, which can be derived from the NLO parton model and the FFNS results \cite{Nason:1989zy, Beenakker:1988bq, Bojak:2001fx} in an appropriate calculation scheme. We will adopt the scheme first presented in Ref.~\cite{Kramer:2001gd} in the context of $\gamma\gamma$ collisions. It can be presented in the following way
\begin{equation}
d\hat\sigma(m_c) = d{\hat\sigma}^\mathrm{FFNS}(m_c) - \lim_{m_c\to 0}d{\hat\sigma}^\mathrm{FFNS}(m_c) + d{\hat\sigma}^\mathrm{ZM}\,.
\label{eqn:gmvfns}
\end{equation}
The subtraction of the zero-mass limit of the FFNS result avoids a double counting with the ZM part, which contains contributions of charm quarks in the initial state. Terms proportional to $m_c^2/p_T^2$, on the other hand, are retained in the partonic cross section.
This procedure constitutes a certain scheme choice, since the zero-mass limit of the FFNS result is not equal to the ZM one~\cite{Mele:1990cw}. This is due to the fact that the ZM calculation is performed in the \msbar scheme, which implies a dimensional regulator $\varepsilon$, while in the FFNS, the mass effectively regulates the collinear divergences. These two schemes do not necessarily have the same limits for $\varepsilon\to 0$ and $m_c\to 0$, respectively.
Finally, the massive partonic cross sections are convoluted with PDFs and FFs as written in the factorization formula (\ref{eqn:fact}). In fact, the explicit form is similar to the one in eq.~(\ref{eqn:explFact}), except that one has to take into account that, for a massive final-state hadron, the definition of the scaling variable $z$ has to be adapted, since $p^2\neq P^2$, which makes the definition~(\ref{eqn:zdef}) of the variable $z$ unsuitable.
We choose $z$ to be the factor between the large light-cone component of the parton $p^+=(p^0+|\vec{p}|)/\sqrt{2}$ and that of the hadron $P^+$. This change of definition leads to a phase space factor~\cite{Kniehl:2015fla} in the cross section in eq.~(\ref{eqn:explFact}),
\begin{equation}
d\sigma\to\frac{d\sigma}{R^2}\,,\quad R=1-\frac{m_h^2-z^2m_c^2}{(P^0+|\vec{P}|)^2-z^2m_c^2}\,.
\end{equation}
Furthermore, the definitions of the kinematic variables $v$ and $w$ need to be changed accordingly,
\begin{equation}
v=1+\frac{t-m_c^2}{s}\,,\quad w=\frac{-u+m_c^2}{s+t-m_c^2}\,.
\end{equation}
For the FFs we use the set {\texttt{KKKS08}} that has been fitted at NLO to $e^+e^-$ data in the context of the GM-VFNS approach~\cite{Kneesch:2007ey}.

An analysis of charmed-hadron production at the LHC using the GM-VFNS has been performed in Ref.~\cite{Kniehl:2012ti}. In this work, we extend this procedure to be viable at very small $p_T$ in order to apply it to charm production in the atmosphere, where a significant contribution appears in the very forward region.

Due to the form of the factorized cross section for inclusive heavy-meson hadro-production, there appear three independent scale parameters, namely the renormalization scale $\mu_r$ and the factorization scales $\mu_i$ and $\mu_f$, corresponding to the initial and final states, respectively. A natural choice for these scales is to set them all equal to each other to $\mu_r=\mu_i=\mu_f=\sqrt{p_T^2+m_c^2}$.  
However, following this procedure leads to a badly behaved differential cross section for $p_T\to 0$. This is related to contributions with the heavy quark appearing in the initial state, calculated using the massless scheme. It is, therefore, necessary to develop a method to suppress these contributions in the aforementioned limit and to retain the FFNS result, appropriately describing the cross section at small $p_T$. Recently, it has been suggested to use the freedom of choice for the scale parameters to this end \cite{Kniehl:2015fla}. Specifically, one uses the fact that the heavy-quark PDFs vanish for a scale $\mu_i<m_c$. By setting the factorization scale for initial states to the transverse mass multiplied by a parameter $\xi_i<1$, it becomes smaller than the heavy-quark mass for small enough $p_T$:
\begin{equation}
\mu_i=\xi_i\sqrt{p_T^2+m_c^2}< m_c\quad\Leftrightarrow\quad p_T< m_c\sqrt{\frac{1}{\xi_i^2}-1}\,.
\end{equation}
In this way, the contributions with the heavy quark in the initial state are switched off for small $p_T$, and only the FFNS contributions with the heavy quark just in the final state remain.

For our predictions, we are using a \texttt{FORTRAN} code that performs the necessary numerical integrations and yields the cross sections differential in the required variables. The integrator is an implementation of \texttt{VEGAS} \cite{Lepage:1977sw} as it is provided in the \texttt{CUBA} package \cite{Hahn:2004fe}.

Throughout this paper, we will consider up to 4 flavors in the initial state and evaluate $\alpha_s(\mu_r)$ at NLO with $\Lambda^{(4)}_{\overline{\text{MS}}}=328\,$MeV. The charm-quark pole mass is taken to be $m_c=1.3\,$GeV as is appropriate for the {\texttt{CT14nlo}} PDF \cite{Dulat:2015mca} that we are using as our standard.\footnote{In order to evaluate PDF uncertainties, we use the 5-flavor version of these PDFs, neglecting contributions from bottom quarks as initial state partons.}
We observe that using the 5-flavor strong-coupling constant $\alpha_s$ and including the contribution of bottom initial states for energies above the bottom threshold would cause modifications of our predictions by some percent. However, this is not particularly relevant in the context of this paper, where QCD and astrophysical uncertainties of many ten percents dominate our results for prompt-neutrino fluxes, as shown in the following. 

\begin{figure}[ht]
\centering
\includegraphics[width=120mm]{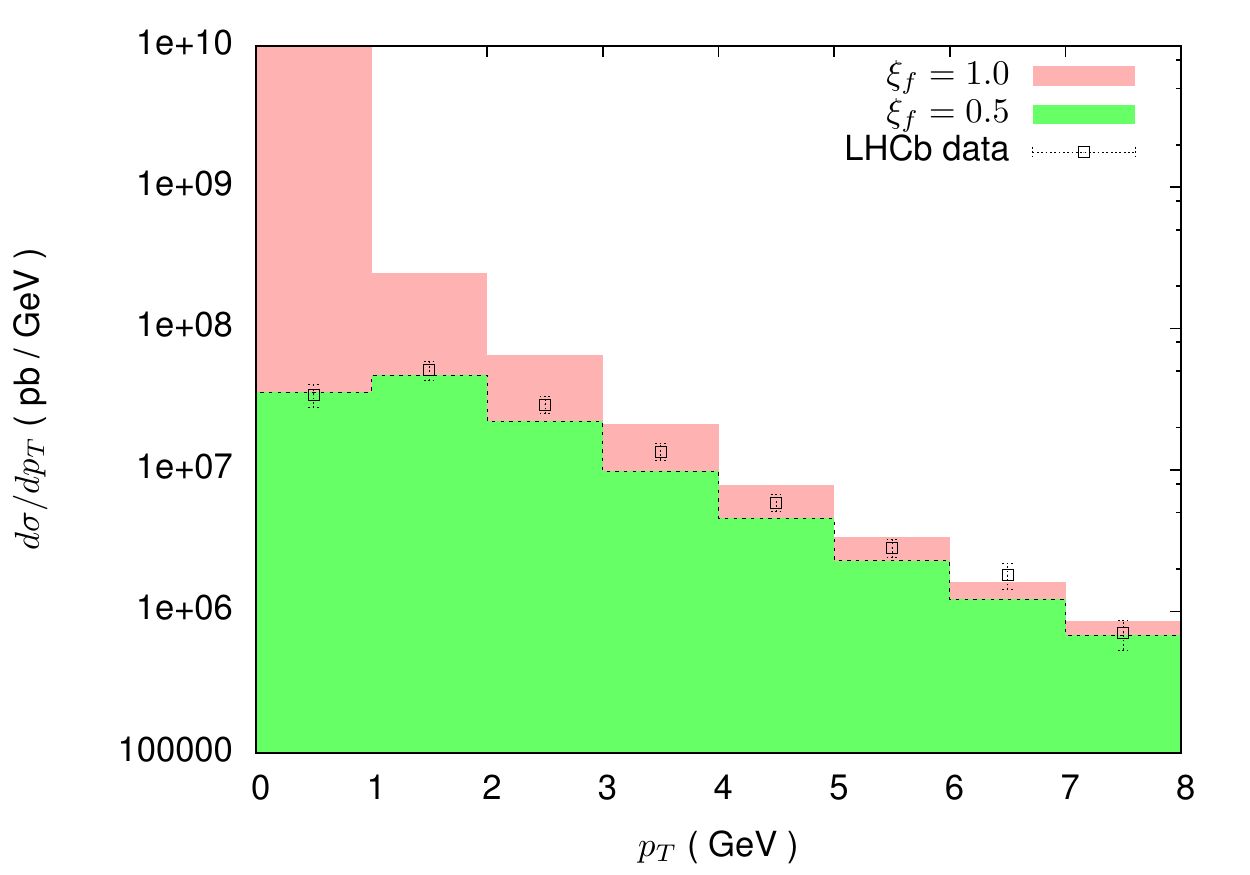}
\caption{($D^++D^-$) differential cross sections d$\sigma$/d$p_T$ in $pp$ collisions at $7\,$TeV in the $3.0<y<3.5$ rapidity range. The histograms correspond to the choices $\mu_r=\sqrt{p_T^2+4m_c^2}$ and $\mu_i=\mu_f=\xi\mu_r$ with $\xi=0.5$ or $\xi=1.0$. The experimental data are taken from Ref.~\cite{Aaij:2013mga}.}
\label{fig:xii}
\end{figure}

\begin{figure}[ht]
\centering
\includegraphics[width=73mm]{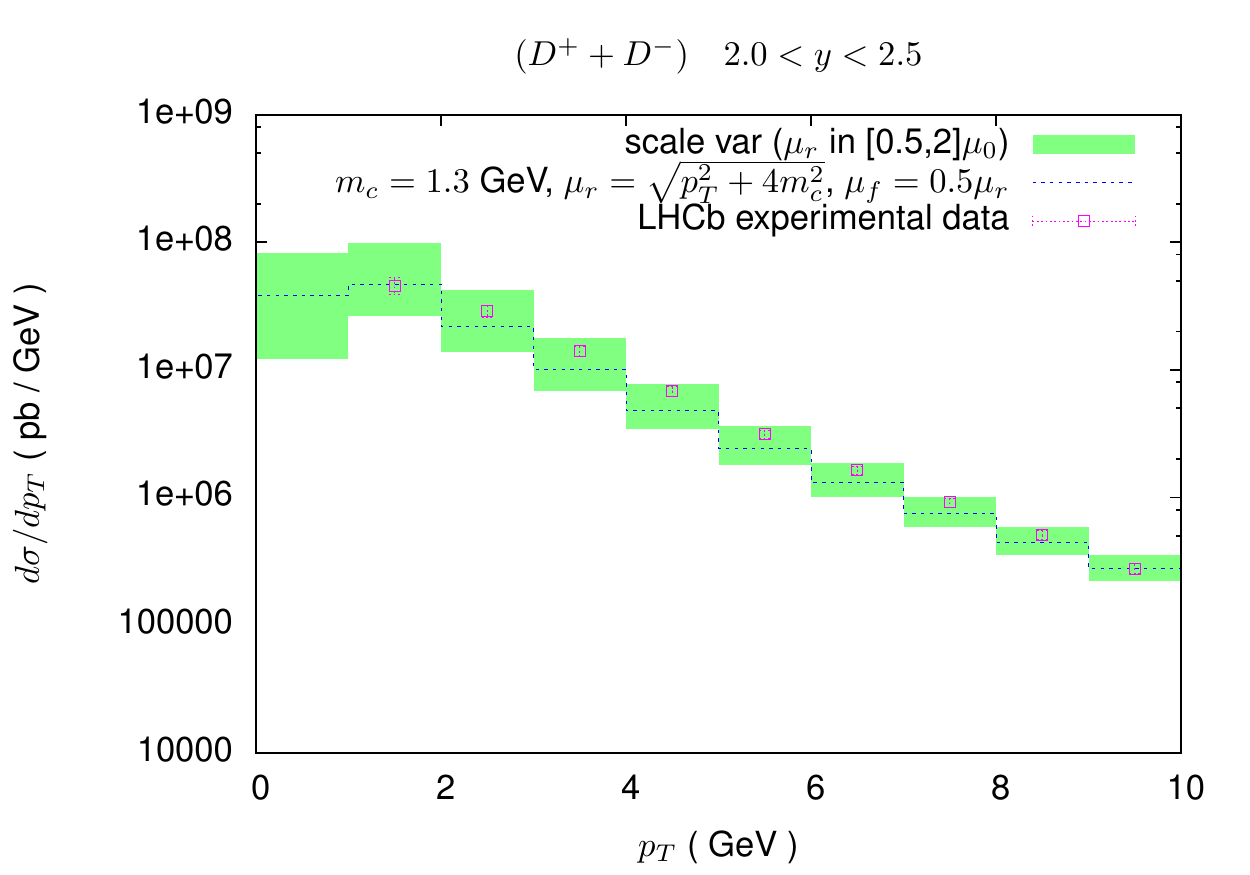}\quad\includegraphics[width=73mm]{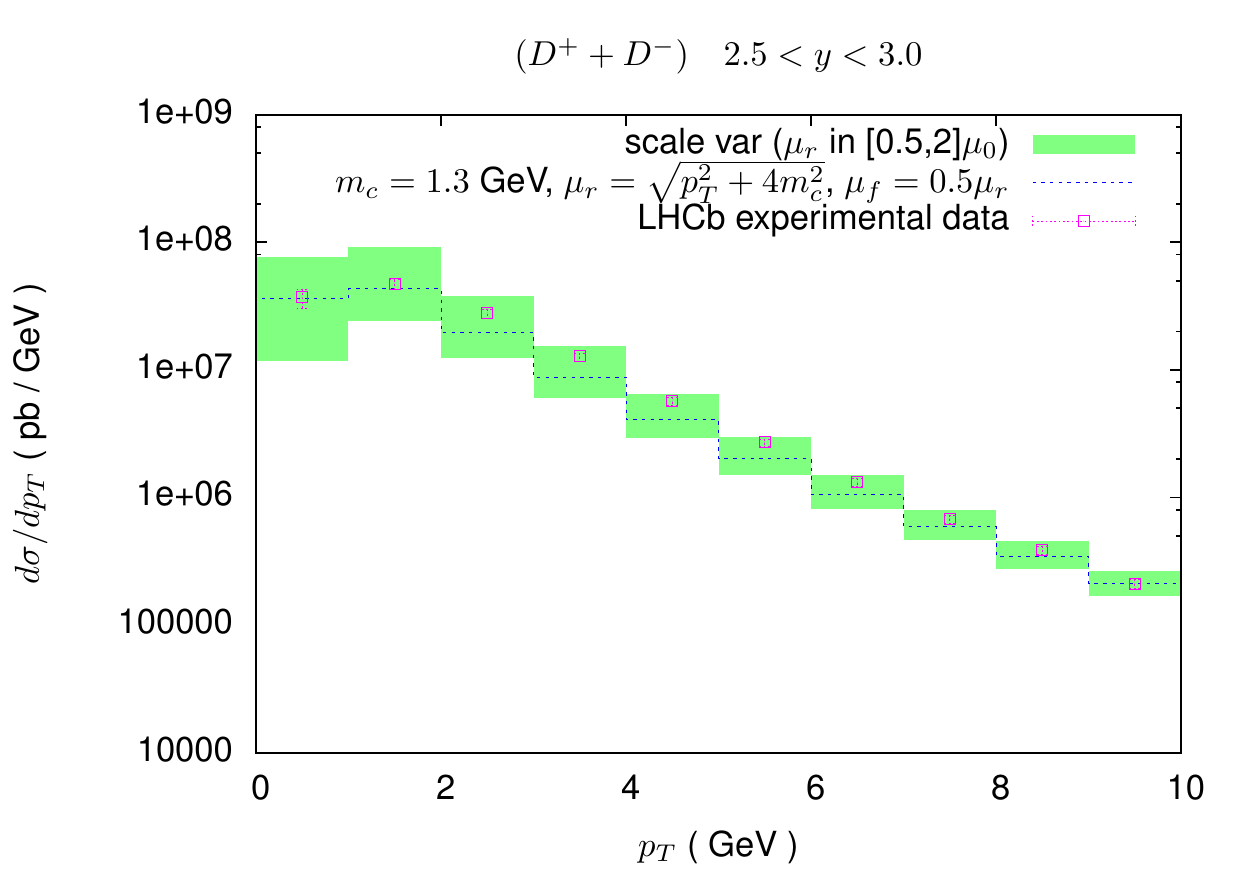}\\
\includegraphics[width=73mm]{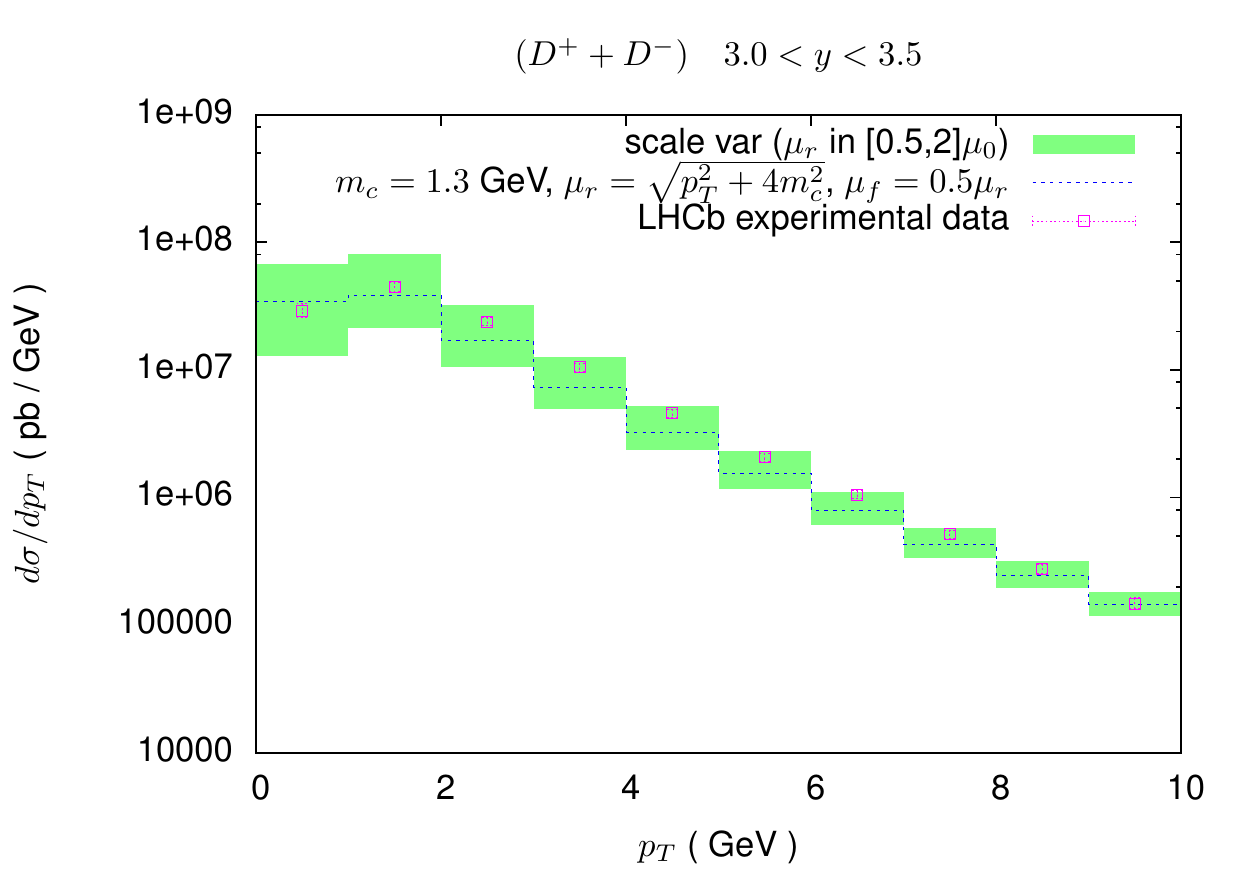}\quad\includegraphics[width=73mm]{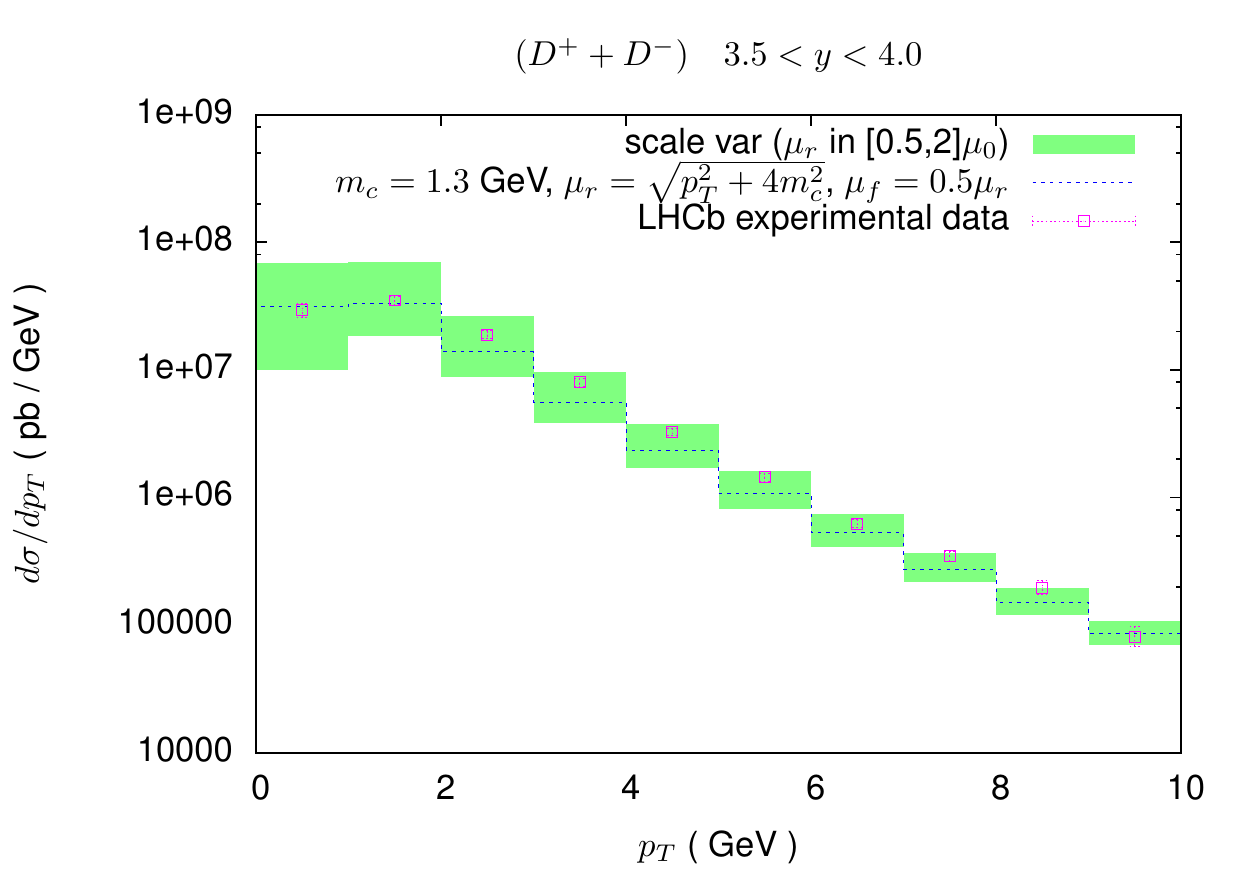}\\
\includegraphics[width=73mm]{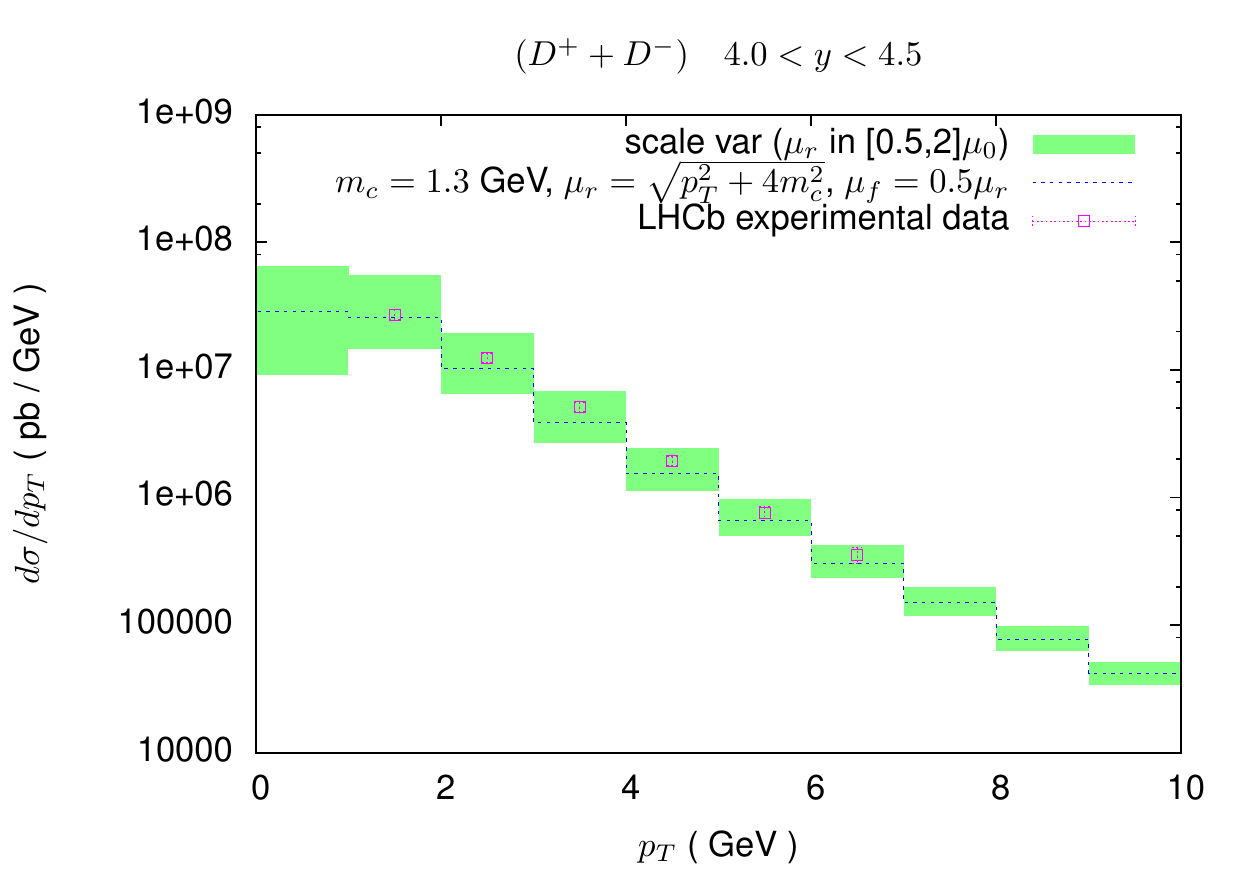}
\caption{\label{fig2} Our GM-VFNS predictions for ($D^+$ + $D^-$) transverse-momentum distributions for $pp$ collisions at $\sqrt{S}$ = 5 TeV vs. LHCb experimental data of Ref.~\cite{Aaij:2016jht}. Each panel corresponds to a different rapidity bin in the interval 2 $<$ $y$ $<$ 4.5. The renormalization scale $\mu_r$ is chosen as $\sqrt{p_T^2+4m_c^2}$, and varied in the [0.5, 2] interval around this central value, whereas the factorization scales $\mu_i$ and $\mu_f$ are fixed at $\xi_{i/f}\sqrt{p_T^2+4m_c^2}$, with $\xi_{i,f} = 0.5$. 
}
\label{fig:lhcbcompare5}
\end{figure}

\begin{figure}[ht]
\centering
\includegraphics[width=73mm]{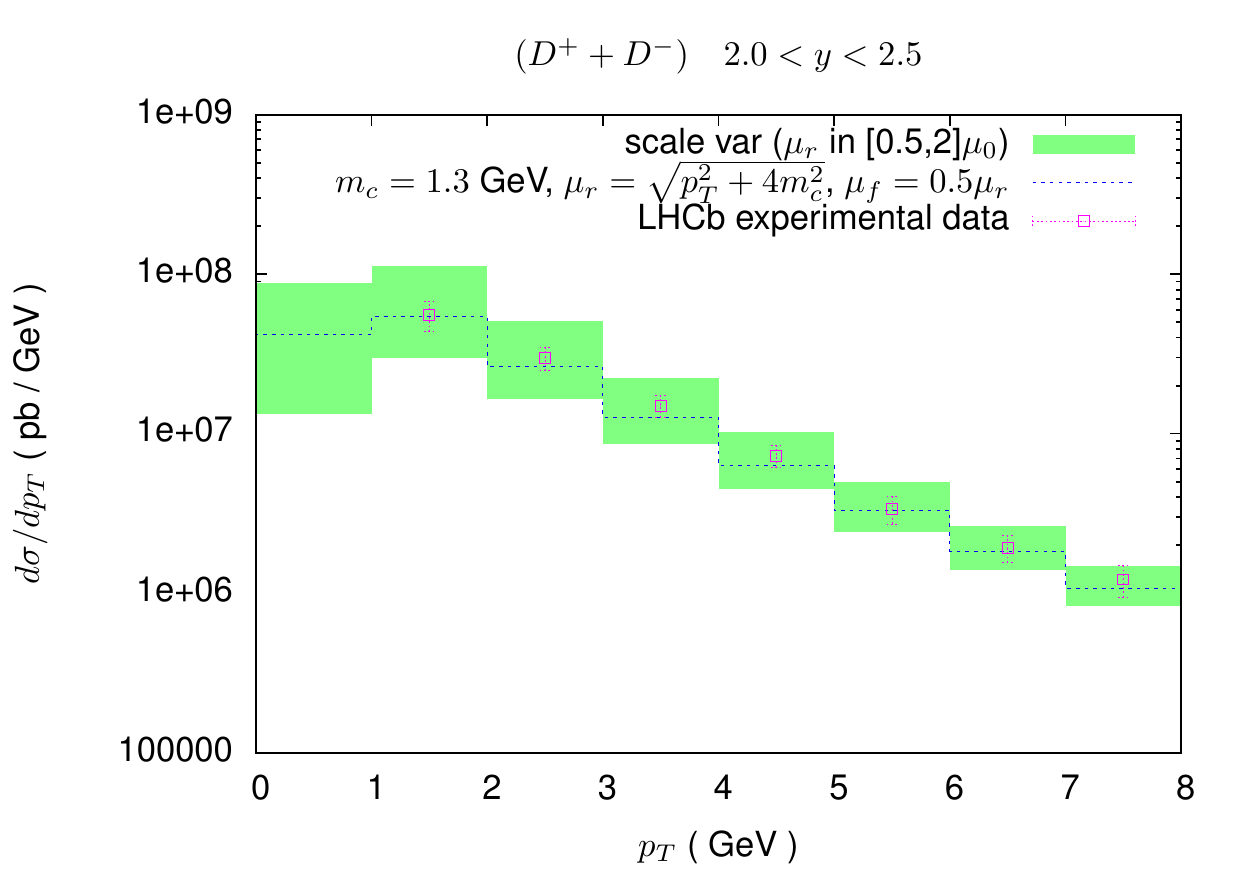}\quad\includegraphics[width=73mm]{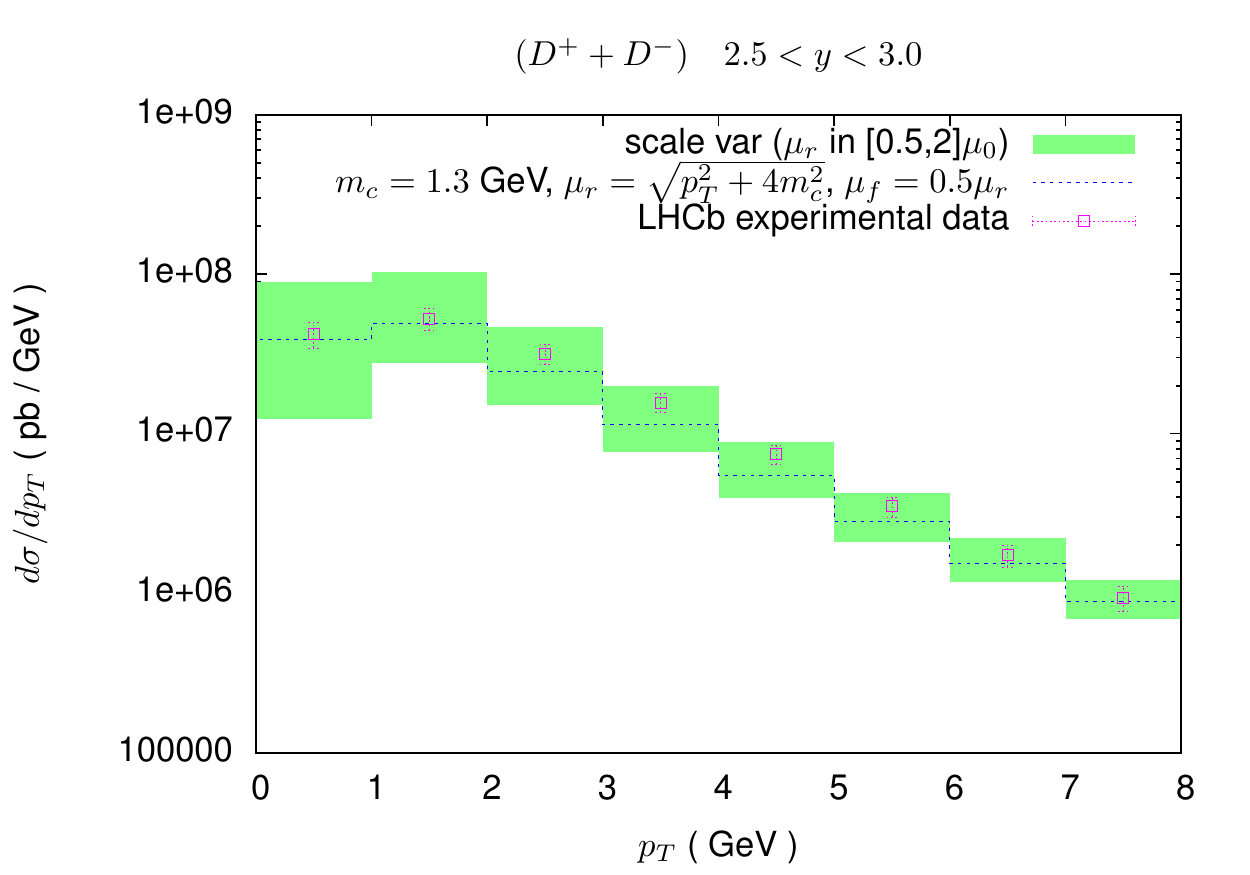}\\
\includegraphics[width=73mm]{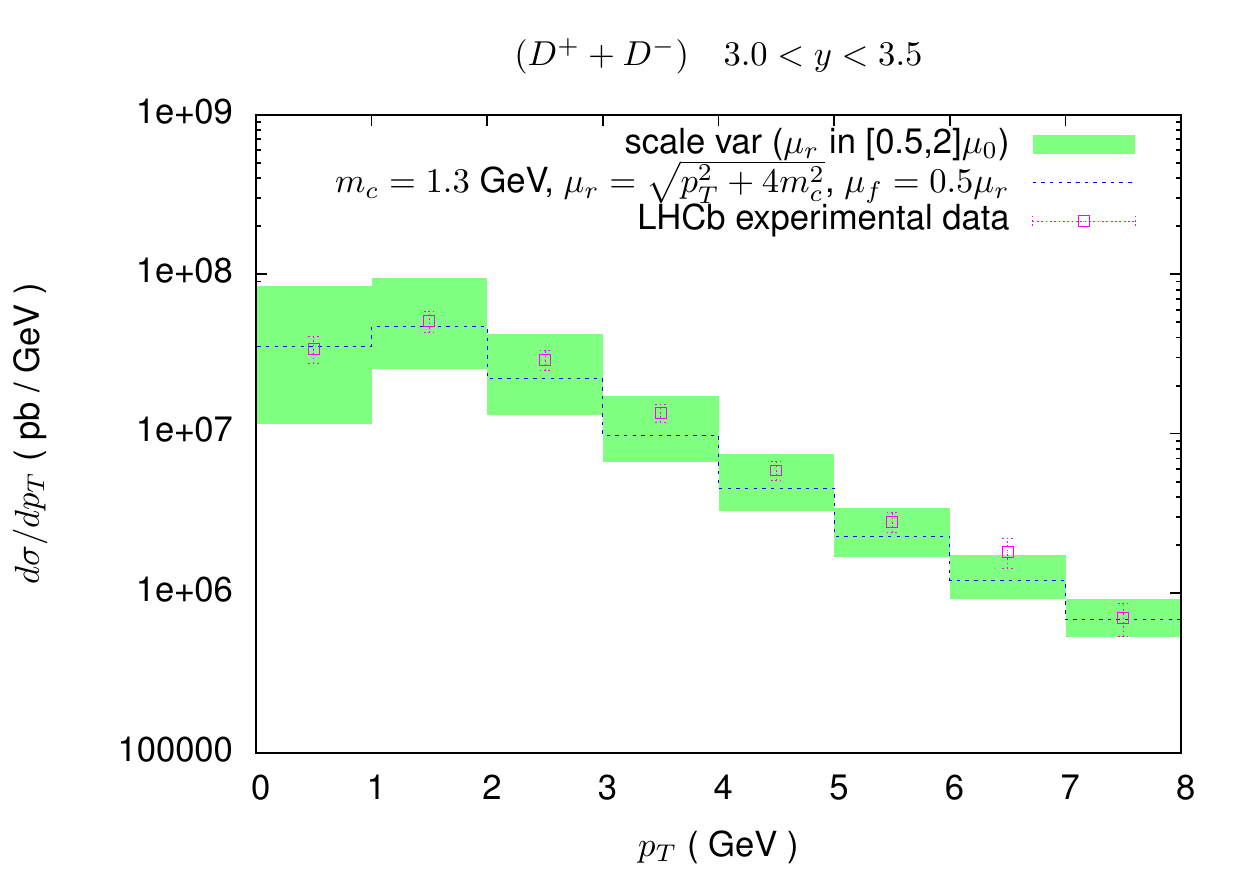}\quad\includegraphics[width=73mm]{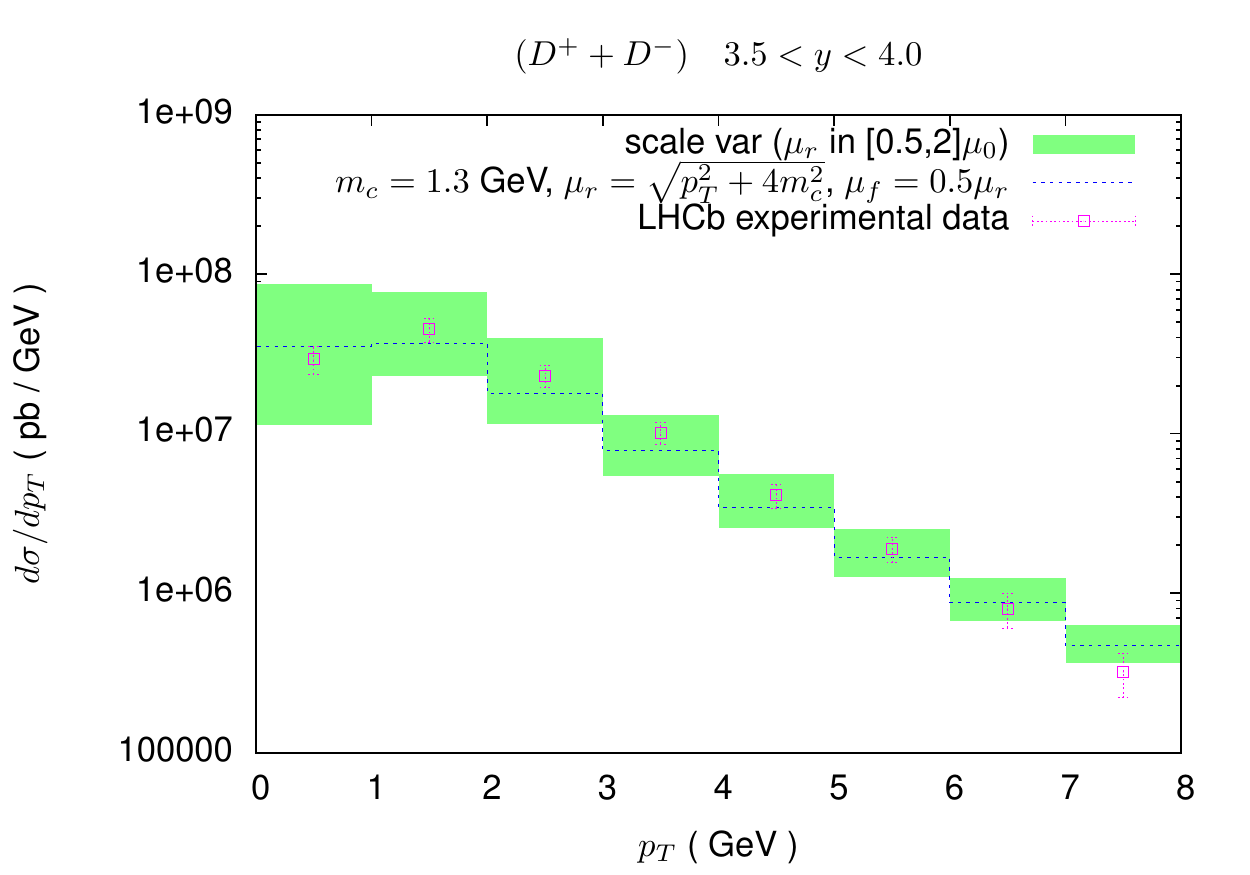}\\
\includegraphics[width=73mm]{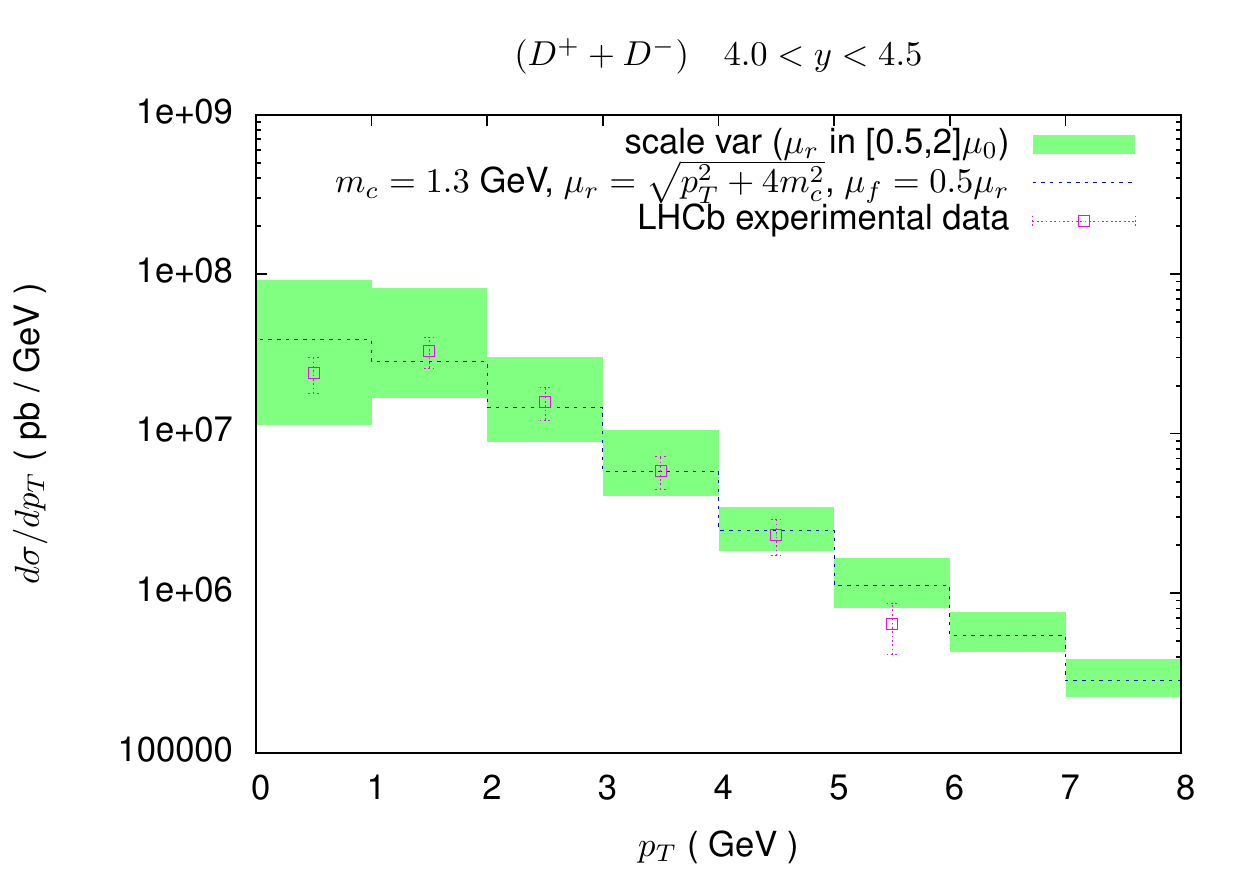}
\caption{Our GM-VFNS predictions for ($D^+$ + $D^-$) transverse-momentum distributions for $pp$ collisions at $\sqrt{S}$ = 7 TeV vs. LHCb experimental data of Ref.~\cite{Aaij:2013mga}. Each panel corresponds to a different rapidity bin in the interval 2 $<$ $y$ $<$ 4.5.
The scales are chosen as in figure~\ref{fig2}. 
}
\label{fig:lhcbcompare7}
\end{figure}

\begin{figure}[ht]
\centering
\includegraphics[width=73mm]{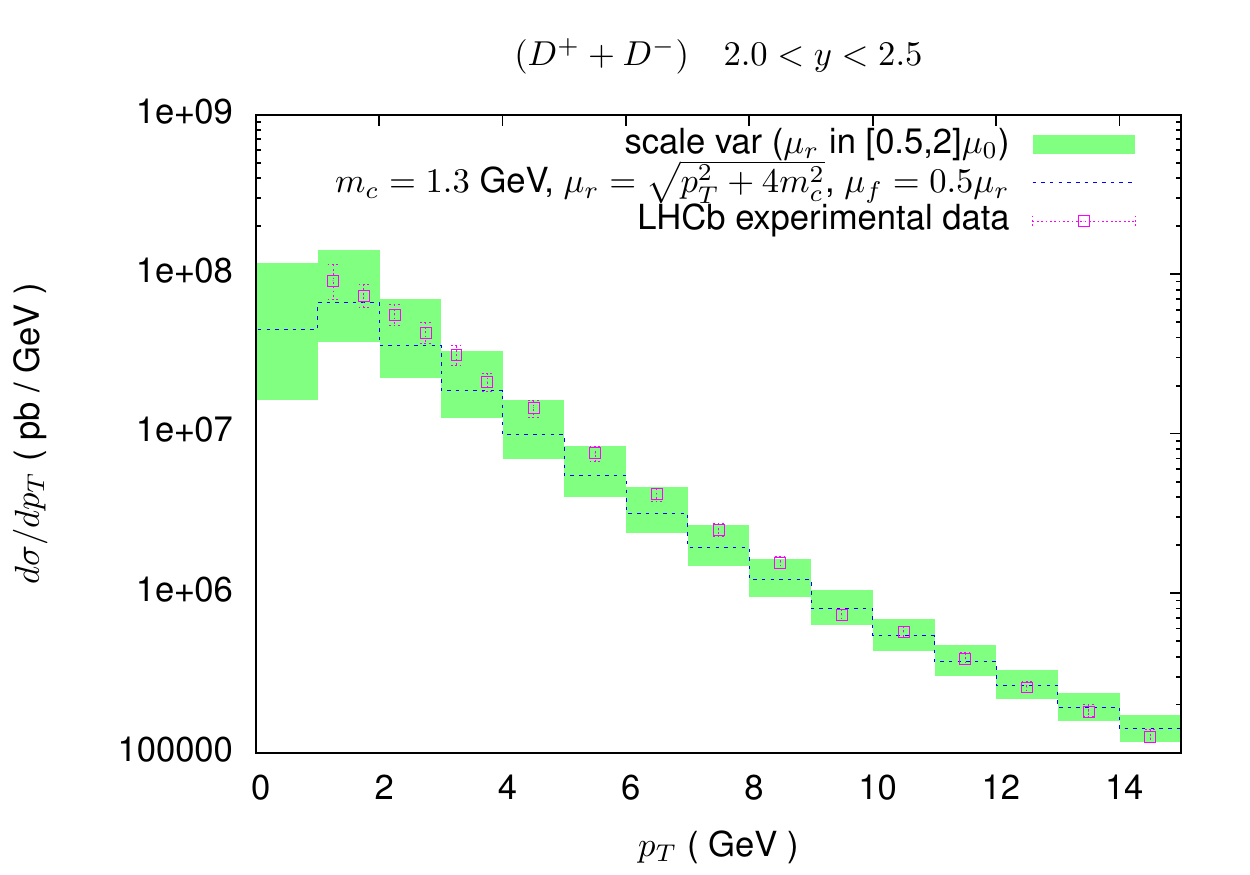}\quad\includegraphics[width=73mm]{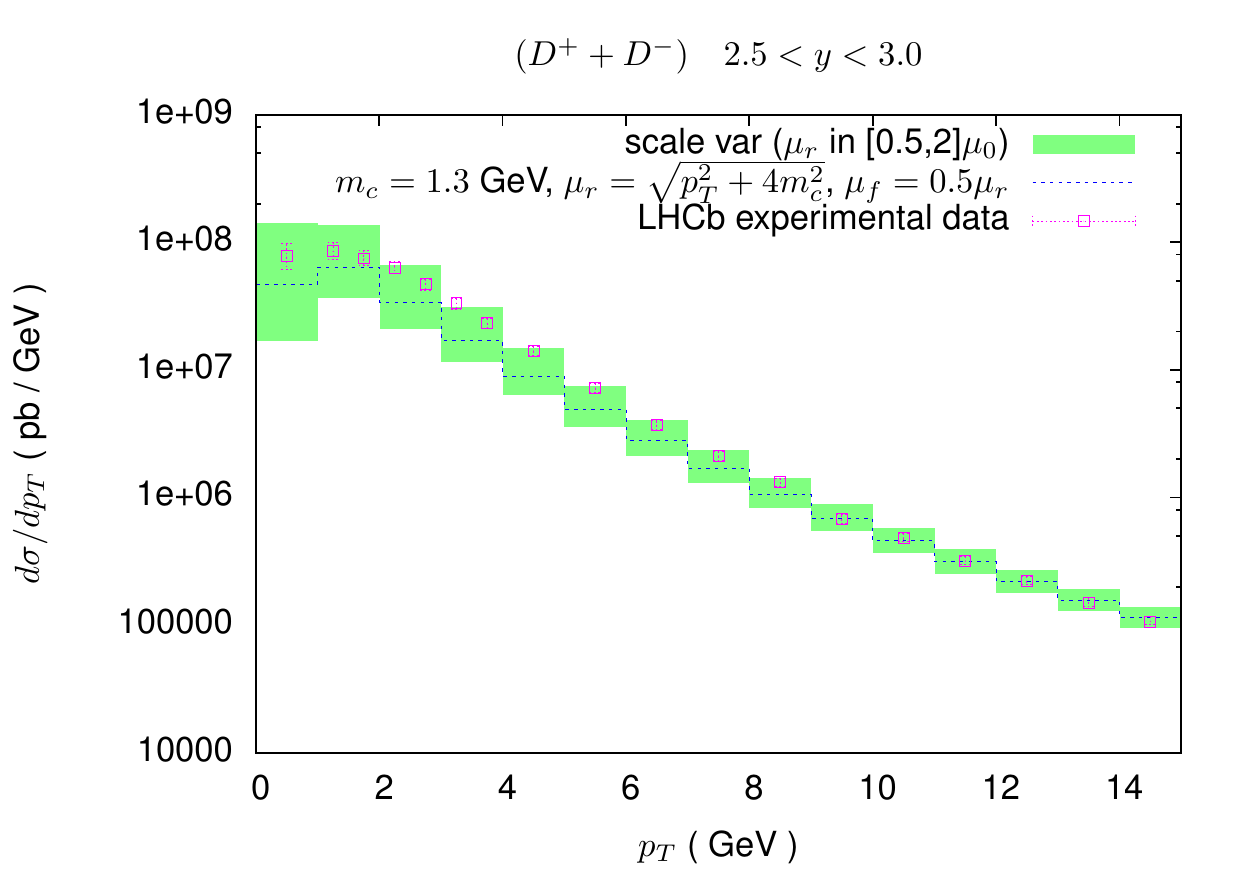}\\
\includegraphics[width=73mm]{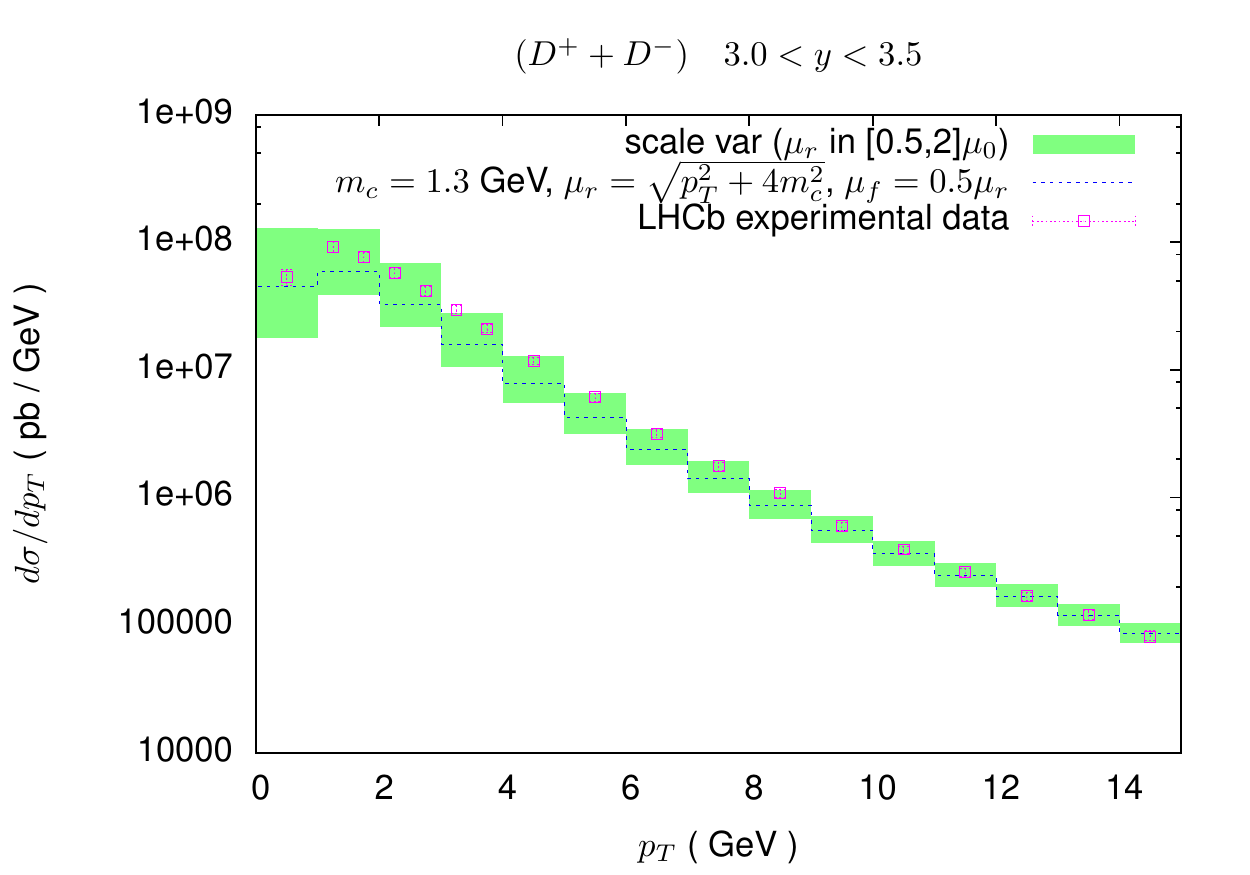}\quad\includegraphics[width=73mm]{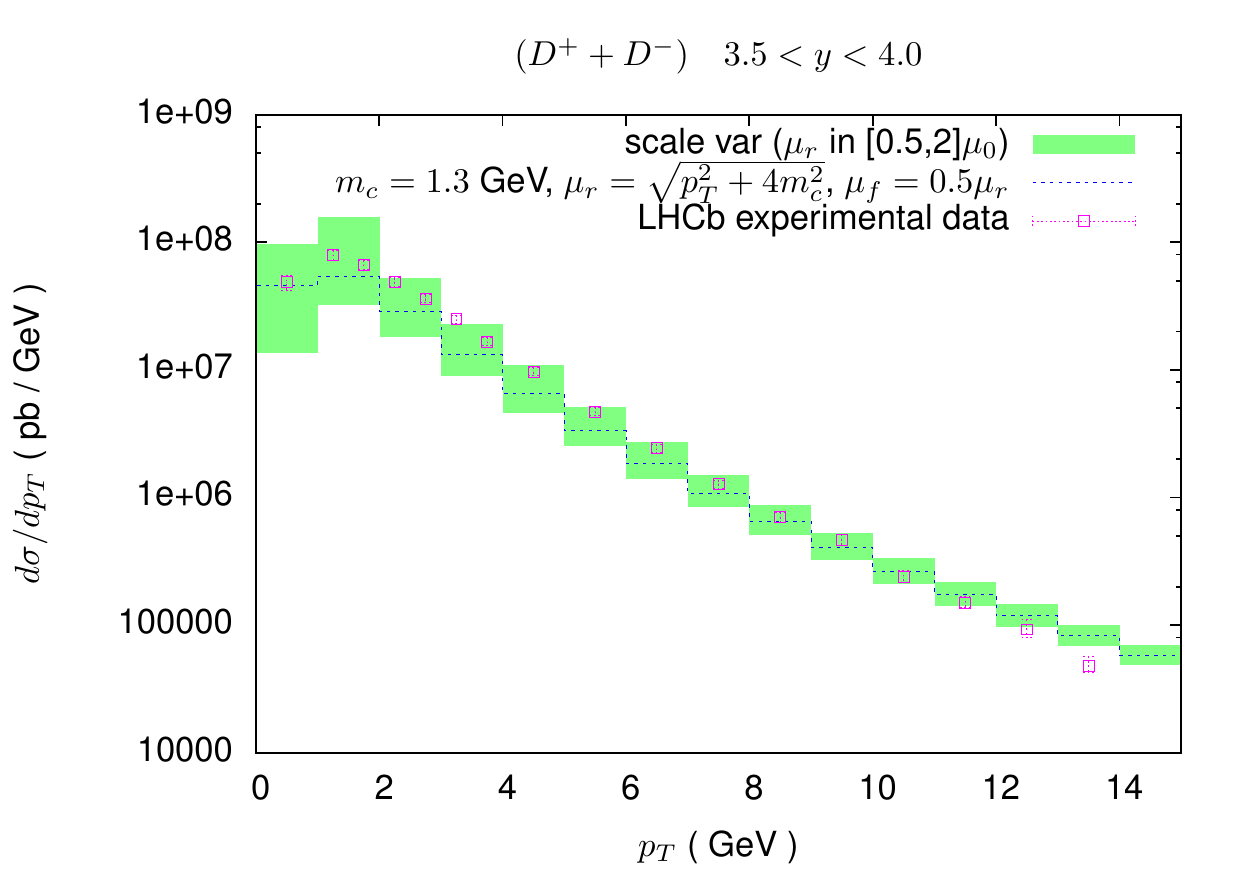}\\
\includegraphics[width=73mm]{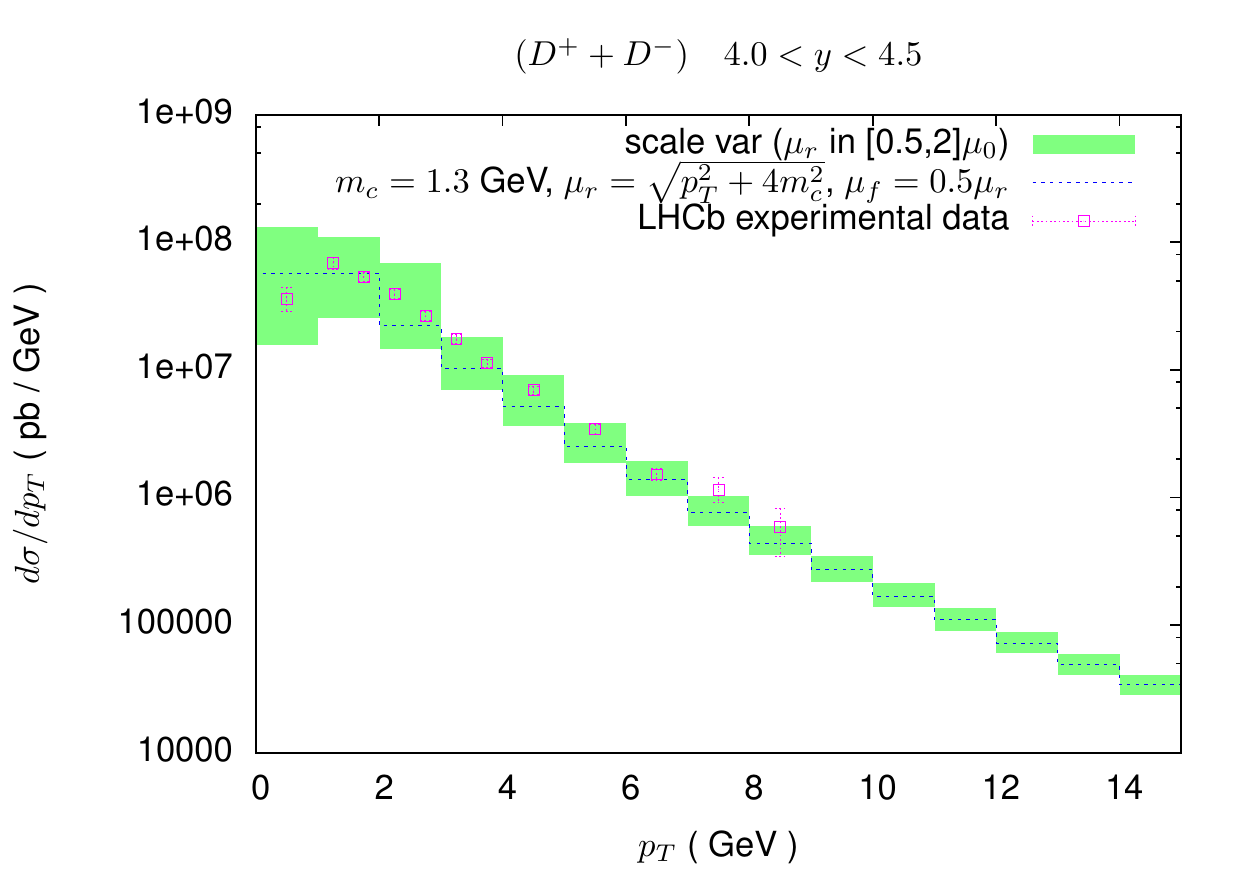}
\caption{Our GM-VFNS predictions for ($D^+$ + $D^-$) transverse-momentum distributions for $pp$ collisions at $\sqrt{S}$ = 13 TeV vs. LHCb experimental data of Ref.~\cite{Aaij:2015bpa}. Each panel corresponds to a different rapidity bin in the interval 2 $<$ $y$ $<$ 4.5. 
The scales are chosen as in figure~\ref{fig2}. 
}
\label{fig:lhcbcompare13}
\end{figure}

In figure \ref{fig:xii} our results for different choices of the parameters $\xi_i=\xi_f$, at fixed $\mu_r$, are compared to each other and to LHC experimental data at $7\,$TeV. Using $\mu_f=\mu_r/2$ leads to a suppression of the cross sections in the first bin, as observed in the experiment, while this suppression is not observed when adopting the $\mu_f$~=~$\mu_r$ choice. Additionally, it turns out that the data are better reproduced when using the $\mu_r=\sqrt{p_T^2+4m_c^2}$ functional form instead of the $\mu_r=\sqrt{p_T^2+m_c^2}$ one. This was already observed in case of FFNS calculations, where it can be motivated by the fact that charm quarks are always produced in pairs in the hard interaction, while in the ZM-VFNS a single charm can come out of the proton.
As a result of this method, the uncertainty due to scale variations is determined by varying only the renormalization scale $\mu_r$ but keeping the initial- and final-state factorization scales fixed at their best value. For our choice of parameters, we compare differential distributions for $(D^++D^-)$ hadroproduction in different rapidity bins and at $\sqrt{S}=5$, $7$ and $13\,$TeV to LHCb experimental data in figure \ref{fig:lhcbcompare5}, \ref{fig:lhcbcompare7} and \ref{fig:lhcbcompare13}, respectively.
We observe that the LHCb data are generally well reproduced, also for $\sqrt{S}=5$ and $13\,$TeV when taking into account the latest revisions of Refs.~\cite{Aaij:2016jht,Aaij:2015bpa}. PDF uncertainties will be discussed later. The corresponding plots including their effects can be found in Appendix \ref{sec:appendix}.

\begin{figure}[ht]
\centering
\includegraphics[width=73mm]{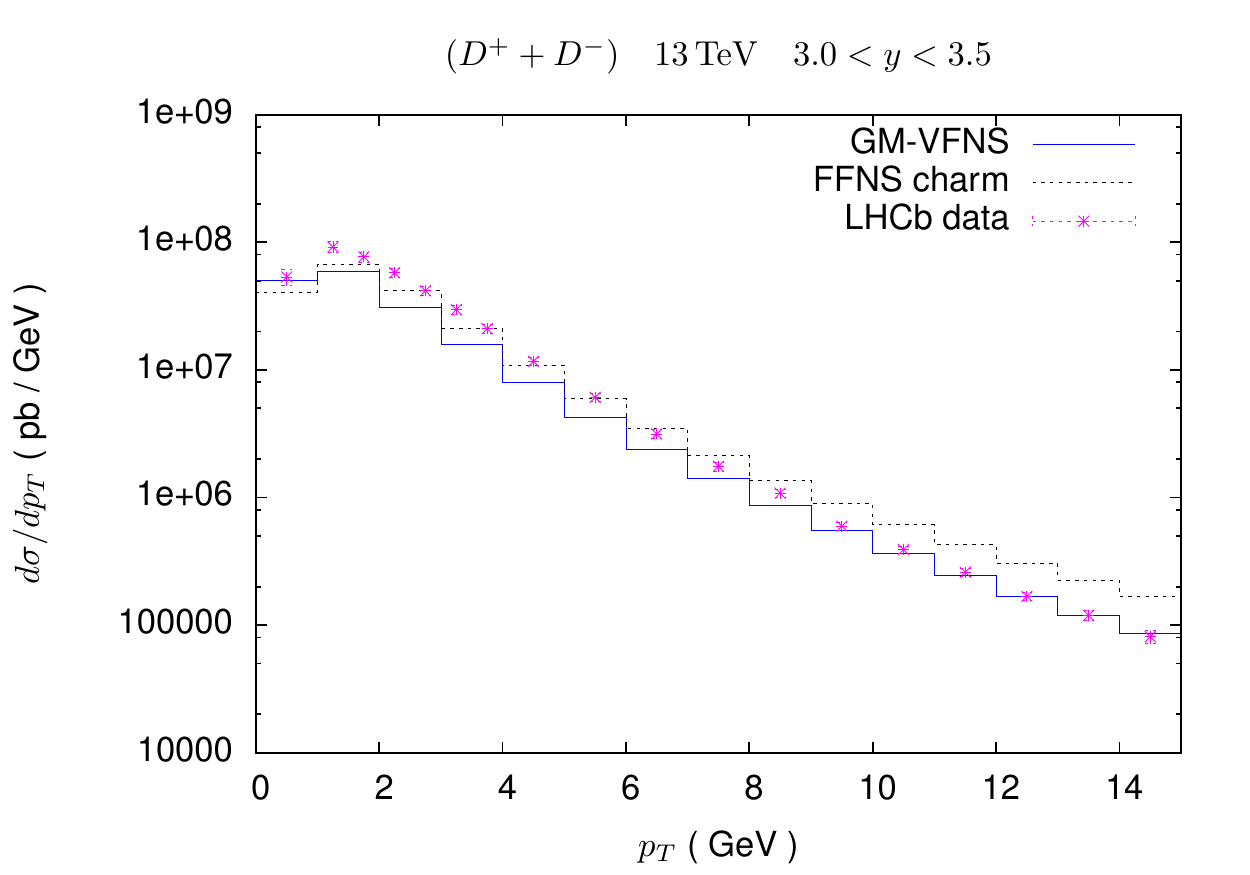}\quad\includegraphics[width=73mm]{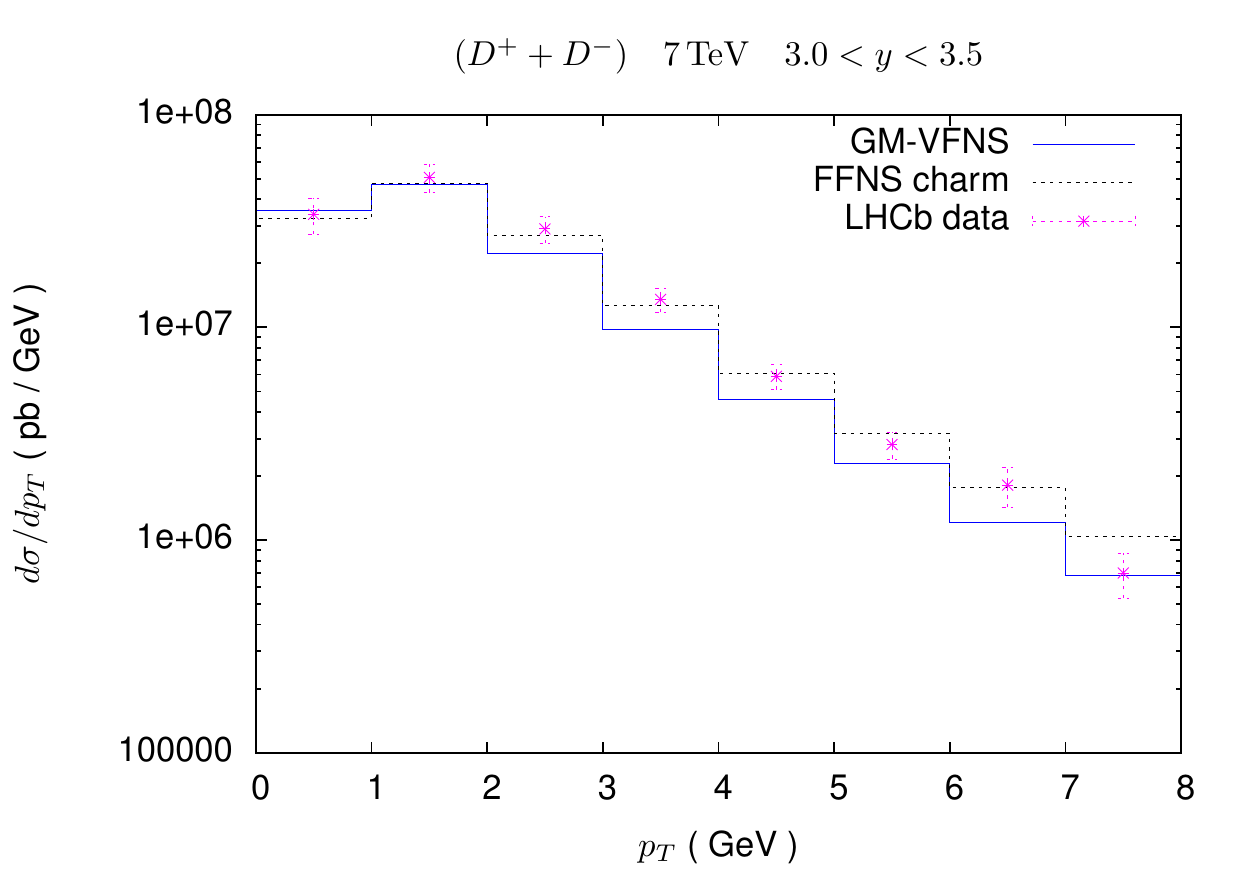}
\caption{Our GM-VFNS predictions for ($D^+$ + $D^-$) transverse-momentum distributions compared to the FFNS calculation of charm hadroproduction multiplied with the branching fraction for $c\to D^+$ and $D^-$ (see text for explanation). The figure refers to the $3<y<3.5$ rapidity bin with $\sqrt{S}=13\,$TeV (left) and $7\,$TeV (right).
}
\label{fig:gmnfnvsffn}
\end{figure}

In order to isolate the effect of the usage of a GM-VFNS instead of the FFNS, we compare in figure~\ref{fig:gmnfnvsffn} two of our GM-VFNS $p_T$ distributions to the corresponding ones obtained by using the FFNS (the same trend is observed for the distributions in the other LHCb rapidity bins).
In practice, the latter are calculated by switching off the the ZM contribution as well as the subtraction terms in eq.~(\ref{eqn:gmvfns}) and using the appropriate \texttt{CT14nlo\_NF3} PDF set. Furthermore, we do not include a fragmentation function in the FFNS calculation, since there exists no factorized expression corresponding to the one in eq.~(\ref{eqn:fact}) which would allow a systematic resummation of logarithms related to final-state collinear singularities. Instead, the FFNS result is multiplied with the branching fraction of the relevant $D$ meson. 
For the initial-state factorization scale, we adopt the natural value of $\mu_f=\sqrt{p_T^2+4m_c^2}$ in the FFNS calculation.
We note that both scheme choices yield similar results for very low $p_T$, while for larger $p_T$ they increasingly deviate (by about $50\,\%$ in the last bins). At $\sqrt{S}=13\,$TeV the experimental data at high $p_T$ agree much better with the GM-VFNS result, than with the FFNS result. This can be explained by the combined effect of resummation (through DGLAP evolution) and of the use of the non-pertubative FF with parameters fitted to experimental data at fixed scales.
It is possible to also introduce an FF in the FFNS, in order to phenomenologically account for hadronization effects. 
This purely phenomenological FF, lacking the universality that characterizes the GM-VFNS FFs, is thus far always determined by fitting a new non-perturbative input distribution each time the evolution for the systematic resummation of logarithms related to final-state collinear singularities crosses a heavy-flavor threshold.
Convoluting such a function with the partonic cross section tends to decrease the contributions at large $p_T$ while increasing the ones at small $p_T$, reflecting the fact that a FF gives rise to mesons produced at momentum fractions $z$ smaller than one. The exact shape of this phenomenological FF is determined by fitting a simple parametrization (e.g. Peterson) to experimental data at a certain energy scale. By using an appropriate choice of the parameters, the FFNS result can be made to resemble the GM-VFNS very closely, even though the FF in the FFNS is not run to the relevant energy by DGLAP evolution. Therefore, the effect of resummation cannot be disentangled from the effects of hadronization by comparing the GM-VFNS predictions to the ones obtained in the FFNS complemented by a fragmentation function.

Before presenting our results for the prompt-neutrino fluxes, we will briefly review in the next section how the inclusive meson production cross section and atmospheric fluxes can be related.

\section{Astrophysical application of the GM-VFNS approach to the determination of prompt-neutrino fluxes}
\label{sec:astro}

\subsection{Methodology for computing fluxes and astrophysical input}
\label{sec:equa}

The evolution of particle fluxes in the atmosphere can be described by a system of coupled differential equations \cite{gaisserNewBook,Gaisser:1990vg,Lipari:1993hd,Pasquali:1998ji,Garzelli:2015psa,Gauld:2015kvh}, also known as cascade equations,
\begin{eqnarray}
\frac{d\phi_j(E_j,X)}{dX} & = & - \frac{\phi_j(E_j,X)}{\lambda_j^\mathrm{int}(E_j)} 
- \frac{\phi_j(E_j,X)}{\lambda_j^\mathrm{dec}(E_j)} + \nonumber\\
& + & \sum_{k\ne j} S_\mathrm{prod}^{k\rightarrow j}(E_j, X) + \sum_{k\ne j} S_\mathrm{decay}^{k\rightarrow j}(E_j, X) + S_\mathrm{reg}^{j \rightarrow j}(E_j,X) \,.
\end{eqnarray}
Here, $j$ denotes a particle species with flux $\phi_j$ and $X (l,\theta)$ = $\int_l^{+\infty} dl^\prime \rho[h(l^\prime,\theta)]$ is the slant depth traversed by the particle while moving from the top of the atmosphere along a trajectory with an angle $\theta$ with respect to the zenith, down to a point with a distance $l$ from the Earth's surface. The atmospheric profile as a function of the altitude is supposed to have an exponential form $\rho(h) = \rho_0$ exp(-$h$/$h_0$) (isothermal model), with scale height $h_0$ = 6.4 km and $\rho_0$ = 2.03~$\cdot$~10$^{-3}$ g/cm$^3$, as appropriate for the stratosphere. $E$ is the particle energy, $\lambda_j^\mathrm{int}$ and $\lambda_j^\mathrm{dec}$ are its interaction and decay lengths, $S_\mathrm{prod}$, $S_\mathrm{decay}$, and $S_\mathrm{reg}$ denote the generation functions for production, decay and regeneration of this particle, which, under the assumption that the $X$ dependence of fluxes factorizes from the energy dependence, can be rewritten in terms of $Z$ moments as
\begin{align}
S_\mathrm{prod}^{k \rightarrow j} (E_j, X) &\simeq Z_{kj}^\mathrm{int}(E_j) {\phi_k(E_j, X)}/{\lambda_k^\mathrm{int}(E_j)}\,,\nonumber\\ 
S_\mathrm{decay}^{k \rightarrow j} (E_j, X) &\simeq~Z_{kj}^\mathrm{dec}(E_j) {\phi_k(E_j, X)}/{\lambda_k^\mathrm{dec}(E_j)}\,, \nonumber\\
S_\mathrm{reg}^{j \rightarrow j} (E_j, X) &\simeq~Z_{jj}^\mathrm{int}(E_j) {\phi_j(E_j, X)}/{\lambda_j^\mathrm{int}(E_j)}\,. 
\end{align}
The $Z$ moments for production and decay are defined as
\begin{eqnarray}
Z_{kj}^\mathrm{int}(E_j) & = & \int_{E_j}^{+\infty} d E^\prime_k \frac{\phi_k(E_k^\prime, 0)}{\phi_k(E_j, 0)} \frac{\lambda_k^\mathrm{int}(E_j)}{\lambda_k^\mathrm{int}(E_k^\prime)}\frac{dn(kA \rightarrow j X; E_k^\prime, E_j)}{dE_j} \, ,\label{eqn:Zint}\\
Z_{kj}^\mathrm{dec}(E_j) & = & \int_{E_j}^{+\infty} d E^\prime_k \frac{\phi_k(E_k^\prime, 0)}{\phi_k(E_j, 0)} \frac{\lambda_k^\mathrm{dec}(E_j)}{\lambda_k^\mathrm{dec}(E_k^\prime)}\frac{dn(k \rightarrow j X; E_k^\prime, E_j)}{dE_j} \,.
\end{eqnarray}
In these expressions, $dn$ is the number of particles with energy between $E_j$ and $E_j + dE_j$ produced during the interaction/decay of particle $k$ with energy $E_k^\prime$, and $A$ denotes the mass number of an air nucleus. 
In this paper, it is assumed that the average mass number of an air nucleus is $\langle A\rangle$ = 14.5 and that the interaction of a primary CR with an air nucleus leading to the production of a hadron $h$ can be approximated by a linear superposition of $pp$ interactions according to the superposition model, neglecting shadowing effects. 
This approach is supported by the observation that the nuclear modification factor for $D$-meson production in proton-lead collisions is, in fact, close to unity, as measured by the ALICE Collaboration~\cite{Adam:2016ich}. For lighter nuclei, an even smaller effect is expected.
Thus $dn/dE_j$ of eq.~(\ref{eqn:Zint}) can be rewritten in terms of $pp$ interaction cross sections according to the formula ${dn(pA \rightarrow h X; E_k^\prime, E_j)}/{dE_j}$~=~$({\langle A\rangle}/{\sigma_{pA}(E_k^\prime)})\cdot {d\sigma(pp\rightarrow h X; E_k^\prime, E_j)}/{dE_j}$.
In\-tro\-du\-cing the scaling variable $x_E$ = $E_j/E_k^\prime$ and considering the limit where the interaction lengths are energy independent, the $Z$ moments for the production of the hadron $h$ can be rewritten as
\begin{eqnarray}
Z_{ph}^\mathrm{int} (E_j) = \int_0^1 \frac{d x_E}{x_E} \frac{\phi_k(E_j/x_E)}{\phi_k(E_j)}
\frac{\sigma_{pA}(E_k^\prime)}{\langle A\rangle} \frac{d\sigma_{pp \rightarrow hA} (E_j/x_E)}{d x_E}\,.
\end{eqnarray}
Analogously, the $Z$ moment for the decay of the hadron $h$ producing a lepton $\ell$ with energy $E_\ell$ can be written as an integral over $x_E^\prime = E_\ell/E_j$, 
\begin{eqnarray}
Z_{hl}^\mathrm{dec} (E_\ell) = \int_0^{1} d x_E^\prime \frac{\phi_h(E_\ell/x_E^\prime)}{\phi_h(E_\ell)}F_{k\rightarrow l}(x_E^\prime)\,,
\end{eqnarray}
taking into account that $dn(h \rightarrow l X; E_\ell, E_j)/dE_\ell = F_{h\rightarrow l} (E_\ell/E_j)/E_j$, where $F_{h\rightarrow l}$ is the energy spectrum of $\ell$ in the rest frame of the hadron $h$.
Prompt-neutrino fluxes originate from the decay of heavy hadrons. In this work, we focus on the case of charmed hadrons, by considering prompt neutrinos generated by the decay of $h$~=~$h_c$~=~$D^0, \bar{D^0}, D^\pm, D_s^\pm, \Lambda_c^\pm$ states, produced in the reactions $pp$ $\rightarrow$ $h_c$ $X$. 
Other charmed hadrons have even smaller branching fractions and can be neglected, as well as the contribution of bottom-flavored states promptly decaying into charmed ones, considering that the cross section for inclusive hadroproduction of $b$ quarks is by far smaller than that for $c$ quarks. The regeneration $Z$ moments, $Z_{pp}$ and $Z_{hh}$, also play a role in the evolution equations. In this work, we compute them as in our previous work \cite{Garzelli:2015psa}. 

The evolution equations admit analytic solutions in the limit where the energy of the intermediate hadron $h$ with mass $m_h$ and proper lifetime $\tau_{0,h}$  
is either very small or very large with respect to its critical energy.
This critical energy represents the energy above which the hadron decay probability is suppressed with respect to its interaction probability, thus separating the low-energy and high-energy regimes. It is defined in vertical direction as $E_\mathrm{crit}^h = m_h c^2 h_0/c\tau_{0,h}$ and depends on the atmospheric density profile through $h_0$. The solutions are
\begin{eqnarray}
\phi_\ell^{h, \mathrm{low}} (E_\ell) & = & Z_{hl}^\mathrm{low} (E_\ell) \frac{Z_{ph} (E_\ell)}{1-Z_{pp}(E_\ell)} \phi_p(E_\ell,0) \, , \\
\phi_\ell^{h, \mathrm{high}} (E_\ell) & = & Z_{hl}^\mathrm{high} (E_\ell) \frac{Z_{ph} (E_\ell)}{1-Z_{pp}(E_\ell)} 
\frac{E_\mathrm{crit}^h}{E_\ell} \frac{\ln(\Lambda_h(E_h)/\Lambda_p(E_p))}{1-\frac{\Lambda_p(E_p)}{\Lambda_h(E_h)}}f(\theta)\phi_p(E_\ell,0)\,. 
\end{eqnarray}
Here $\phi_p(E,0)$ is the flux of primary CR protons entering the upper layer of the atmosphere ($X$~=~0),  $\Lambda_i(E)$ is an effective interaction length, defined as $\Lambda_i(E)=\lambda_i(E)/[1-Z_{ii}(E)]$, and $Z_{hl}^\mathrm{low}$ and $Z_{hl}^\mathrm{high}$ are the limits of $Z_{hl}$ for $E \ll E_\mathrm{crit}$ and $E \gg E_\mathrm{crit}$. 
While the low-energy solution is isotropic, the high-energy solution has a dependence on the zenith angle encoded in the function $f(\theta)$. The solution in the intermediate energy range can be approximated by the geometrical interpolation $\phi^h_\ell (E_\ell)=(\phi_\ell^{h,\mathrm{low}}(E_\ell) \phi_\ell^{h, \mathrm{high}}(E_\ell))/(\phi_\ell^{h,\mathrm{low}}(E_\ell) + \phi_\ell^{h,\mathrm{high}}(E_\ell))$. The total neutrino flux is obtained by summing the contributions due to all intermediate hadron production and decay processes $\phi_\ell (E_\ell)$ = $\sum_h \phi^h_\ell (E_\ell)$.

As for primary CR fluxes, we consider the fits provided by Gaisser et al. in Ref.~\cite{Gaisser:2013bla} and one of the more recent fits by the Nijmegen group~\cite{Thoudam:2016syr}. The Gaisser~et~al. fits include spectra labeled in the literature as GST-3 and GST-4, H3a and H3p. The H3p and H3a fits include three populations of CRs, two of galactic and one of extra-galactic origin, characterized by different rigidities, and involving protons and different nuclear groups (He, CNO, Mg-Si, Fe) with different spectral indices. They differ because of the composition of the third population, which, in the case of H3p, is supposed to be made of protons only. 
The GST-3 fit includes three populations as well, involving the $p$, He, C, O, Fe nuclear groups, whereas the GST-4 fit involves an additional fourth population of extra-galactic origin, including only $p$ with large rigidity. On the other hand, the Nijmegen group made a study of the sources and propagation of CRs by means of astrophysical models. This study takes into account very recent CR data, provided by different experiments (KASCADE, IceTop, Tibet III, HiRes-II and the Pierre Auger Observatory). Among the different variants of the fit presented in Ref.~\cite{Thoudam:2016syr}, we consider the one (labeled in~\cite{Thoudam:2016syr} as ``WR-CRs (C/He=0.4) + EG-UFA") with two galactic components, one produced in supernova remnants and the other produced by the explosion of Wolf-Rayet stars (with a Carbon/Helium ratio of 0.4), and an extra-galactic component according to the extra-galactic ankle model by Unger et al.~\cite{Unger:2015laa}. This variant predicts the CR composition between the second knee and the ankle in good agreement with results from the Pierre Auger Collaboration~\cite{Aab:2015bza}. In particular, in the energy region between $10^6$ and $10^8\,$GeV the composition predicted by this fit is dominated by Helium and other light elements. Additionally, we consider the broken-power-law all-nucleon spectrum, $\phi_p(E,0) = 1.7\, (E/\text{GeV})^{-2.7}\,$cm$^{-2}$s$^{-1}$sr$^{-1}$GeV$^{-1}$ for $E < 5 \cdot 10^6\,$GeV and $\phi_p(E,0) = 174\, (E/\text{GeV})^{-3}\,$cm$^{-2}$s$^{-1}$sr$^{-1}$GeV$^{-1}$ for $E > 5 \cdot 10^6\,$GeV, introduced se\-veral years ago, as a reference spectrum for comparison with previous works, although its high-energy part is nowadays known to overestimate the CR flux measured by Extended Air Shower experiments (EAS).
  
Another input entering the $Z_{ph}$ moments is the total inelastic $p$-Air cross section as a function of the collision energy. For the latter we consider the {\texttt{QGSJet0.1c}} \cite{Kalmykov:1993qe} predictions, as available inside the {\texttt{CORSIKA}} package \cite{Heck:1998vt}, which turn out to be in agreement with the measurement at the Pierre Auger Observatory at $\sqrt{S}$~=~57~TeV \cite{Collaboration:2012wt}. We observe that reasonable variations in this input, obtained e.g. by considering other models in {\texttt{CORSIKA}}, such as {\texttt{SYBILL2.1}} \cite{Ahn:2009wx}, have a minimal impact on the spectra of prompt-neutrino fluxes, with variations within very few percent~\cite{Garzelli:2015psa}.  

The core of the computation of prompt-neutrino fluxes for neutrino energies in the range [100 GeV, $10^8$ GeV] is represented by the estimate of the $d\sigma/dx_E$ distributions for $pp$ $\rightarrow$ $h$ + $X$ in the $E_{p,\,\mathrm{lab}}$ energy range [100 GeV, 5 $\cdot$ $10^{10}$ GeV]. For this purpose the GM-VFNS approach described in Section~\ref{sec:gmvfns} is used in this paper. 
Predictions for $d\sigma/dx_E$ are presented and discussed in Subsection~\ref{sec:spectra}, whereas we report predictions for prompt-neutrino fluxes in Section~\ref{sec:fluxes}. 

\subsection{QCD input in the GM-VFNS}
\label{sec:spectra}

Our predictions are based on the numerical integration of the factorization formula~(\ref{eqn:fact}) using the {\texttt{CT14nlo}} PDF fit and the {\texttt{KKKS08}} NLO FFs. In the context of high-energy physics at colliders, the cross sections are usually given as differential in the transverse momentum $p_T$ and the rapidity $y$ of the hadron evaluated in the center-of-mass frame. For the use in the cascade equations, we need to consider the laboratory frame, where one initial proton with mass $m_p$ is at rest while the other has the energy $E_{p,\,\mathrm{lab}}\approx S/(2m_p)$, which corresponds to the shift
\begin{equation}
y\to y+\frac{1}{2}\ln\frac{m_p^2}{S}\,,
\end{equation}
and to perform a variable transformation to the final hadron energy $E_h$ and the polar angle $\theta$ of the final hadron momentum with respect to the beam axis,
\begin{align}
E_h =&\, \sqrt{p_T^2+m_h^2}\cosh y\,, \\
\tan\theta =&\, \frac{p_T}{\sqrt{p_T^2+m_h^2}\sinh y}\,.
\end{align}
Finally, we have to integrate over all angles $\theta$. 

\begin{figure}[ht]
\centering
\includegraphics[width=120mm]{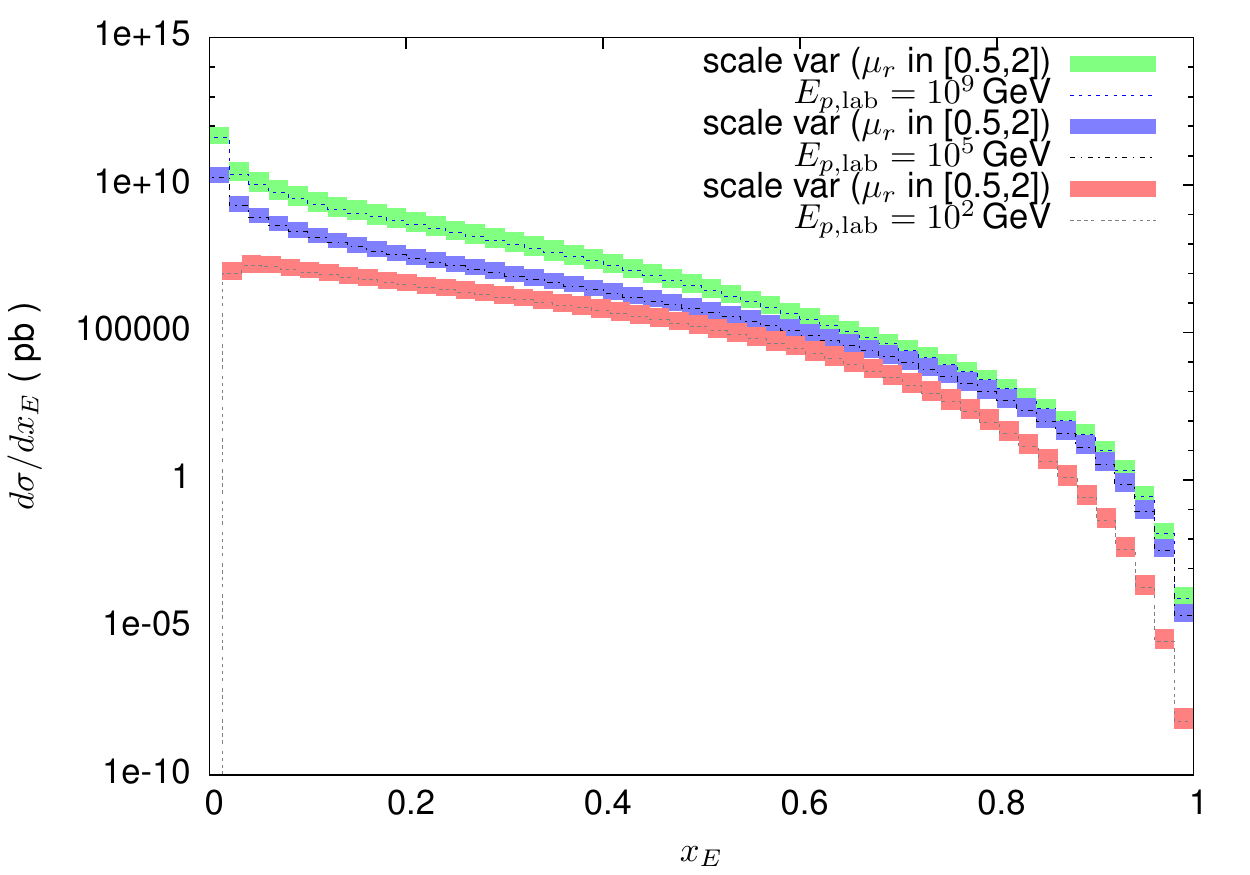}
\caption{Differential cross sections d$\sigma$/d$x_E$ for inclusive $D^0$ hadroproduction, $pp$ $\rightarrow$ $D^0 + X$, for initial-state protons with laboratory energies $E_{p,\,\text{lab}}=10^2\,$GeV, $E_{p,\,\text{lab}}=10^5\,$GeV and $E_{p,\,\text{lab}}=10^9\,$GeV. Central values correspond to the choice of scales $\mu_r=\sqrt{p_T^2+4m_c^2}$ and $\mu_i=\mu_f=\xi\mu_r$ with $\xi=0.5$, the charm mass fixed to $m_c=1.3\,$GeV, and the central set of the {\texttt{CT14nlo}} PDF fit. The uncertainty bands correspond to the $\mu_r$ scale variation explained in the text.}
\label{fig:1E2_1E5_1E9}
\end{figure}

In figure \ref{fig:1E2_1E5_1E9}, we plot the differential cross section $d\sigma/dx_E$ for $D^0$ hadroproduction, where $x_E$ is the ratio of the energy of the final-state meson and that of the incoming protons, all evaluated in the laboratory frame, for three different incoming-proton energies. The uncertainty bands correspond to the variation of the renormalization scale in the [$1/2$, $2$]$\mu_0$ interval around the central value $\mu_0$ = $\sqrt{p_T^2+4m_c^2}$.

\begin{figure}[ht]
\centering
\includegraphics[width=120mm]{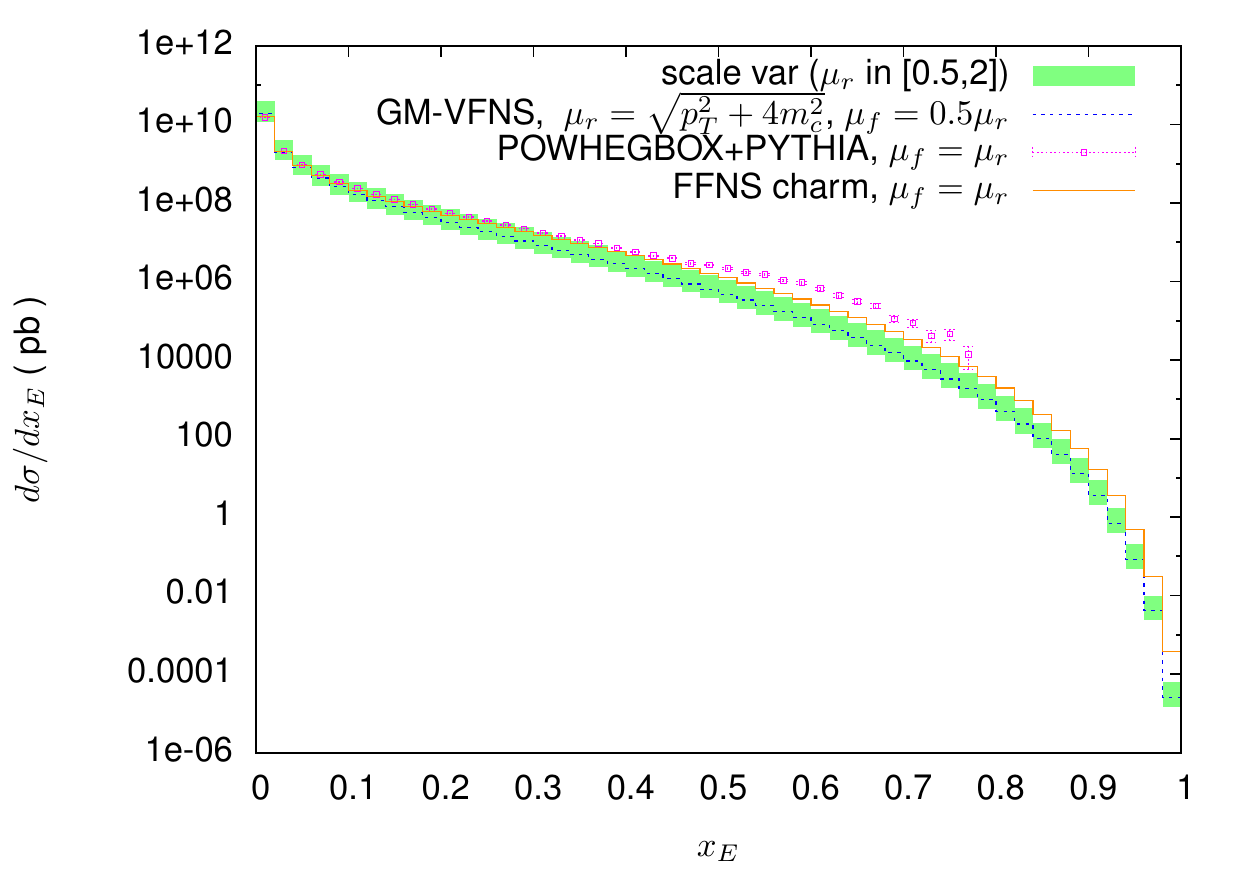}
\caption{Differential cross sections d$\sigma$/d$x_E$ for $D^0$ hadroproduction for colliding protons with a laboratory energy $E_{p,\,\text{lab}}=10^5\,$GeV (corresponding to center-of-mass energy $\sqrt{S}\approx 433\,$GeV). For the GM-VFNS prediction, we use the {\texttt{CT14nlo}} PDF, $m_c=1.3\,$GeV, and factorization scales $\mu_i=\mu_f=0.5\,\mu_r=0.5\sqrt{p_T^2+4m_c^2}$, where $p_T$ is the hadronic transverse momentum. The \texttt{POWHEGBOX+PYTHIA} and the FFNS predictions on the other hand are calculated using the natural scale choice $\mu_i=\mu_f=\mu_r=\sqrt{p_T^2+4m_c^2}$, with $p_T$ representing the charm momentum.
}
\label{fig:D01E5}
\end{figure}

As expected, the cross sections go to zero for $E_h\to E_{p,\,\mathrm{lab}}$ (i.e. $x_E$ $\to$ 1). For small energies, the cross section peaks and then quickly vanishes for $E_h\to m_h$. This is most noticeable at $E_{p,\,\mathrm{lab}}=10^2\,$GeV, while, for larger energies, the normalization of $x_E$ to $1$ moves the peak too close to zero. The location of the peak is related to the fact that, in the laboratory frame, the largest hadron energies $E_h$ correspond only to particles produced at small $p_T$ (i.e. in the forward region), while, at small $E_h$, the whole $p_T$ range contributes.
In figure \ref{fig:D01E5}, we compare the $d\sigma/d x_E$ distribution for $D^0$ hadroproduction in $pp$ collisions at laboratory energy $E_{p,\,\text{lab}}=10^5\,$GeV to the central value of the same distribution obtained using \texttt{POWHEGBOX+PYTHIA} and the FFNS without an FF, as explained in the context of $p_T$ distributions at the end of Section~\ref{sec:gmvfns}.
We note that the predictions start to deviate for large energies, with \texttt{POWHEGBOX+PYTHIA} being the largest.
This is expected, since, if the charm quark is produced in the forward region, it can recombine with parts of the target remnant to form the charmed meson, as already observed in Ref.~\cite{Bhattacharya:2016jce}. Such an effect is not included in the factorized approach using FFs, which are fitted to $e^+e^-$ data. A Monte Carlo event ge\-ne\-ra\-tor, such as \texttt{PYTHIA}, on the other hand, implements such effects in its hadronization model \cite{Norrbin:1998bw,Norrbin:2000zc}. At small energies, there is good agreement between all predictions when one uses the standard choice $\xi_f=1$ in the \texttt{POWHEGBOX+PYTHIA} and FFNS method, while in the GM-VFNS $\xi_f=0.5$ allows one to regulate the divergence, as explained above.
The difference between the GM-VFNS and the FFNS is due to fragmentation and the resummation of logarithms, and can be significantly reduced by use of a phenomenological FF in the FFNS calculation.

Apart from the scale uncertainty, we also consider the uncertainty due to the PDF choice. We restrict ourselves to the use of the {\texttt{CT14nlo}} PDF fit and estimate the PDF uncertainty from its member sets, according to the prescription in Ref.~\cite{Lai:2010vv}. It turns out that PDF uncertainties are most pronounced at large collision energies, since there the smallest $x$ region is probed, where the data constraining PDF fits are scarce or still completely absent. In the following section, we will present our predictions for the prompt-neutrino fluxes and include a thorough discussion of the PDF uncertainties at that level.

\section{Prompt-Neutrino Fluxes}
\label{sec:fluxes}

\subsection{GM-VFNS theoretical predictions and their uncertainties}

\begin{figure}[ht]
\begin{center}
\includegraphics[width=0.49\textwidth]{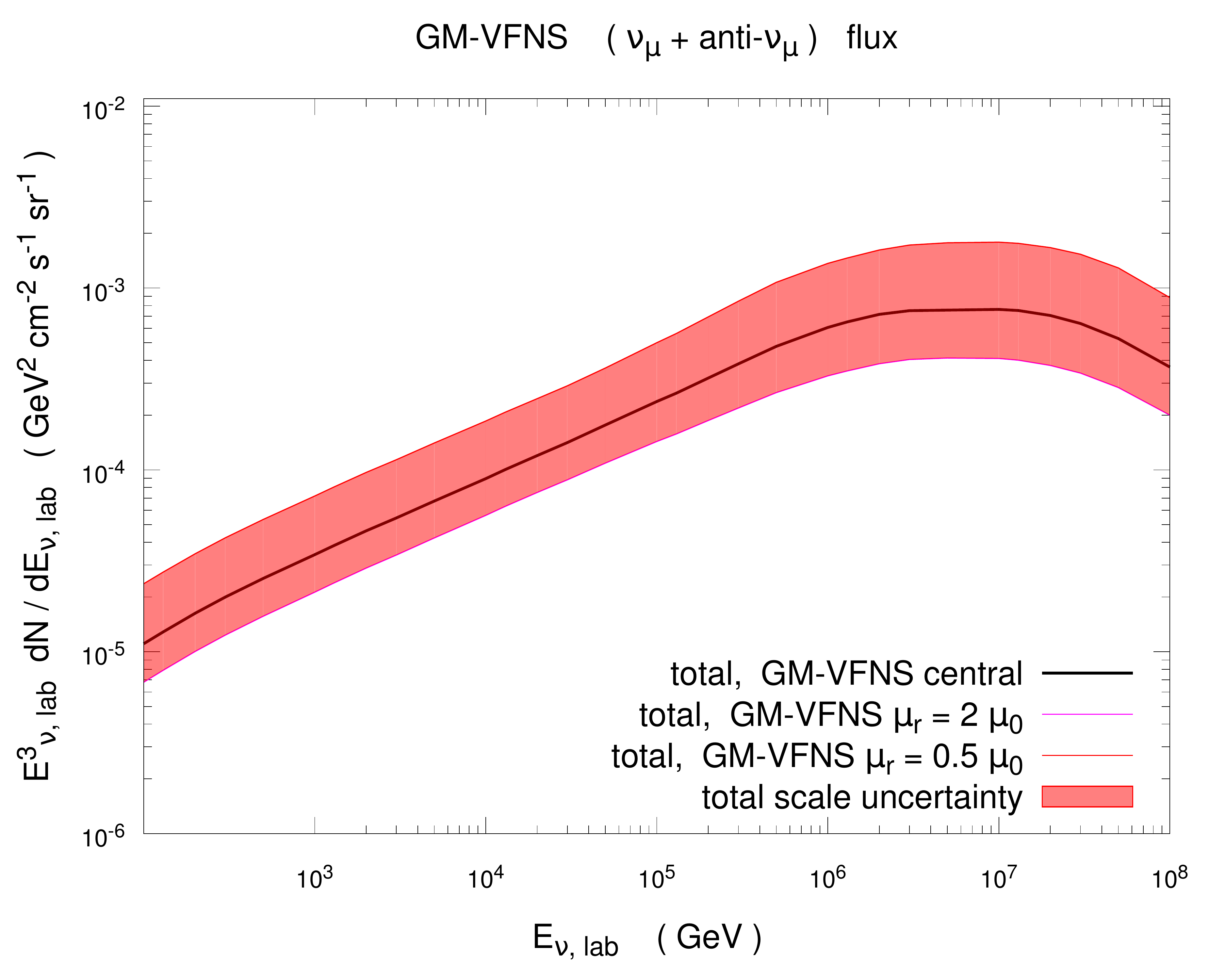}
\includegraphics[width=0.49\textwidth]{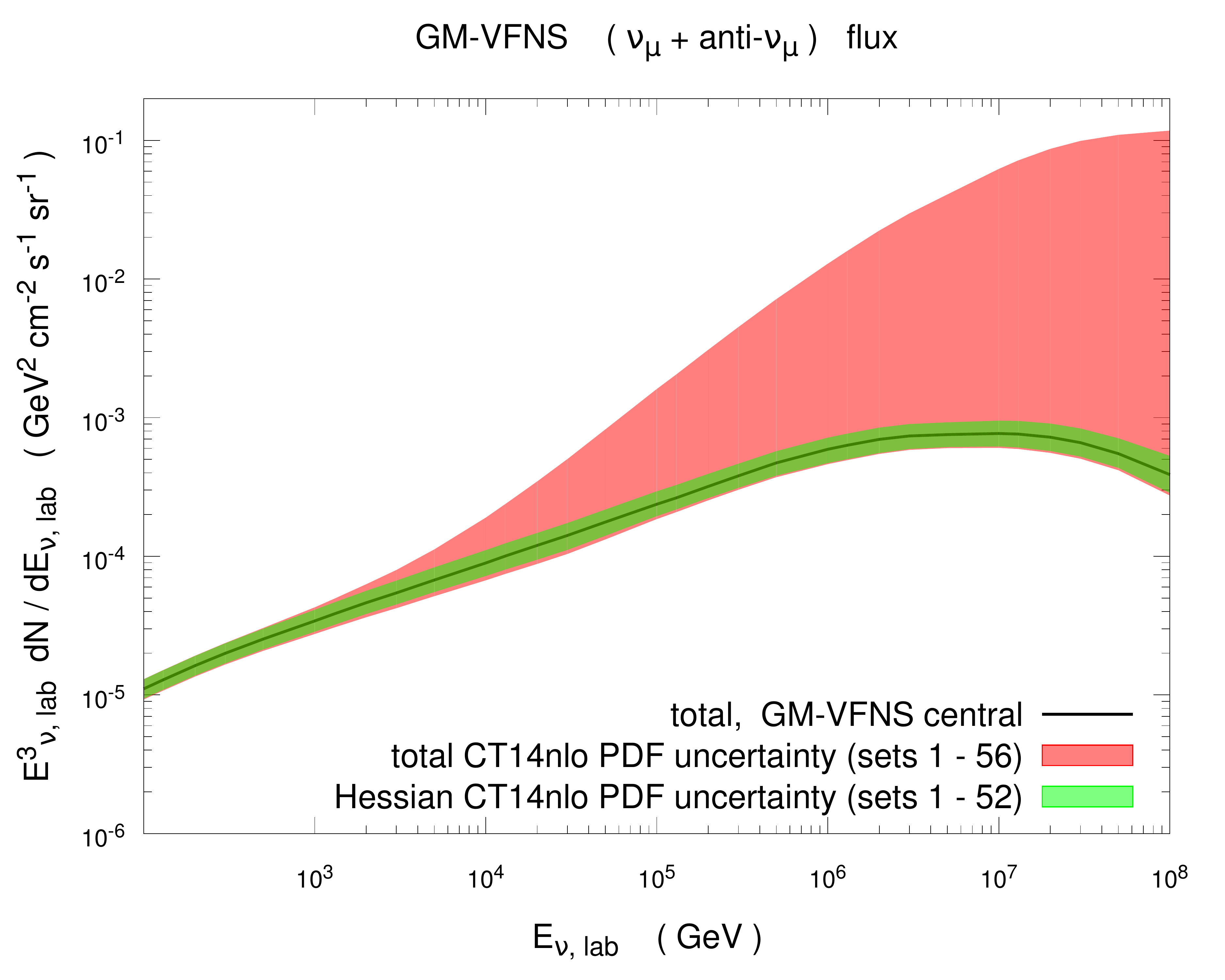}
\caption{Predictions for prompt-($\nu_\mu$ + $\bar{\nu}_\mu$) fluxes as functions of neutrino energy $E_\nu$, according to the GM-VFNS computation presented in this paper. Renormalization scale uncertainties are shown in the left panel, whereas PDF uncertainties are shown in the right panel, separating the contribution due to the Hessian sets (1--52) from the total one due to all 56 PDF sets included in the {\texttt{CT14nlo}} fit. The broken-power-law CR primary input spectrum is used for these plots. See text for more details.}
\label{scalepdf}
\end{center} 
\end{figure}

In figure~\ref{scalepdf}, we show predictions for prompt-($\nu_\mu$ + $\bar{\nu}_\mu$) fluxes using the broken-power-law primary CR flux, with their uncertainties due to scale and PDF variations, respectively. 
PDF uncertainties are evaluated according to the prescription of Ref.~\cite{Lai:2010vv} by considering the 56 sets available in the {\texttt{CT14nlo}} fit~\cite{Dulat:2015mca}, whereas scale uncertainties are evaluated according to the same criterion used for producing the $p_T$ differential cross sections compared with LHCb experimental data in Section~\ref{sec:gmvfns}. It is evident that the width of the scale uncertainty band on logarithmic scale is approximately constant over the  whole $E_\nu$ interval, whereas the width of the PDF uncertainty band increases with energy and actually blows up at the highest energies. This is due to the behavior of the {\texttt{CT14nlo}} PDF fit in the pair of error sets 53 and 54, corresponding to extreme sets for low-$x$ gluons and quarks, complemented by the 55 and 56 pair of sets, corresponding to extreme sets for strange quarks, as illustrated in the right panel of figure~\ref{scalepdf}, where the contribution of the uncertainty due to the 1--52 Hessian error sets is disentangled from the one due to all 56 error sets.\footnote{In fact, the {\texttt{CTEQ}} Collaboration extracted their central PDF set by the minimization of a log-likelihood $\chi^2$ function, quantifying the agreement between theory predictions and the experimental data used in their fit, complemented by additional sensible ``prior'' assumptions about the forms of PDFs. The boundaries of the 90\% C.L. region around the minimum $\chi^2$ were extracted by an iterative diagonalization of the Hessian matrix. This corresponds to the uncertainty encoded in the 1--52 PDF error sets. Considering that experimental data used to build the $\chi^2$ do not cover the low-$x$ region, the Hessian sets were complemented by four additional sets, obtained using the Lagrange multiplier method: one with enhanced, one with suppressed gluon at low $x$ values, one with enhanced and one with suppressed strangeness at low $x$ values.}
The significant uncertainty at the highest energies reflects the fact that experimental data is missing in that region to sufficiently constrain the parton densities.

\begin{figure}
\begin{center}
\includegraphics[width=0.49\textwidth]{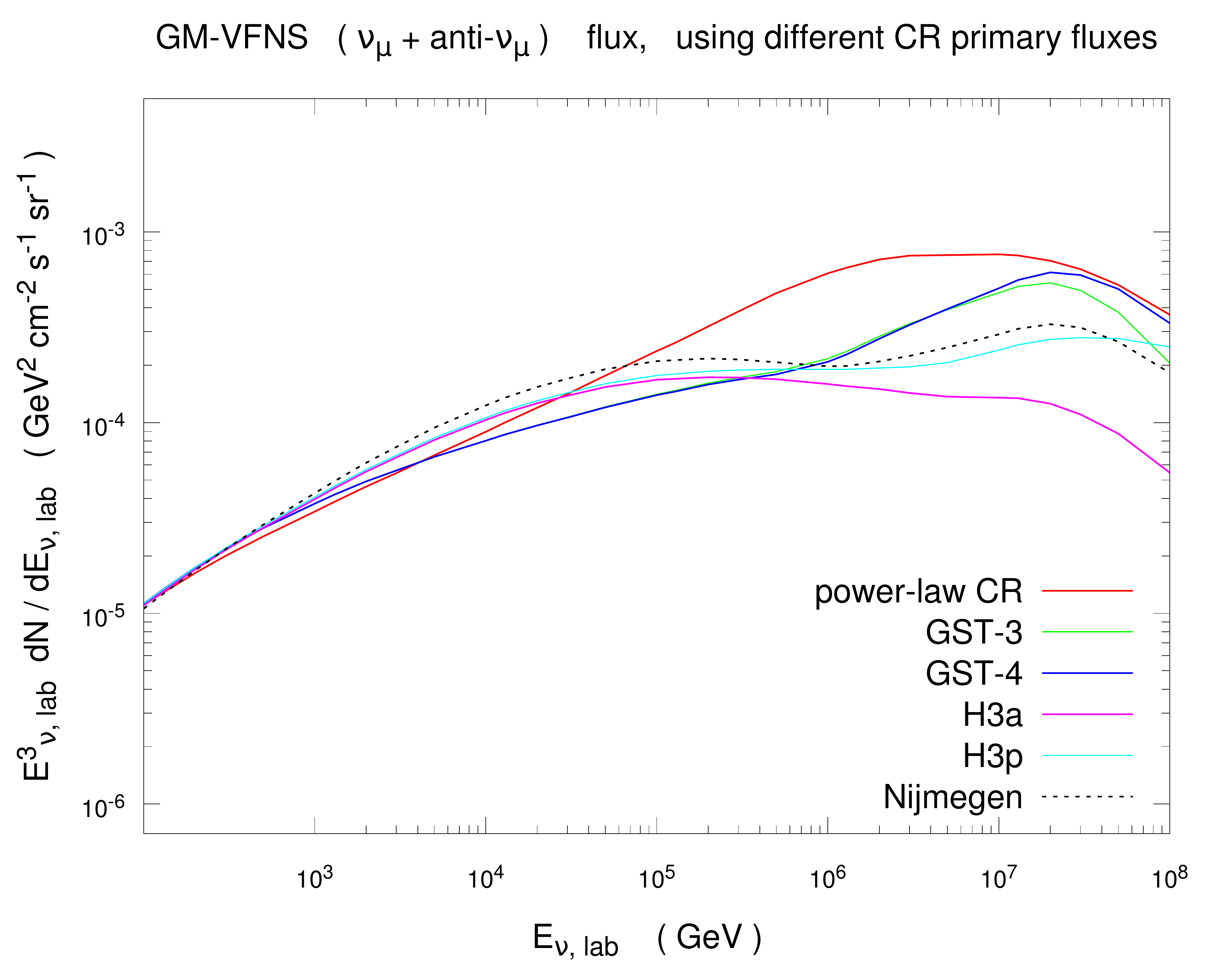}
\includegraphics[width=0.49\textwidth]{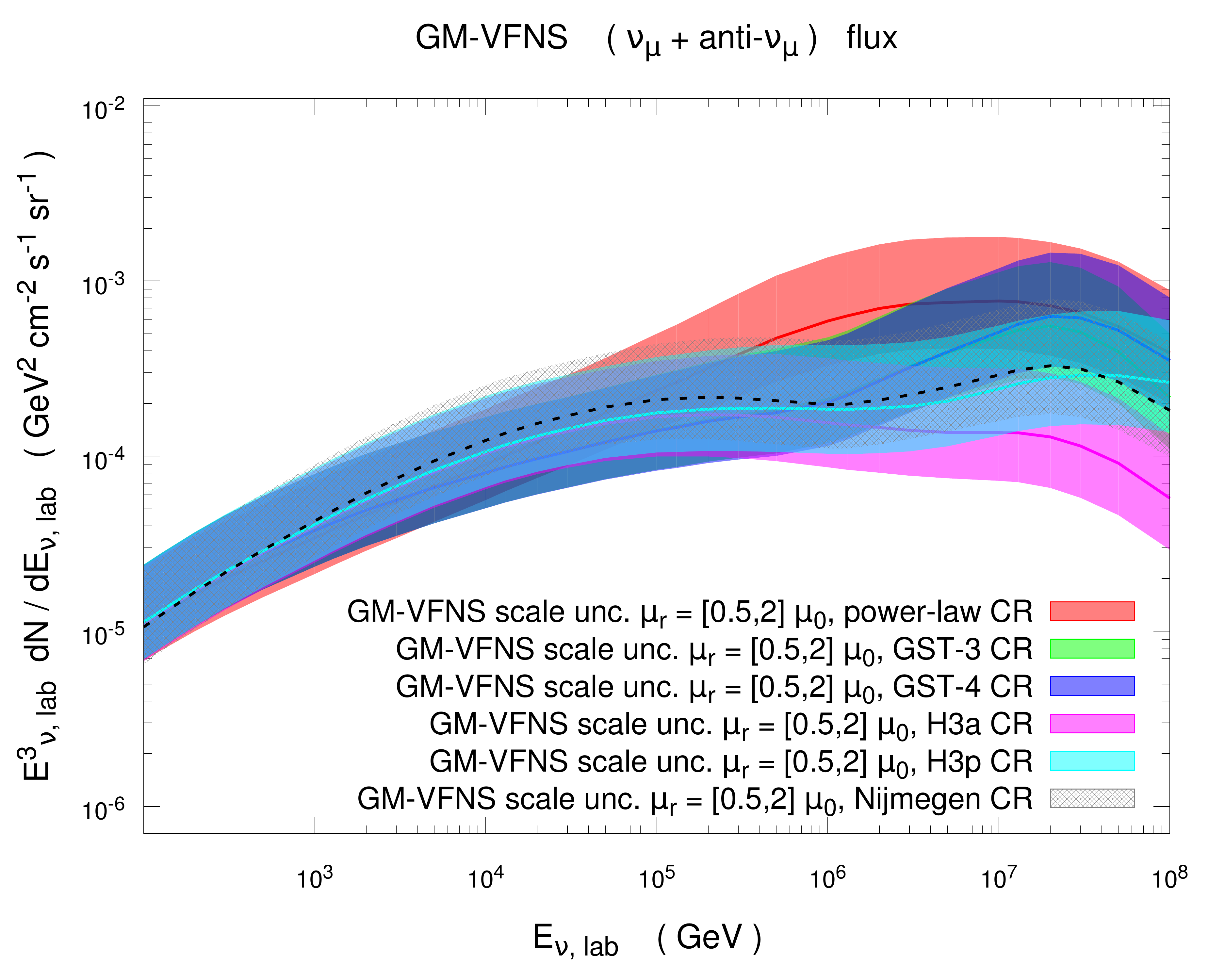}\\
\includegraphics[width=0.49\textwidth]{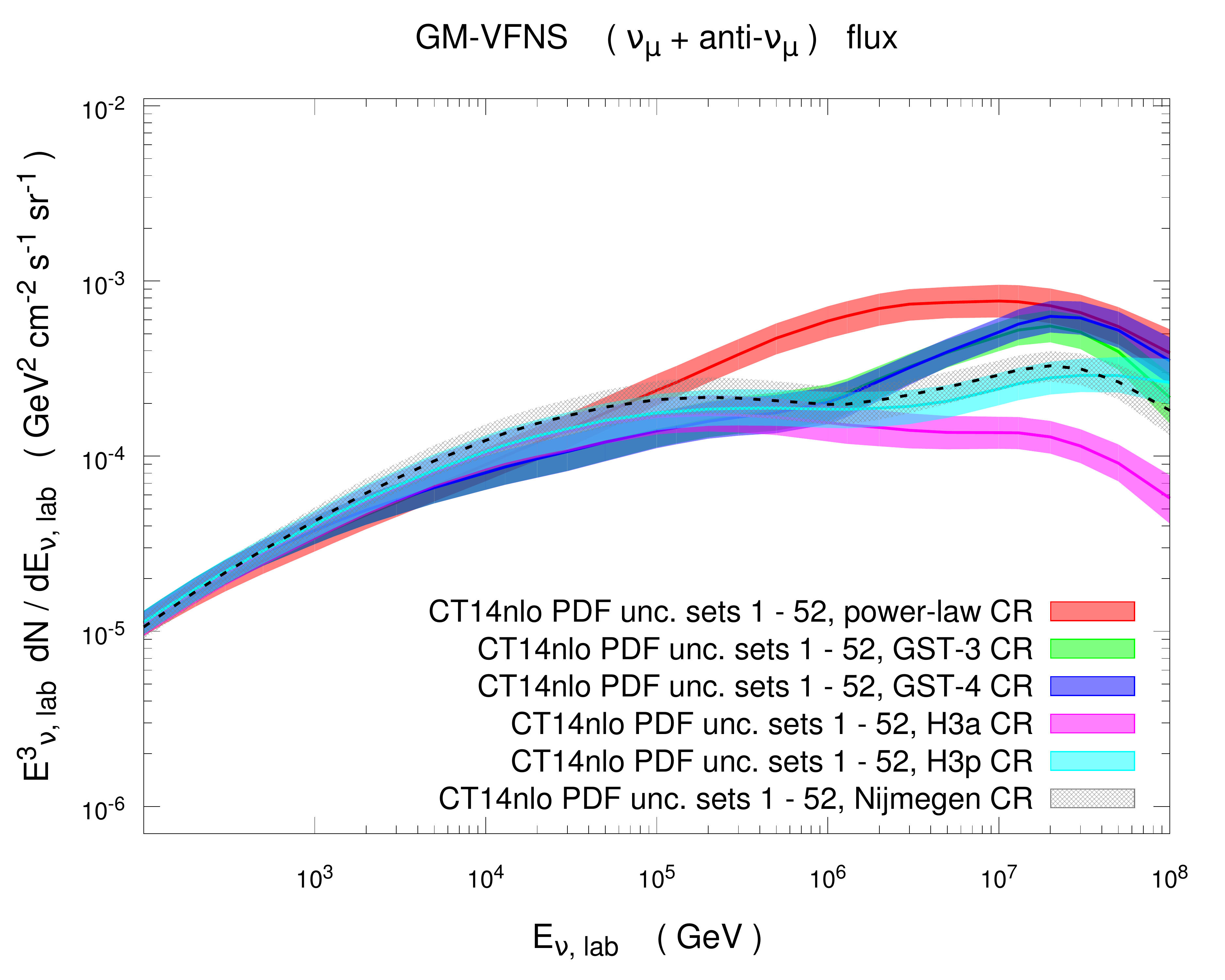}
\includegraphics[width=0.49\textwidth]{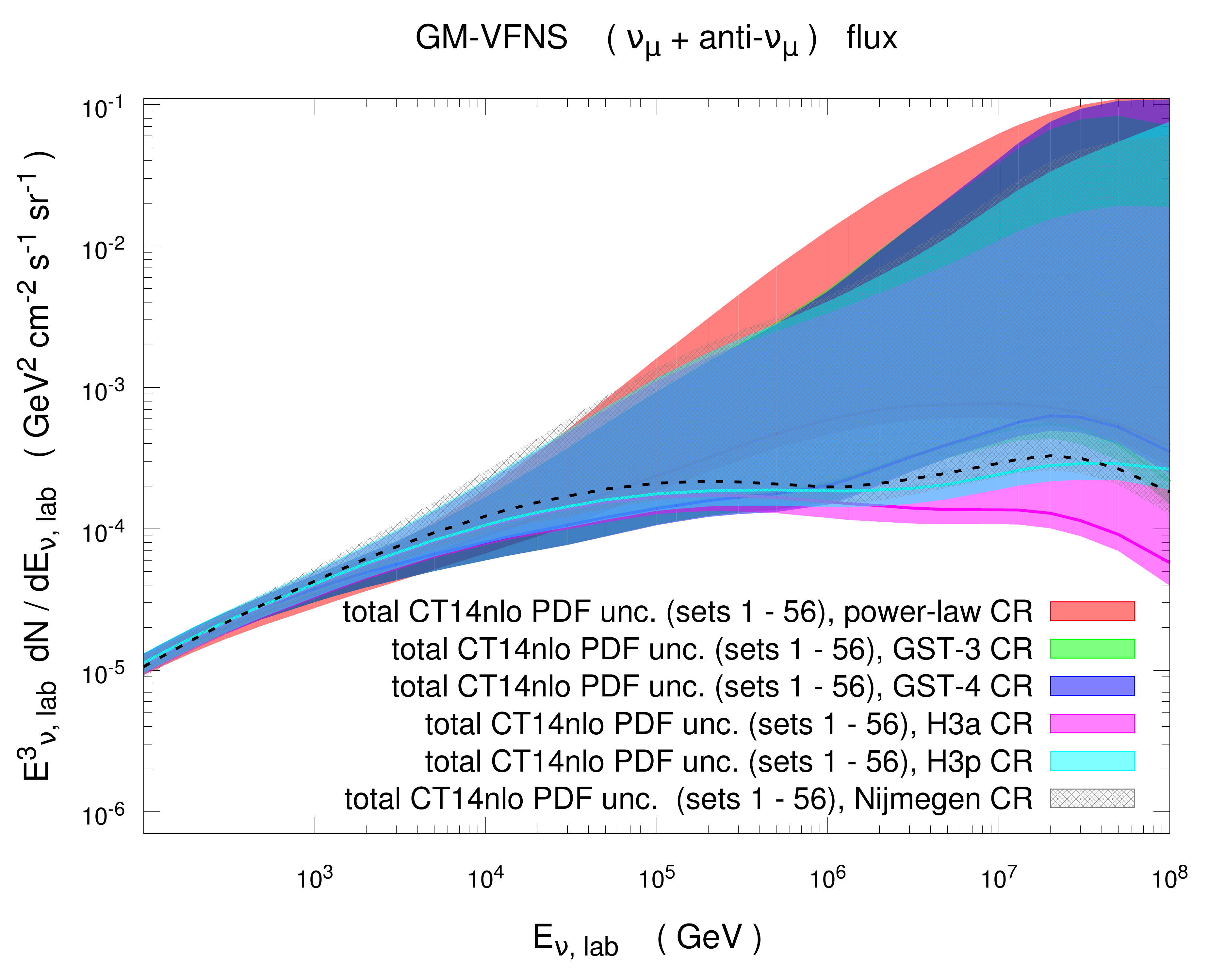}
\caption{\label{flux01234} Predictions for prompt-($\nu_\mu$ + $\bar{\nu}_\mu$) fluxes as functions of neutrino energy $E_\nu$, according to the GM-VFNS computation presented in this paper, for different CR primary fluxes used as input. Central predictions are shown in the upper left panel, renormalization scale uncertainties are shown in the upper right panel, whereas PDF uncertainties are shown in the lower panels, separating the contribution due to the Hessian sets (1--52) (lower left) from the total one due to all 56 PDF sets (lower right) included in the {\texttt{CT14nlo}} fit. See text for more details.}
\end{center}
\end{figure}

In figure~\ref{flux01234}, we show predictions for prompt-($\nu_\mu$ + $\bar{\nu}_\mu)$ fluxes obtained by using as input the six different CR primary spectrum choices described in Section~\ref{sec:equa}.
It is evident from the upper left panel that, for $E_{\nu,\,\text{lab}}$ energies larger than $10^5$--$10^6$~GeV, the uncertainty on the composition of CR spectra becomes increasingly important, with predictions from fits involving a light extra-galactic composition being larger than predictions from those corresponding to heavier compositions. In particular, the H3p predictions are larger and start to deviate from the H3a ones for $E_{\nu,\,\text{lab}}~\gtrsim~10^5\,$GeV, and the GST-4 predictions start to be larger than the GST-3 ones for $E_{\nu,\,\text{lab}}$ $\gtrsim$ $10^7\,$GeV. The most recent Nijmegen CR fit gives rise to a flux slightly larger than the H3p flux, except at the highest energies ($E_{\nu,\,\text{lab}}$ $\gtrsim$ 5 $\cdot$ $10^7\,$GeV). 
However, these effects are largely washed out when considering the QCD uncertainties affecting these predictions, as shown in the other panels of the same figure.
In particular, QCD uncertainties related to PDF variation seem to dominate our computation at the highest energies, being even larger than the astrophysical uncertainties on the CR primary spectrum. As already mentioned, this is due to the functional form of the sets 53--56 in the {\texttt{CT14nlo}} fit, and would not be true if PDF uncertainties would be restricted to the Hessian sets 1--52, as follows from the comparison of the two lower panels of figure~\ref{flux01234}. One has to emphasize that this is an effect related to the use of PDFs of this family.\footnote{Analogous considerations apply to the case of the {\texttt{CT10nlo}} fit, the predecessor of the {\texttt{CT14nlo}} one.} On the other hand, as explicitly shown in Refs.~\cite{Garzelli:2015psa,Garzelli:2016xmx}, the {\texttt{ABM11}}~\cite{Alekhin:2012ig} and {\texttt{PROSA}}~\cite{Zenaiev:2015rfa} PDF fits do not lead to prompt-neutrino fluxes with uncertainty bands so large. In fact, the {\texttt{PROSA}} PDF fit incorporates LHCb data on $D$ and $B$ meson hadroproduction at 7 TeV, which helps constraining the gluon distribution in the low-$x$ region, down to $x$ $\gtrsim$~$10^{-6}$, whereas standard PDF fits as available in the \texttt{{LHAPDF}} interface, do not yet include this information. On the other hand, the {\texttt{ABM11}} fit was performed by using data which allowed to constrain PDFs for $x$ $\gtrsim$ $10^{-4}$ plus HERA neutral-current deep-inelastic scattering data concerning the longitudinal structure function $F_L$, which allowed to probe even slightly smaller $x$ values ($x$ $\gtrsim$ 5 $\cdot$ $10^{-5}$). The size of the uncertainty bands affecting partonic distributions in the {\texttt{ABM11}} fit at lower $x$ values follows from an extrapolation, according to the functional form/parametrization of the structure functions adopted. 
However, these PDF sets are available in the context of the FFNS. While it is in principle possible to still use them in the GM-VFNS by switching to different flavors at the appropriate scales, one would miss the effects of the resummed logarithms.
Even using the global VFNS PDF fits {\texttt{MMHT2014}} \cite{Harland-Lang:2014zoa} and {\texttt{NNPDF3.0}} \cite{Ball:2014uwa}, as available in the present public {\texttt{LHAPDF 6.1.6}} interface, widely adopted in collider phenomenology, does not lead to predictions with uncertainty bands smaller than for the {\texttt{CT14nlo}} ones \cite{Accardi:2016ndt}. On the other hand, it might be worth investigating the effect of other, more recent VFNS PDF fits, in particular future revisions of the extension {\texttt{NNPDF3.0}+\texttt{LHCb}}~\cite{Gauld:2016kpd} obtained through an a-posteriori Bayesian reweighting \cite{Ball:2011gg} of the original {\texttt{NNPDF3.0}} fit \cite{Ball:2014uwa} by taking into account recent LHCb data on charm meson hadroproduction at 5, 7 and 13 TeV, when combined with our GM-VFNS framework. We leave this for future work, taking into account that, although we find the results of Ref.~\cite{Gauld:2016kpd} quite pro\-mi\-sing and encouraging, we believe that the robustness of this fit still deserves a deeper investigation.\footnote{Even though the latest update of {\texttt{NNPDF3.0}+\texttt{LHCb}} now includes all the latest revisions of the LHCb data sets (see errata in \cite{Aaij:2016jht,Aaij:2015bpa}), it still gives rise to negative PDFs for very small $x$ at low scales. This in turn makes the $x_E$ differential cross sections become negative for large energies at large $x_E$, which is unphysical. Unfortunately, this makes them unsuitable for the phase space region we are interested in here. For more details, see Appendix \ref{sec:nnpdf}.}

\begin{figure}
\begin{center}
\includegraphics[width=0.47\textwidth]{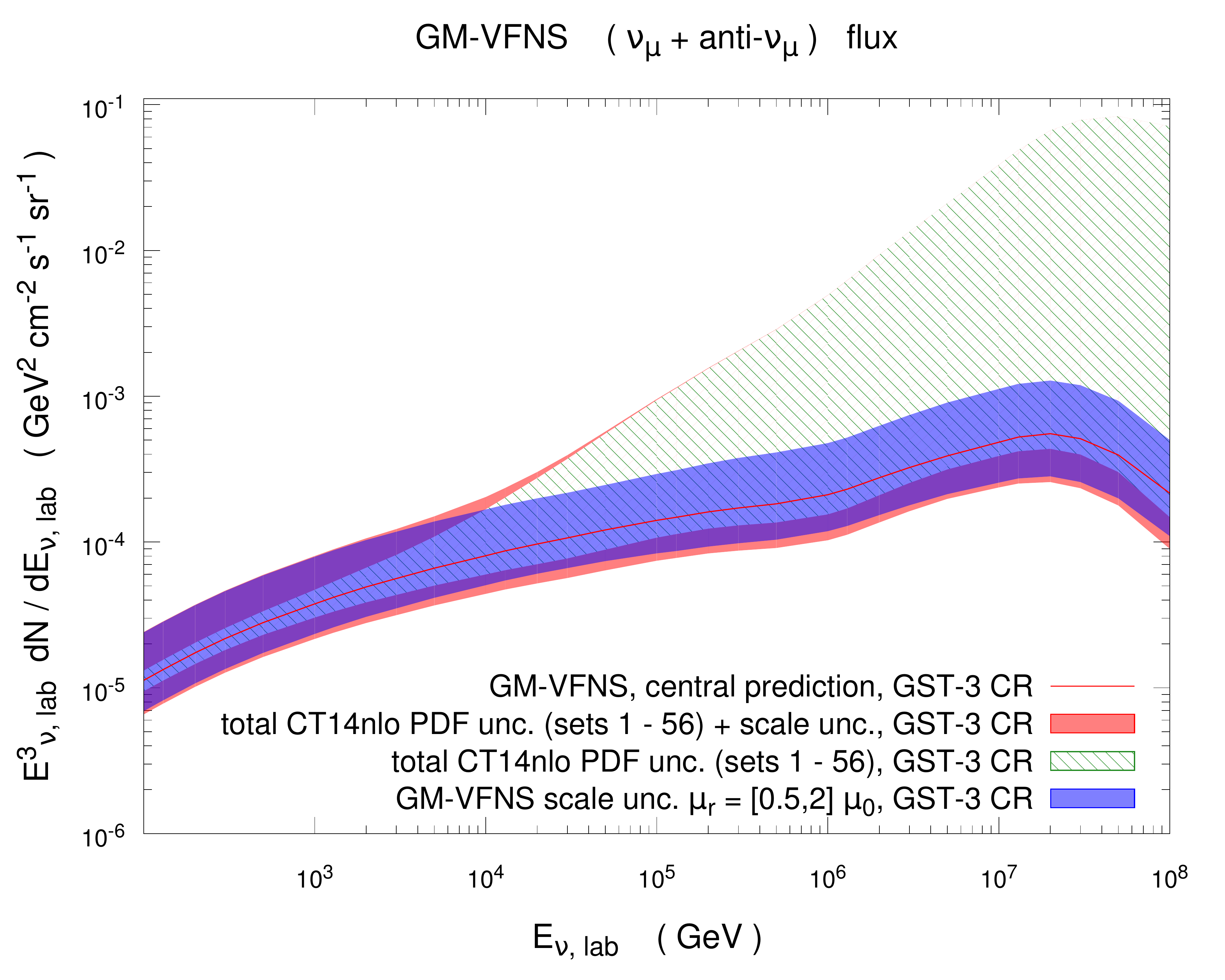}
\includegraphics[width=0.47\textwidth]{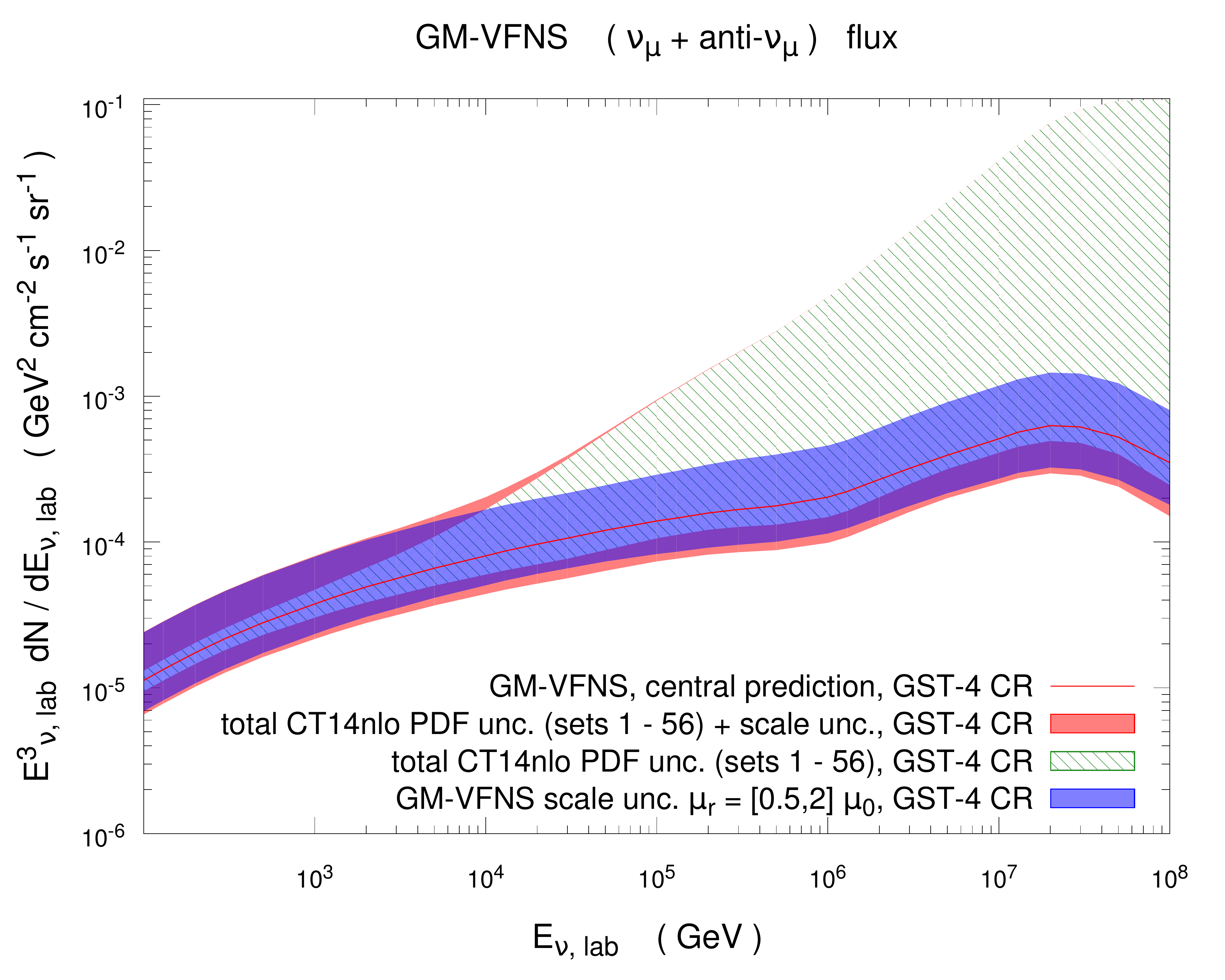}
\includegraphics[width=0.47\textwidth]{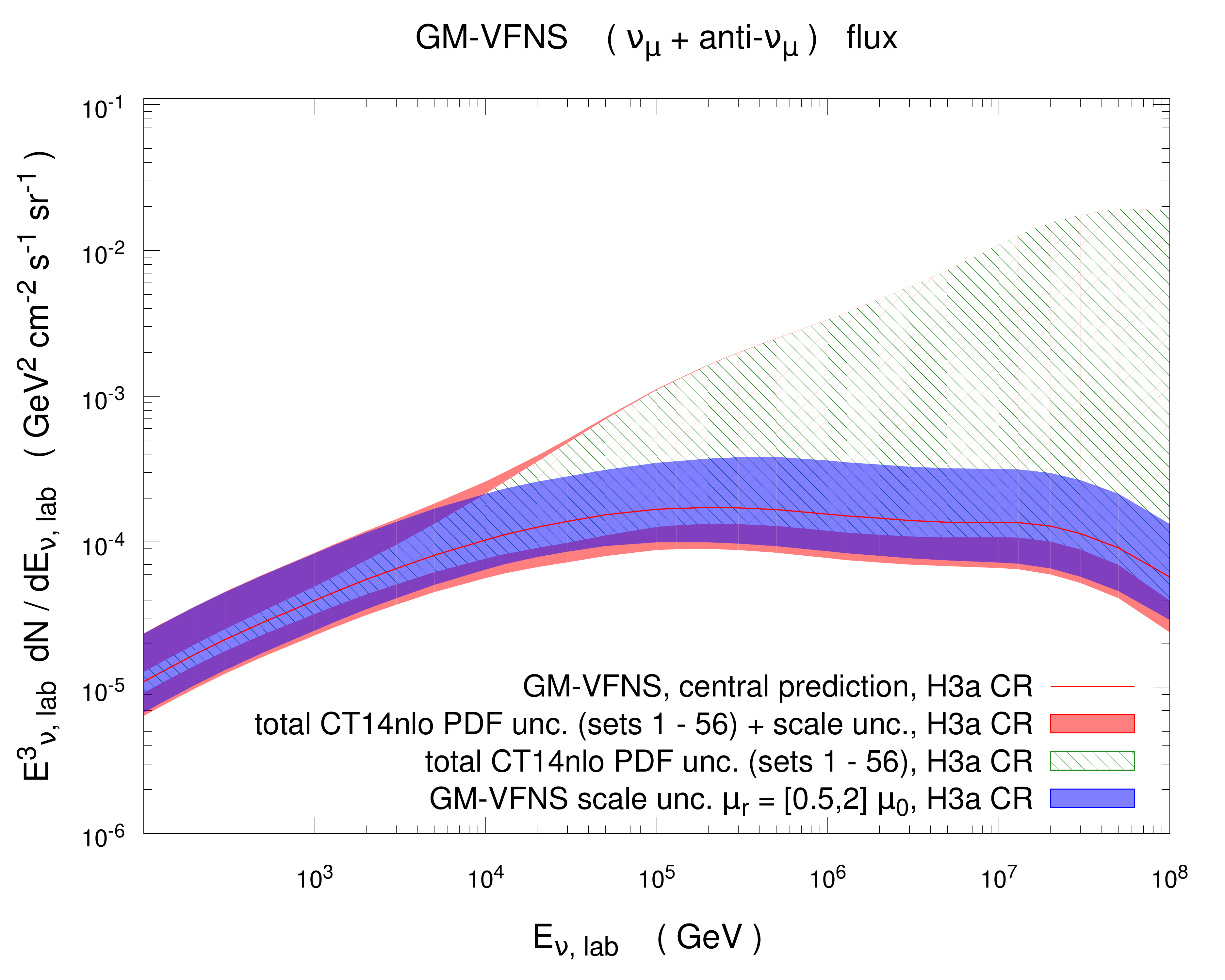}
\includegraphics[width=0.47\textwidth]{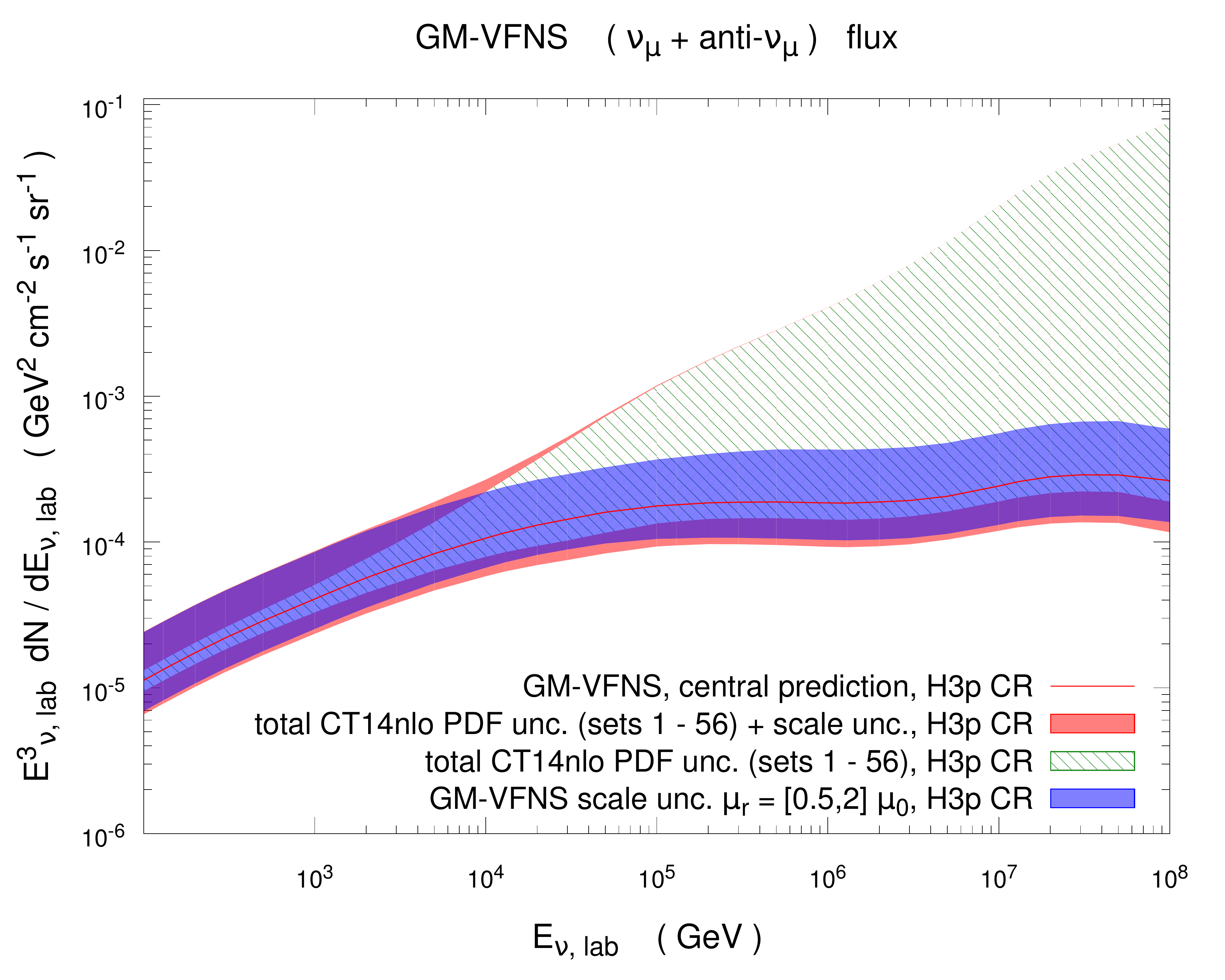}
\includegraphics[width=0.47\textwidth]{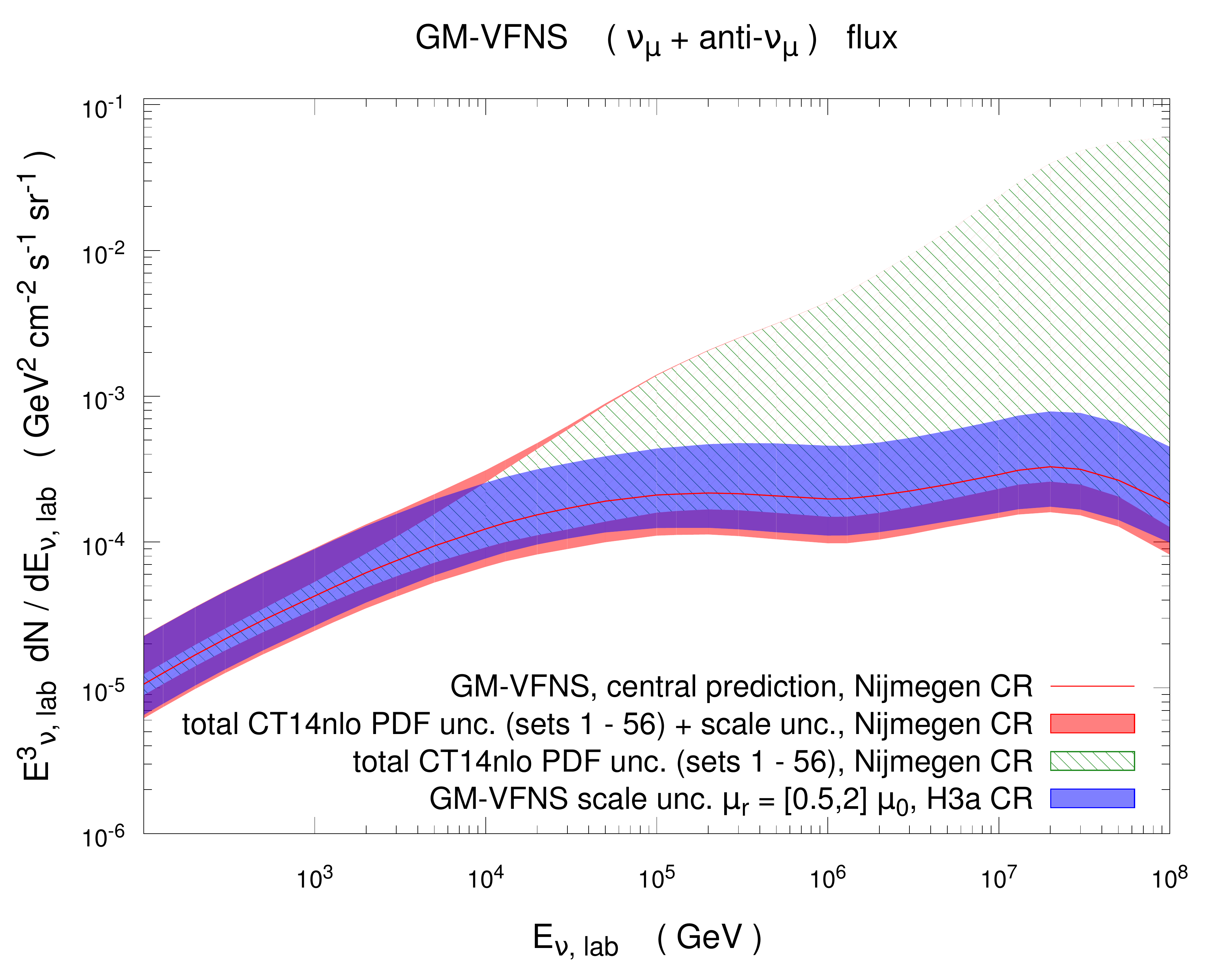}
\caption{\label{total56} Prompt-($\nu_\mu$ + $\bar{\nu}_\mu$) fluxes for different CR primary spectra. Scale (blue) and PDF (hatched green) uncertainties, and the result obtained by summing them in quadrature (pink), are shown separately in each panel. Each panel corresponds to a different CR primary spectrum (GST-3, GST-4, H3a, H3p and Nijmegen).}
\end{center}
\end{figure}

\begin{figure}
\begin{center}
\includegraphics[width=0.47\textwidth]{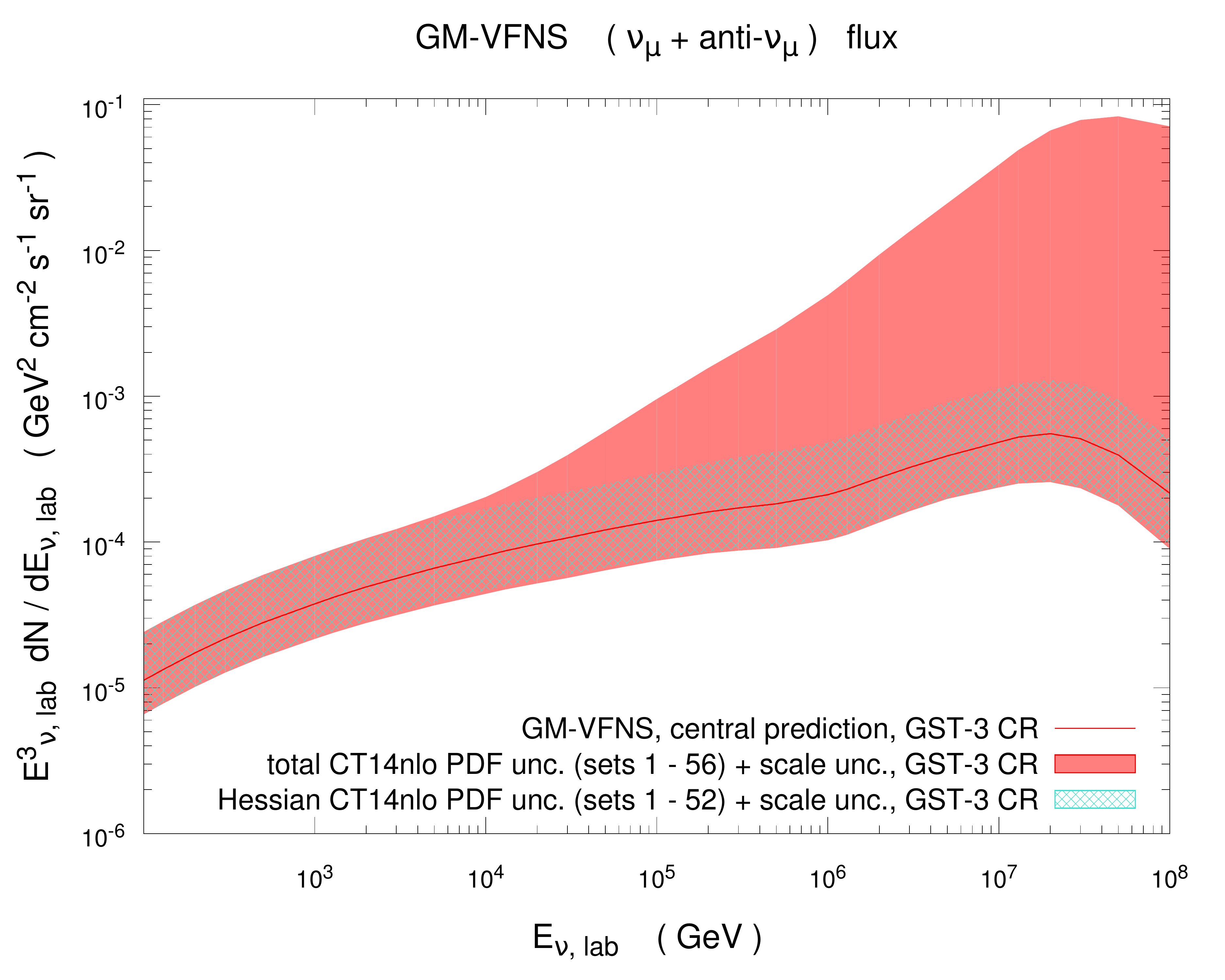}
\includegraphics[width=0.47\textwidth]{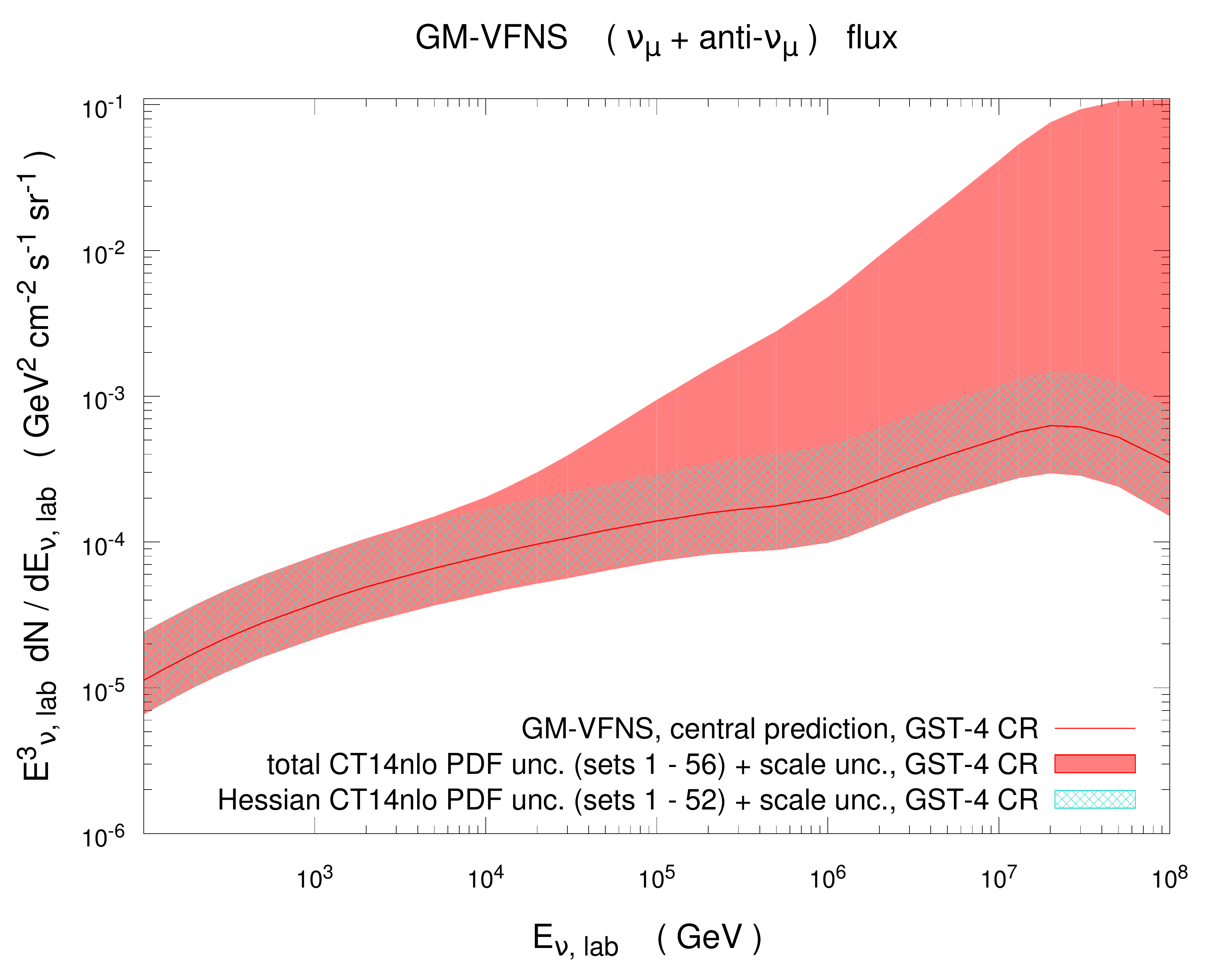}
\includegraphics[width=0.47\textwidth]{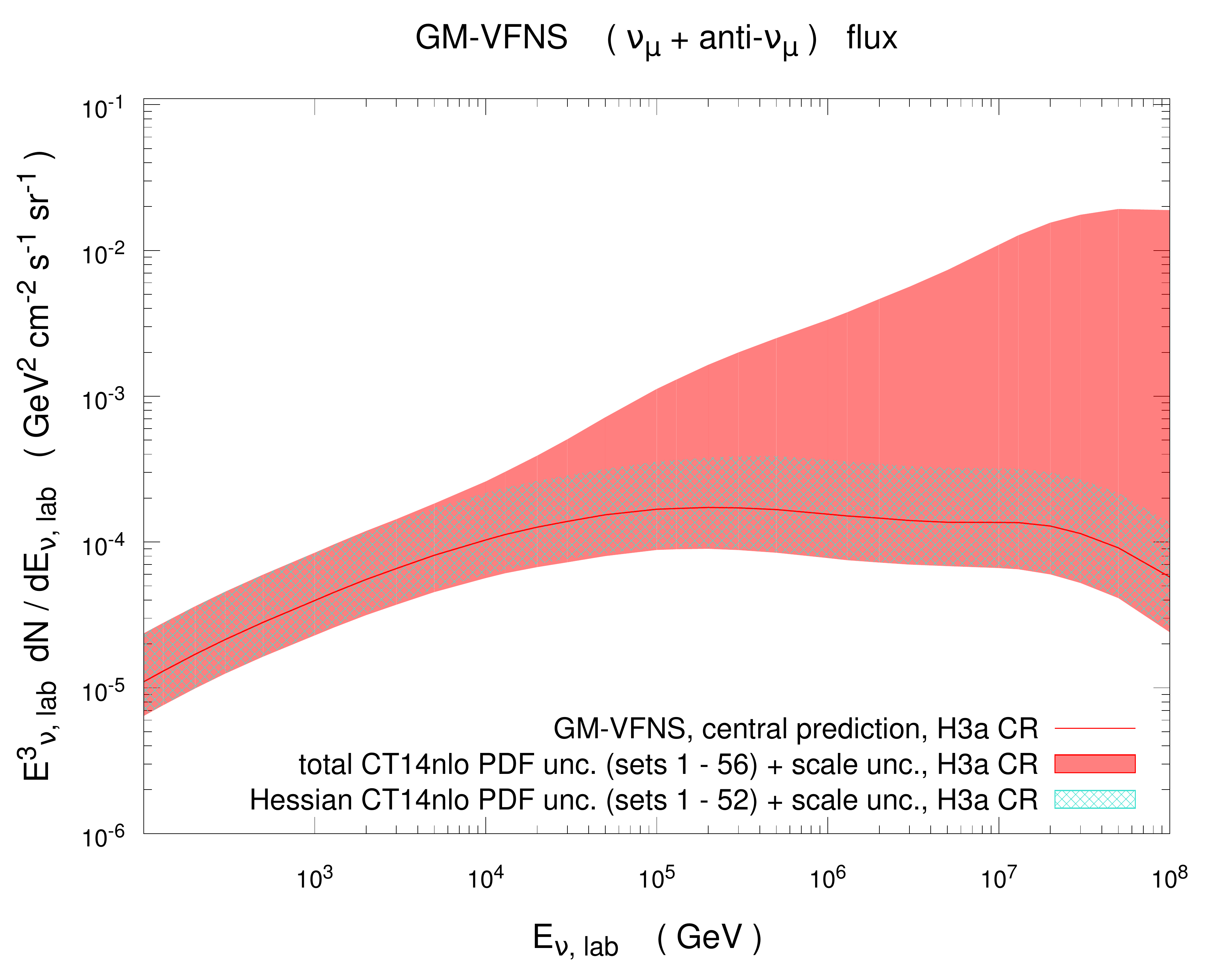}
\includegraphics[width=0.47\textwidth]{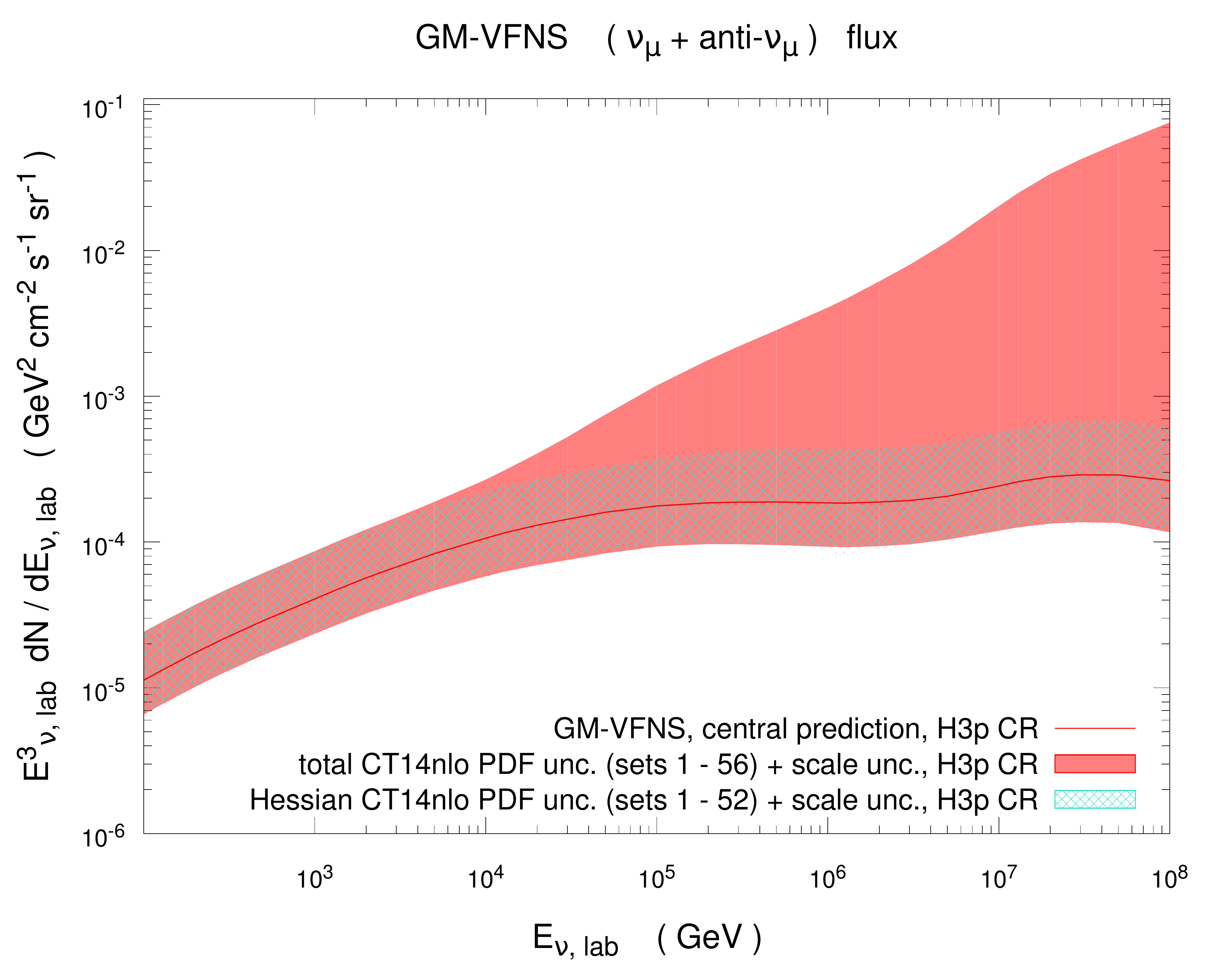}
\includegraphics[width=0.47\textwidth]{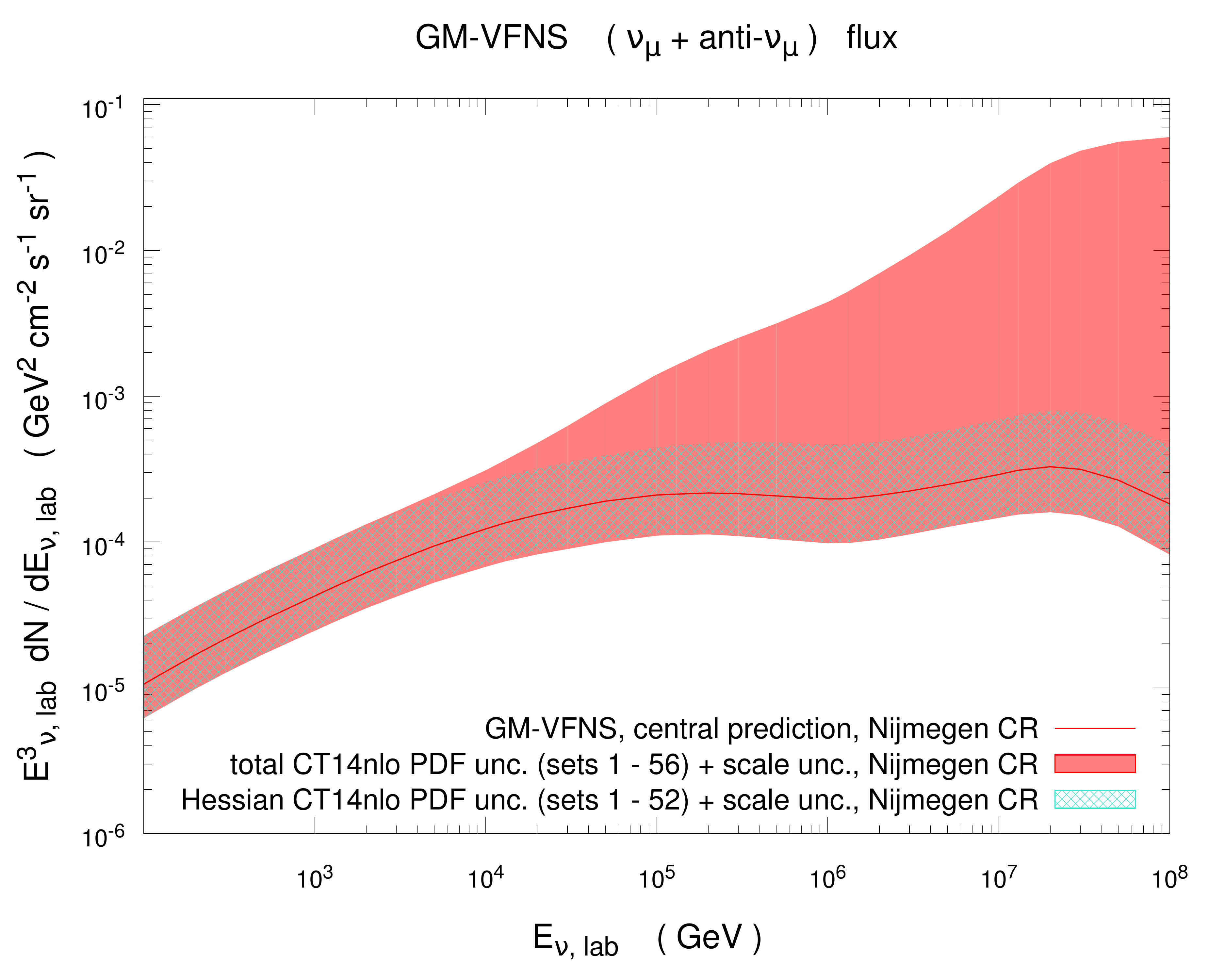}
\caption{\label{total52} Total uncertainties in prompt-($\nu_\mu$ + $\bar{\nu}_\mu$) fluxes due to scale and PDF variations, computed by considering all 56 PDF sets of the {\texttt{CT14nlo}} fit (pink solid bands), as compared to those arising when restricting the PDF variation to the Hessian sets (light-blue hatched bands). Each panel refers to a different CR primary spectrum (GST-3, GST-4, H3a, H3p and Nijmegen).}
\end{center}
\end{figure}

Finally, the PDF and scale uncertainties summed in quadrature, are shown in figure~\ref{total56}, for each of the five CR primary spectra: GST-3, GST-4, H3a, H3p and Nijmegen. They are compared in figure~\ref{total52} with those we get when restricting the PDF uncertainty to the Hessian sets 1--52.

\subsection{Comparison with predictions from {\texttt{POWHEGBOX + PYTHIA}}}

\begin{figure}
\begin{center}
\includegraphics[width=0.65\textwidth]{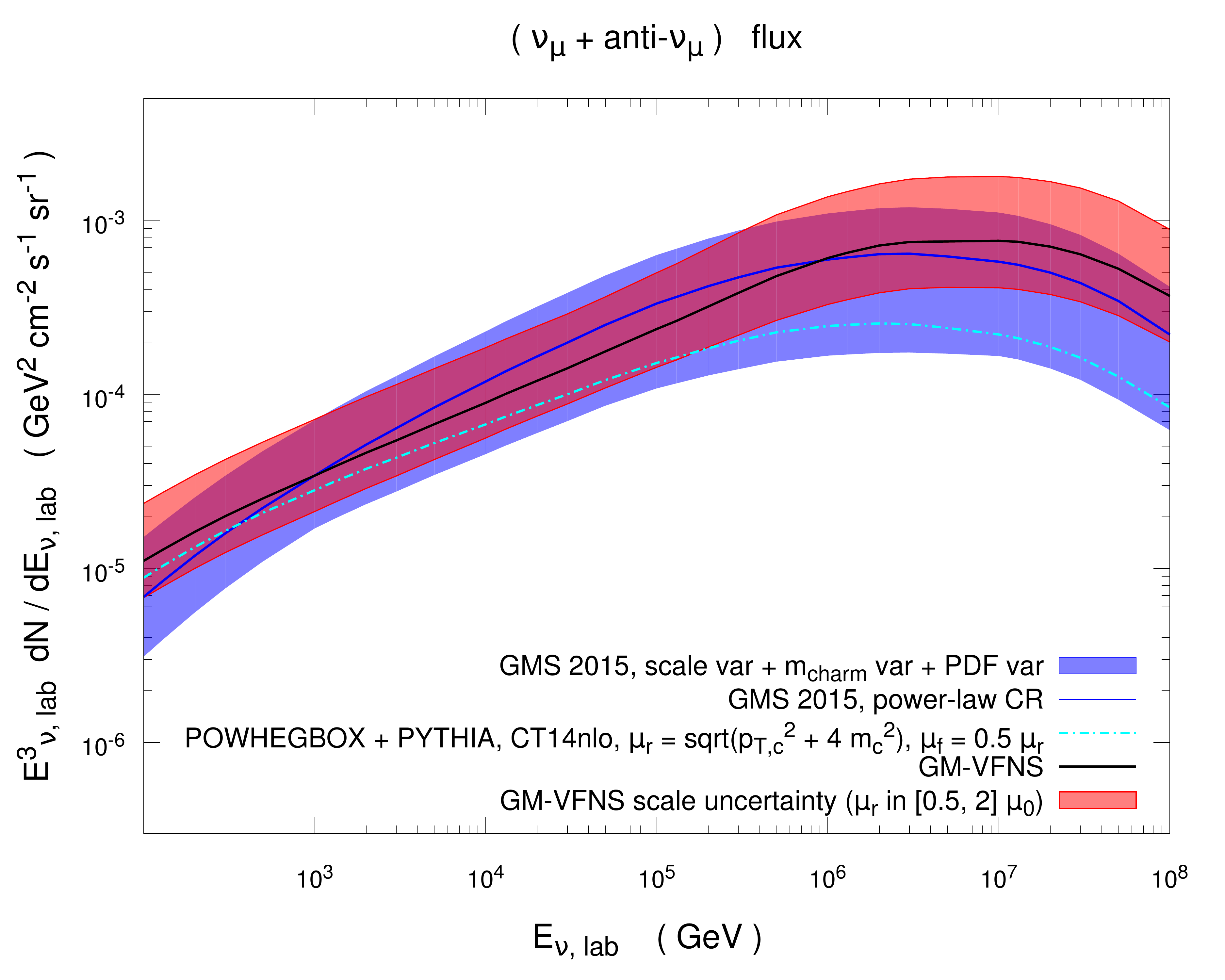}
\caption{Prompt-($\nu_\mu$ + $\bar{\nu}_\mu$) fluxes as functions of neutrino energy $E_{\nu,\,\text{lab}}$. Predictions in the GM-VFNS according to the computation presented in this paper (red band) are compared with predictions by {\texttt{POWHEGBOX}} + {\texttt{PYTHIA}} (blue band), with hard scattering in the FFNS, according to the computation in Ref.~\cite{Garzelli:2015psa}. 
In particular, the blue line and band refer to central values and QCD uncertainties of the predictions obtained in Ref.~\cite{Garzelli:2015psa} (GMS 2015) using central scales $\mu_r$ = $\mu_f$ = $\mu_0$ = $\sqrt{p_{T,c}^2 + 4 m_c^2}$, with variations ($\mu_r$, $\mu_f$)=[(0.5, 0.5), (2, 2), (0.5, 1), (1, 0.5), (1, 2), (2,1)] ($\mu_0$, $\mu_0$), the 28 uncertainty sets of {\texttt{ABM11nlo}-3fl} PDF, and $m_c$ = (1.4 $\pm$ 0.15) GeV. 
The light-blue dot-dashed line refers to predictions by {\texttt{POWHEGBOX}}~+~{\texttt{PYTHIA}} using $\mu_r$ = $\sqrt{p_{T,c}^2 + 4 m_c^2}$, $\mu_f$ = $\mu_r/2$, $m_c$ = 1.3 GeV and the central set of the {\texttt{CT14nlo-3fl}} PDFs. The broken-power-law CR primary spectrum is used as input in all predictions.
}
\label{compagms}
\end{center}
\end{figure}

Our predictions in the GM-VFNS are compared to those obtained in Ref.~\cite{Garzelli:2015psa} in fi\-gure~\ref{compagms}. 
The differences in the central predictions are due to the use of different FNSs, scales, charm mass values, PDFs, and fragmentation methods. In particular, in Ref.~\cite{Garzelli:2015psa}, the charm mass was fixed to $m_c=1.4\,$GeV, the {\texttt{ABM11nlo} PDFs~\cite{Alekhin:2012ig} were adopted, while both the $\mu_r$ and $\mu_f$ central values were fixed to $\mu_0$~=~$\sqrt{p_{T,c}^2 + 4 m_c^2}$. Furthermore, the hadronization process, according to the phenomenological Lund string model implemented in the {\texttt{PYTHIA}} generator \cite{Sjostrand:2014zea}, applied after a $p_T$-ordered parton shower matched to the NLO hard scattering, with matrix elements in the 3 flavor number scheme, allows us to transform partonic distributions into hadronic distributions. A recent version of the Perugia tune~\cite{Skands:2010ak} was adopted to fix various parameters entering the {\texttt{PYTHIA}} computation. On the other hand, in this paper, $m_c=1.3\,$GeV, the $\mu_f$ value was fixed to $\xi\,\sqrt{p_{T,h}^2 + 4 m_c^2}$ = $\xi\,\mu_r$, with $\xi= 0.5$, and the FF fit {\texttt{KKKS08}} was used to describe the transition from NLO hard-scattering partons to charmed hadrons. Running the computation of Ref.~\cite{Garzelli:2015psa} by using scales $\mu_f = 0.5\, \mu_r = 0.5\, \mu_0$, the same $m_c$ value, and PDFs compatible with those used in the GM-VFNS computation (i.e. the {\texttt{CT14nlo}} - 3 flavor for the hard-scattering matrix-elements in the FFNS) produces distributions that, at small energies  ($E_\nu$ $<$ 10$^4$ GeV), have the same shape as the GM-VFNS ones, although being rescaled by an almost constant factor related to the use of $p_{T, c}$ instead of $p_{T, h}$ in the scale definition. On the other hand, at higher energies, the GM-VFNS predictions are characterized by a steeper slope and become increasingly larger than the {\texttt{POWHEGBOX~+~PYTHIA}} ones. The GM-VFNS approach performs a systematic resummation of logarithms of $p_T^2$/$m_c^2$ at NLL precision, while the \texttt{POWHEG} approach~\cite{Nason:2004rx} uses a classical shower.\footnote{Shower algorithms resum a class of logarithms in a practical and effective way, which does not exactly correspond to a resummation procedure with an exact logarithmic accuracy. The deep relation between shower algorithms and traditional resummation techniques is still subject to investigation (see e.g. Ref.~\cite{Nagy:2016pwq} and references therein). It is recognized that the {\texttt{PYTHIA}} shower has at least leading-logarithmic accuracy.}
Resumming the aforementioned logarithms contributes only partially to the reduction of scale uncertainties. A further reduction can indeed be obtained by resumming other kinds of logarithms and including higher fixed-order corrections, which is beyond the scope of this work.
 
Finally, it turns out that the use of accurate FFs as an alternative to the hadronization method does not produce big differences in total predictions for prompt neutrino fluxes. However, the specific contribution of the intermediate production and decay of the $\Lambda_c$ hadron through hadronization yields a different result than FFs. As already mentioned, this can be explained by the fact, that hadronization includes the recombination of the final-state charm quark with initial-state valence quarks, while FFs do not include correlations between initial and final states.
Still, even including these correlations, the $\Lambda_c$ contribution to the fluxes remains sub-dominant with respect to those of the other intermediate charmed hadrons, with $(D^0 + \bar{D^0})$ and $(D^+ + D^-)$ contributions being larger than the $(\Lambda_c^+ + \Lambda_c^-)$ one and dominating at all energies. 
Note that this difference is due to both a different production cross section of the charmed hadrons (the cross sections of $(\Lambda_c^+ + \Lambda_c^-)$ hadroproduction is smaller than that of $(D^0 + \bar{D^0})$ or $(D^+ + D^-)$ by a factor of $\mathcal{O}(10)$) and to the different branching fractions of these hadrons for semi-leptonic decays.

\subsection{Comparison with other predictions available in the literature}
\begin{figure}
\begin{center}
\includegraphics[width=0.49\textwidth]{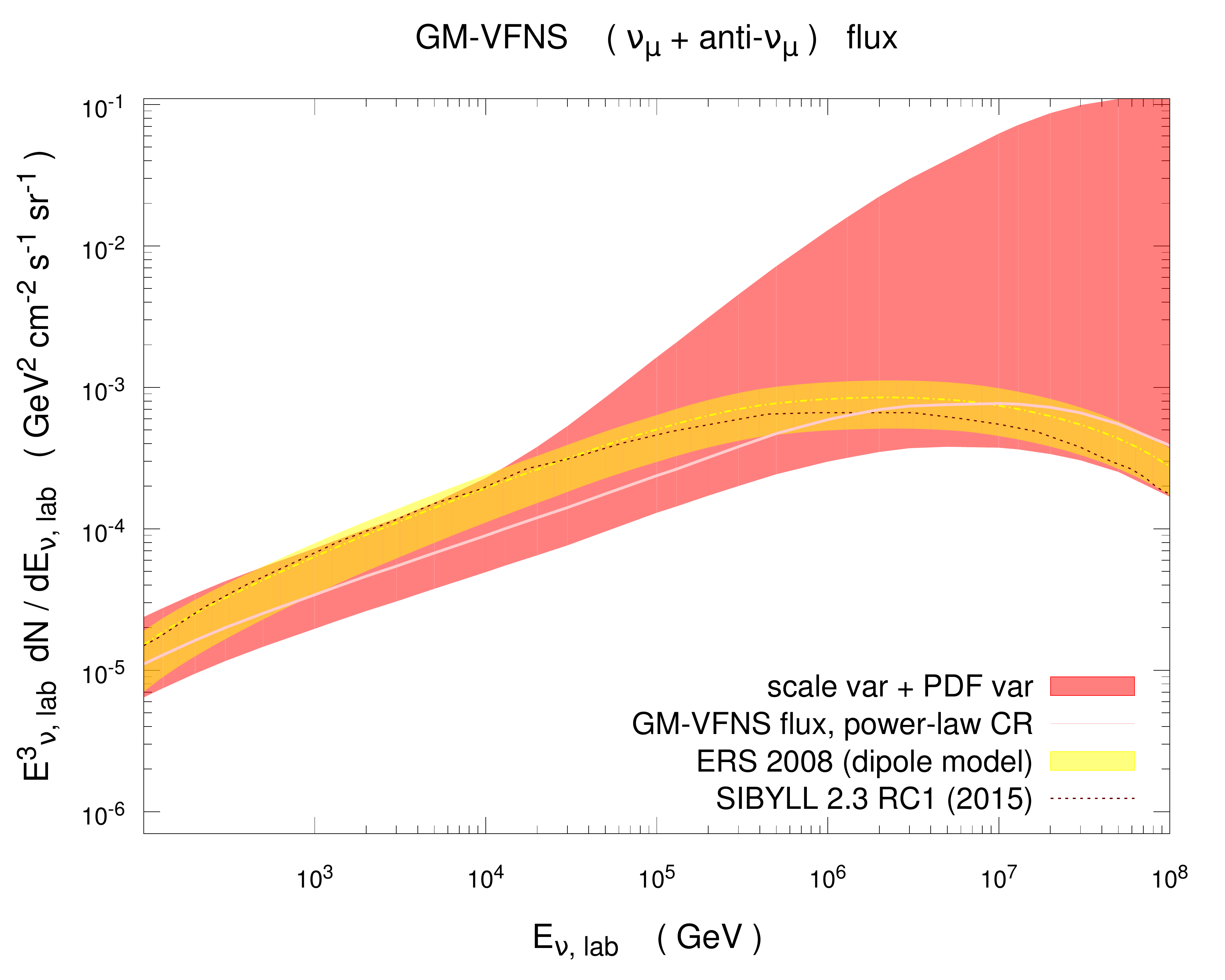}
\includegraphics[width=0.49\textwidth]{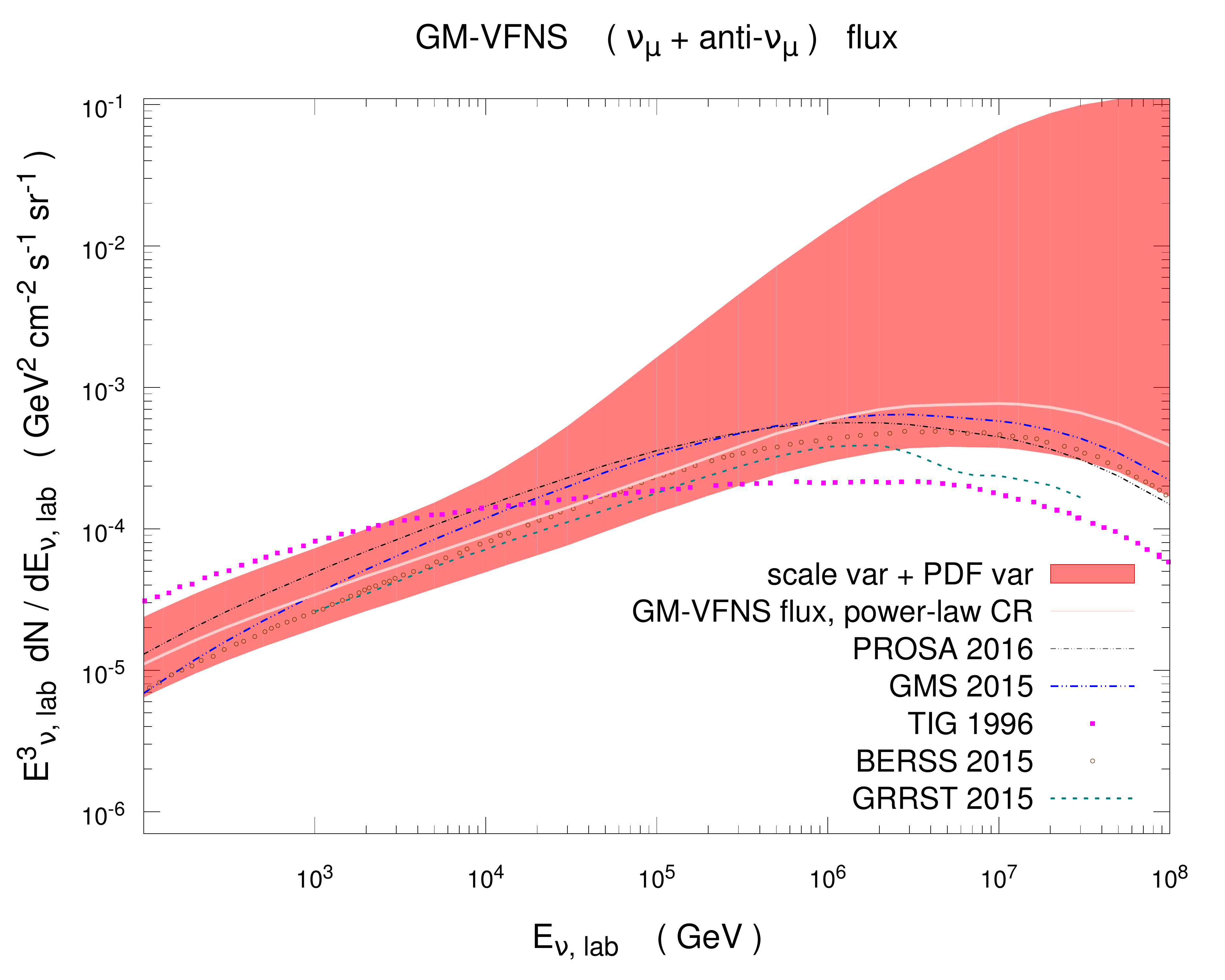}
\caption{\label{compa}
Prompt-($\nu_\mu$ + $\bar{\nu}_\mu$) fluxes as a function of neutrino energy $E_{\nu,\,\text{lab}}$. Predictions according to the GM-VFNS computation of this paper, together with their uncertainty band, are compared to other ones available in the literature~\cite{Garzelli:2015psa,Gauld:2015kvh, Bhattacharya:2015jpa,Gondolo:1995fq, Fedynitch:2015zma, Enberg:2008te}, distinguishing those making use of phenomenological models (the dipole model and a recent version of the \texttt{SYBILL} event generator), shown on the left, from those treating charm hadroproduction at parton-level by means of perturbative QCD, collected in the plot on the right. The broken power-law CR primary spectrum is used as input in all predictions.
} 
\end{center}
\end{figure}
\begin{figure}
\begin{center}
\includegraphics[width=0.49\textwidth]{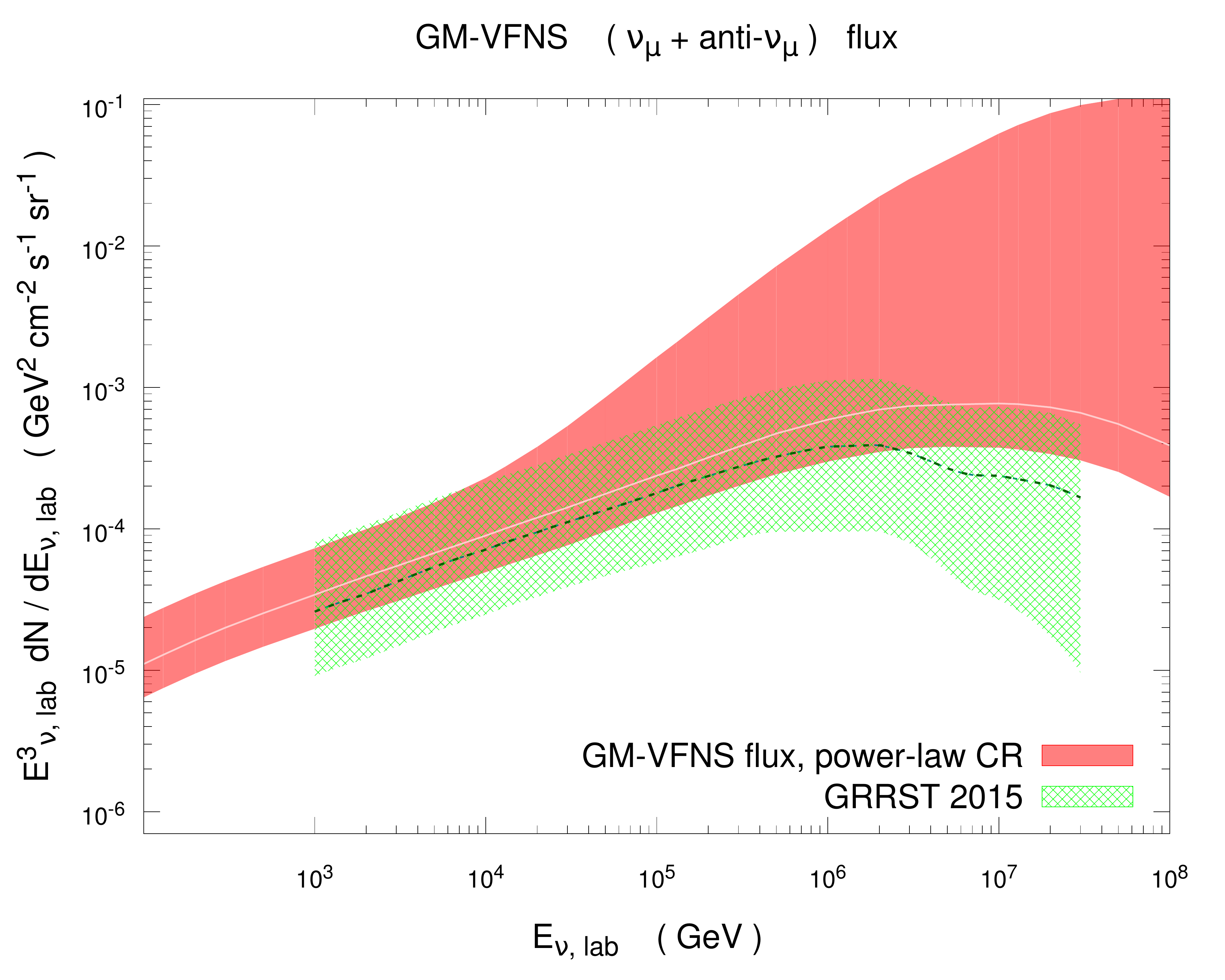}
\includegraphics[width=0.49\textwidth]{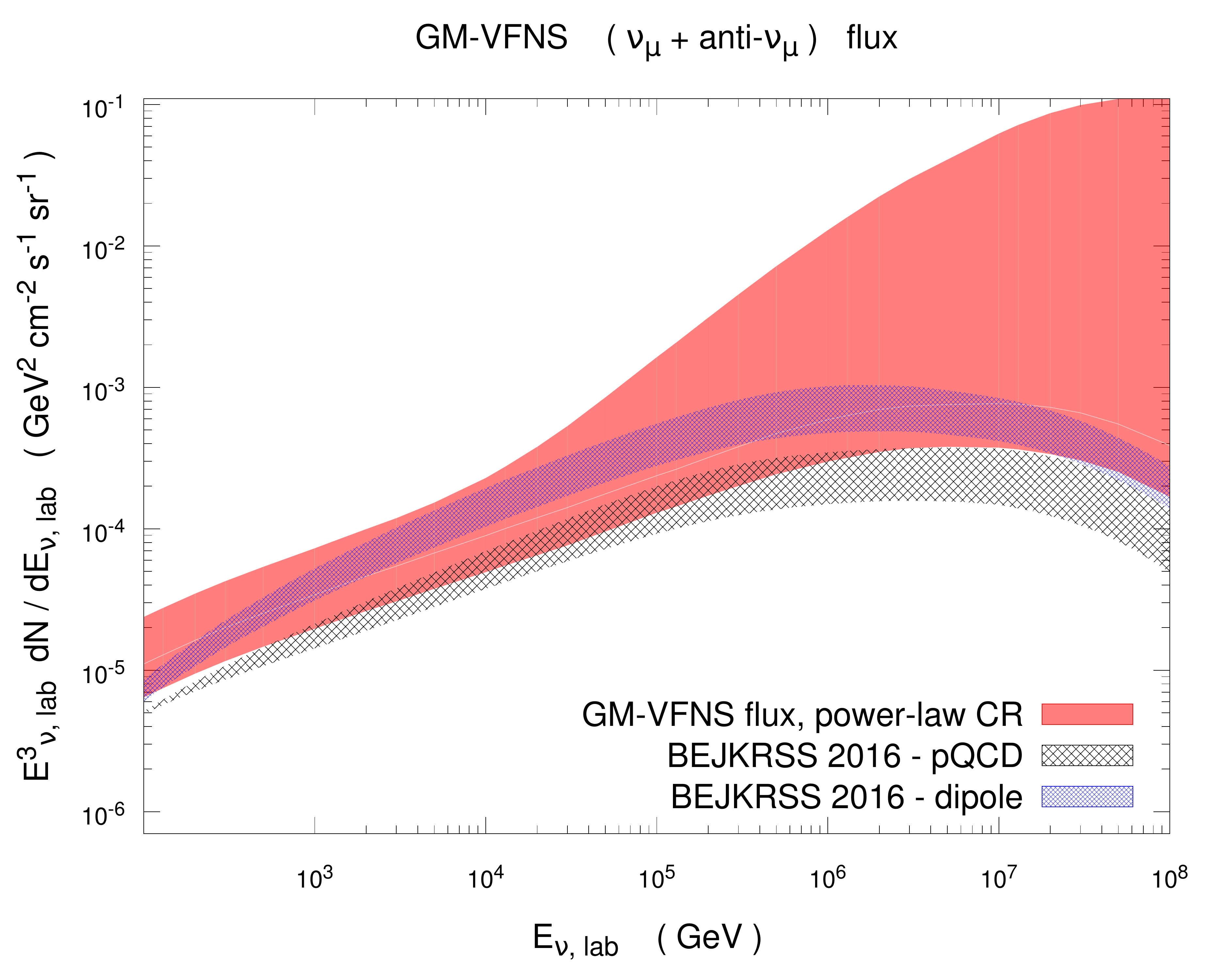}
\caption{\label{compa2}
  Prompt-($\nu_\mu$ + $\bar{\nu}_\mu$) fluxes as a function of neutrino energy $E_{\nu,\,\text{lab}}$. Predictions according to the GM-VFNS computation of this paper, together with their uncertainty band, are compared to other recent selected ones available in the literature~\cite{Gauld:2015kvh, Bhattacharya:2016jce}, taking into account their uncertainty bands. The broken power-law CR primary spectrum is used as input in all predictions.
} 
\end{center}
\end{figure}
In figure~\ref{compa}, we compare our predictions in the GM-VFNS with others available in the literature, making use of pQCD or phenomenological models in the description of charm hadroproduction. We observe that our predictions are compatible, within the uncertainty band, with those from the ERS dipole model~\cite{Enberg:2008te} and from a recent version of the \texttt{SYBILL 2.3} event generator~\cite{Fedynitch:2015zma}, with central GM-VFNS predictions being smaller than those of these models for $E_{\nu,\,\text{lab}}$ values up to a few PeV.  They are also compatible with the BERSS predictions~\cite{Bhattacharya:2015jpa}, using the same PDF central set (however neglecting the PDF uncertainty band), an older set of FFs and a different GM-VFNS implementation (FONLL~\cite{Cacciari:1998it}).
Furthermore, they are consistent with the GMS 2015~\cite{Garzelli:2015psa} and the PROSA 2016~\cite{Garzelli:2016xmx} ones, both on the basis of \texttt{POWHEGBOX~+~PYTHIA}. On the other hand, for high energies, our predictions are larger than the GRRST ones~\cite{Gauld:2015kvh}, with a different shape of the central spectrum. The shape difference can be attributed to the use of different PDFs ({\texttt{CT14nlo} vs. the {\texttt{NNPDF3.0~+~LHCb}} PDF set). However, considering that the uncertainty band of the GRRST predictions, shown on the left of figure~\ref{compa2}, overlaps with the uncertainty band of our predictions, we can conclude that the two results are still compatible.      
Finally, the largest deviations from our GM-VFNS predictions are visible in the TIG 1996 ones~\cite{Gondolo:1995fq}, based on a computation of charm hadroproduction with leading order and leading-logarithmic accuracy, as available in an old standalone version of the {\texttt{PYTHIA6}} parton-shower event generator, and outdated PDFs.
Additionally, on the right of figure~\ref{compa2}, we compare our predictions to those recently published in Ref.~\cite{Bhattacharya:2016jce}. 
We see that our predictions are fully compatible with the results obtained by these authors in three different dipole model frameworks (Soyez~\cite{Soyez:2007kg}, AAMQS~\cite{Albacete:2010sy}, Block~\cite{Block:2014kza}, spanning the area marked in violet in our plot), which represent an update with respect to the ERS ones. On the other hand, our predictions lie above those obtained by the same authors in the pQCD framework, and the uncertainty bands (marked in gray in our plot) overlap only partially, which can be explained as follows.
First of all, nuclear PDFs are used in the BEJKRSS 2016 pQCD computation, instead of the superposition approximation, adopted in our computation. This causes a decrease of the BEJKRSS central predictions with respect to those previously published by the BERSS group in Ref.~\cite{Bhattacharya:2015jpa}, which were much closer to our results. However, the uncertainty bands associated to the nuclear PDFs used in the BEJKRSS 2016 pQCD predictions are not reliable, especially at low $x$ values. Given the fact that a reliable estimate of these uncertainties is missing and probably leads to bands larger than those quoted by the nuclear PDF collaborations, we can conclude that, if these would be taken into account, our uncertainty band would fully overlap with the BEJKRSS 2016 pQCD one.

In summary, we can conclude that predictions on the basis of the GM-VFNS implementation presented in this paper turn out to be compatible, at least within present QCD uncertainties, with other modern predictions, obtained with different methods. However, at high energies, the uncertainties on the fluxes presented in this paper turn out to be larger than those presented in other papers, due to the PDF uncertainties inherent the {\texttt{CT14nlo}} set adopted in our computation. The effects of adopting nuclear PDFs, instead of the superposition approximation, deserve further exploration and require a reliable assessment of current nuclear PDF uncertainties, which is beyond the scope of the present paper.

\section{Implications for VLV\texorpdfstring{$\nu$}{nu}T's}
\label{sec:implications}

So far, the IceCube Collaboration did not report any smoking-gun evidence for the existence of prompt neutrinos. However, upper limits on the prompt component have been inferred on the basis of different analyses, characterized by increasing statistics. These upper limits are model dependent, i.e. they were derived assuming that the shape of the prompt-neutrino spectrum is fixed to the shape of the ERS flux~\cite{Enberg:2008te} used as a baseline for the analyses, and only the normalization is varied. In 2015, a prompt-neutrino upper limit of 3.5 times the ERS flux was quoted in the analysis of the diffuse-muon-neutrino flux reported in Ref.~\cite{Aartsen:2015zva}. This analysis was built on the basis of a set of charged-current ($\nu_\mu$ + $\bar{\nu}_\mu$) events with track topology, reaching the detector from the northern hemisphere,\footnote{The Earth acts as an efficient filter for atmospheric muons at the energies explored in this analysis.} with interaction vertices inside or outside the instrumented volume, collected over three years (from 2009 to 2012). This analysis, already representing an update with respect to previous studies~\cite{Aartsen:2013eka,Aartsen:2015rwa}, was subsequently updated in Ref.~\cite{Aartsen:2016xlq}, thanks to increased statistics, by using $\sim$~350000 events collected over six years, leading to stronger limits. In particular, two different limits were proposed in the last paper, one of 1.06 times the ERS flux and a second one of 0.5 times the ERS flux, with the latter more stringent than the first due to the dependence on the assumptions made in the modeling of the astrophysical neutrino flux. We consider the first limit as a more conservative estimate, as also explained in Ref.~\cite{Aartsen:2016xlq}. A comparison of our predictions for prompt-($\nu_\mu$ + $\bar{\nu}_\mu$) fluxes with this limit is shown in figure~\ref{fig:upperlimit}.
\begin{figure}[ht]
\begin{center}
  \includegraphics[width=0.75\textwidth]{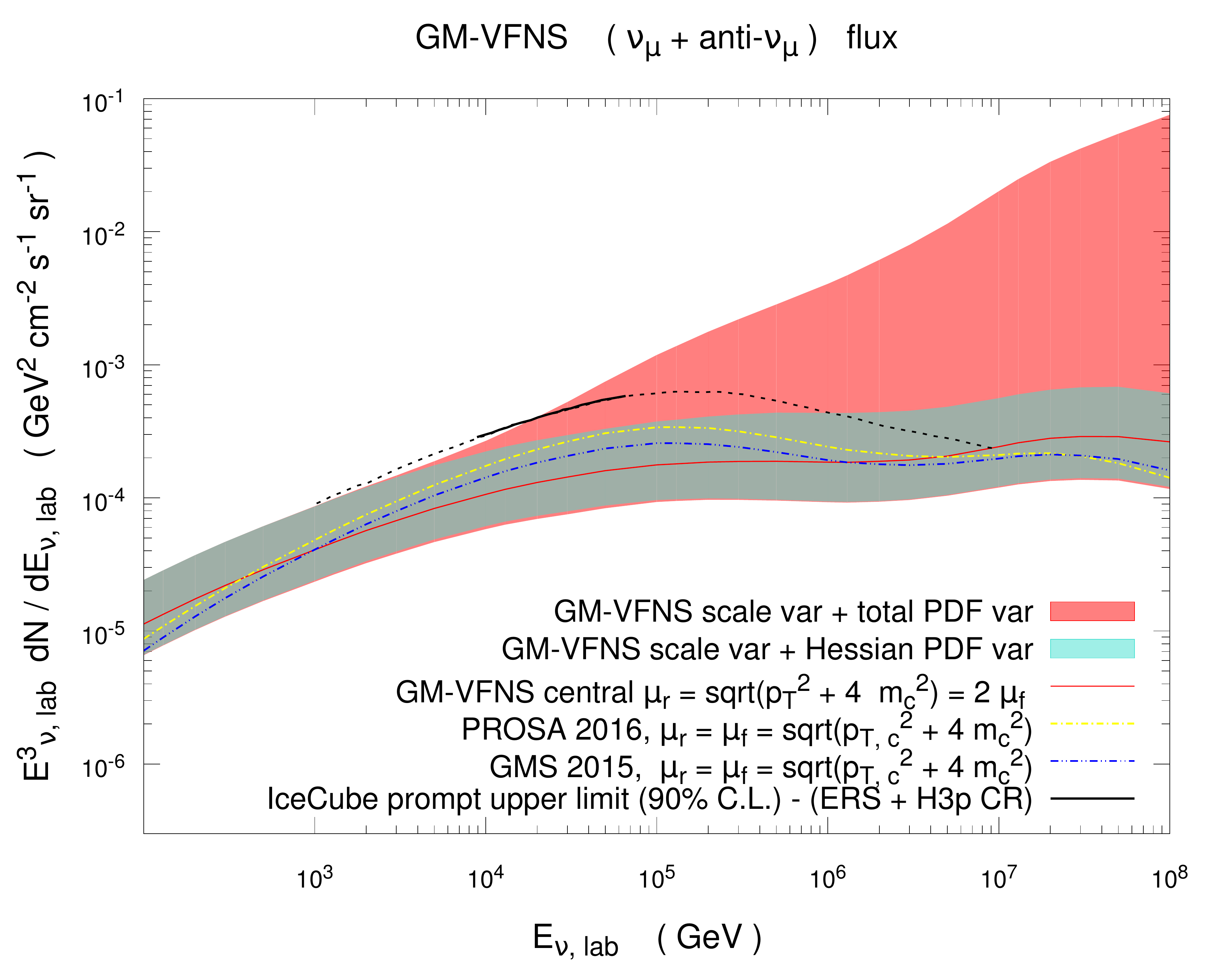}
\end{center}
\caption{\label{fig:upperlimit} 
  Comparison of the prompt-($\nu_\mu + \bar{\nu}_\mu$) flux from the GM-VFNS approach of this paper with the present upper limit on the prompt-neutrino flux at the 90\% confidence level recently obtained by the IceCube experiment~\cite{Aartsen:2016xlq} (solid black line) and its extrapolation (dotted black line), which adopted the ERS model~\cite{Enberg:2008te} as a basis for modeling prompt neutrinos. Central predictions using the scales $\mu_r = \mu_f = \sqrt{p_T^2 + 4 m_c^2}$ with {\texttt{PROSA}} (PROSA 2016) and {\texttt{ABM11}} PDFs (GMS 2015) are also shown. The limit and all predictions refer to the H3p CR flux.} 
\end{figure}

It turns out that the central GM-VFNS predictions are still below this IceCube upper limit, at least for energies up to $E_{\nu,\,\text{lab}}$~$\sim$~$5\cdot 10^6$ GeV, whereas for higher energies the extrapolation of this limit approaches our predictions.
However, one has to take into account that the experimental upper limit was extracted by considering neutrino events with deposited energies between 9~TeV and 69~TeV only. Therefore, results at the highest energies are just the result of an extrapolation, which can still be prone to big uncertainties (related to the shape of the prompt-neutrino and CR primary fluxes). 
On the other hand, one can observe that, in the energy region above 30 TeV, the upper limit of the GM-VFNS flux including the {\texttt{CT14nlo}} PDF uncertainties is already larger than the IceCube upper limit. The difference becomes dramatic with increasing energies, amounting to a factor of $\sim$ 10 at $E_{\nu,\,\text{lab}}$ $\sim$ 1 PeV. We interpret this result as a first evidence of the fact that IceCube  experimental results are already capable of constraining PDFs. In particular, the discrepancy between theoretical predictions and experimental data provides a clear indication that the uncertainty accompanying the {\texttt{CT14nlo}} PDFs, especially related to the 53--56 PDF sets, represents a too conservative estimate, leading to an overestimation of the partonic densities for $x$ $<$ $10^{-4}$.
Thus, IceCube results can complement collider results in constraining PDFs at low $x$ values (as well as other aspects of non-perturbative~QCD).

\begin{figure}[ht!]
\begin{center}
  \includegraphics[width=0.59\textwidth]{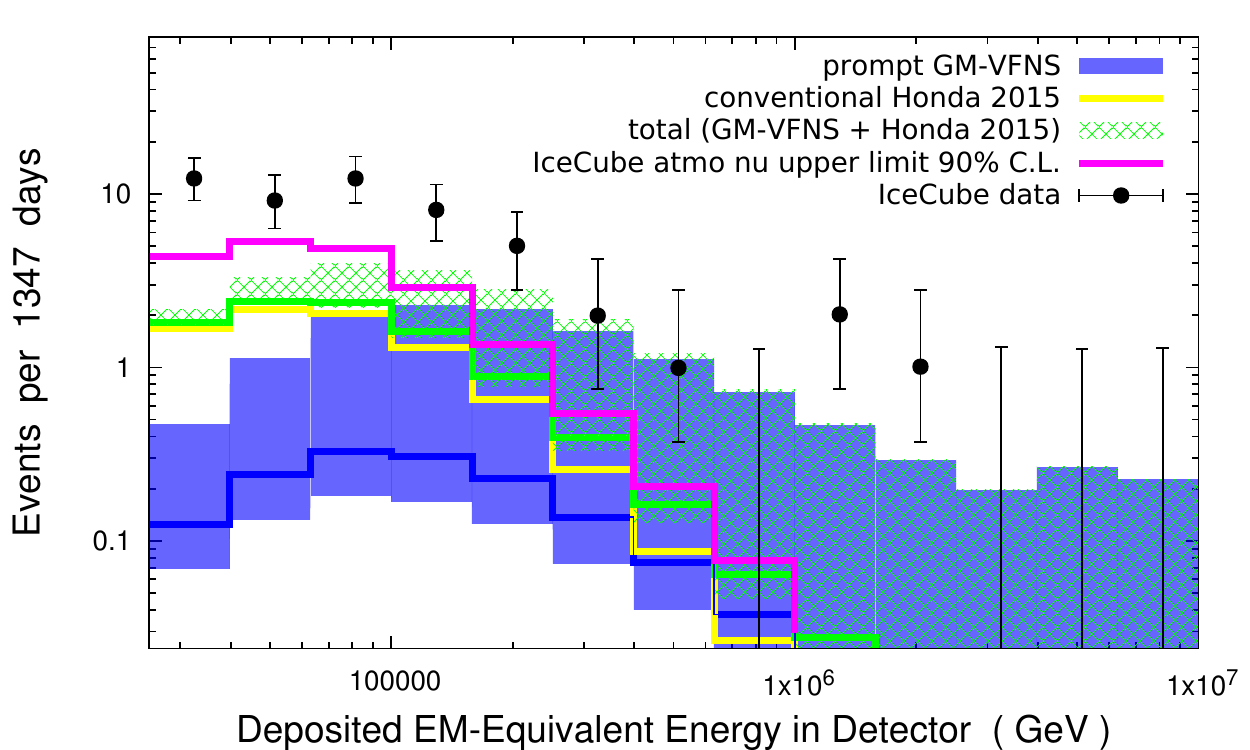}
\includegraphics[width=0.59\textwidth]{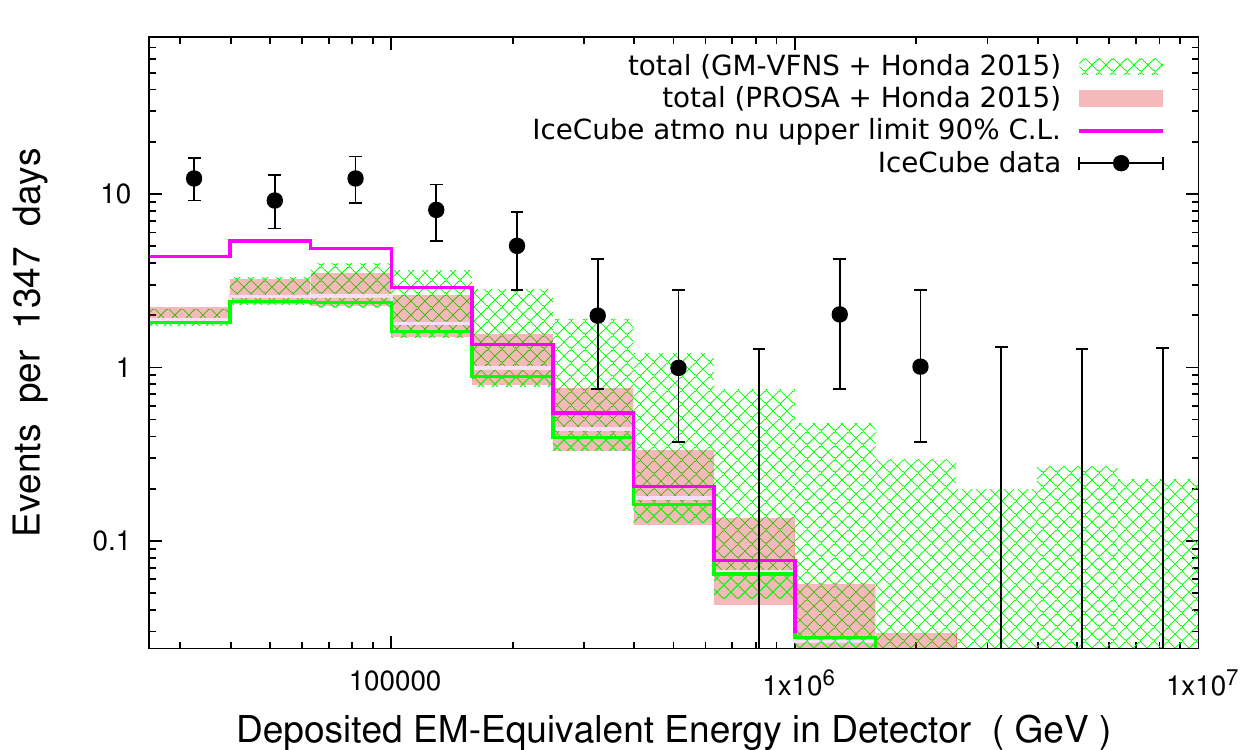}
\caption{\label{heseevents} Top: expected number of atmospheric-neutrino events after 4 years of data taking in comparison with IceCube experimental data~\cite{Aartsen:2015zva}, according to the HESE ana\-ly\-sis setup.
  Theoretical predictions (green) account for both the conventional component (yellow), according to the Honda 2015 model, reweighted for the H3a CR primary spectrum, and a prompt component (blue), computed according to the GM-VFNS method presented in this paper. Predictions for one or more astrophysical-neutrino components are not shown, whereas IceCube experimental data (black) also include neutrinos of non-atmospheric origin. The IceCube upper limit at 90\% C.L. obtained in the analysis of Ref.~\cite{Aartsen:2013eka} and reproduced in Ref.~\cite{Aartsen:2015zva} is also shown (violet). The uncertainty in the total predicted number of events reflects the uncertainty in the prompt component, under the assumption that the conventional component does not lead to any additional uncertainty.  See text for more details.
Bottom: the total expected number of atmospheric neutrino events obtained in this work (green) for the 4-year HESE analysis configuration is compared with an updated version (from the 3-year to the 4-year case) of the one previously published in Ref.~\cite{Garzelli:2016xmx}, on the basis of a different computation of the prompt component, making use of {\texttt{PROSA}} PDFs and {\texttt{POWHEGBOX}}~+~{\texttt{PYTHIA}} predictions for heavy-meson hadroproduction. IceCube HESE experimental data, including neutrinos of both atmospheric and other origins, are also shown. See text for more details.  
}
\end{center}
\end{figure}
Weaker conclusions can be drawn when comparing our theoretical predictions with the results of an analysis on the basis of High-Energy-Starting Events (HESE), i.e. events with interaction vertices inside the detector fiducial volume, characterized by deposited energies between 30 TeV and a few PeV. Also this analysis has been updated over the years, with the most recent results presented in Ref.~\cite{Aartsen:2015zva}, based on the 54 events collected in a 4-year study.
Differently from the analysis of the incoming muon tracks discussed above, the HESE analysis includes events from both the northern and southern hemispheres. As a consequence, muon self-veto techniques~\cite{Gaisser:2014bja} are applied here to veto a part of the expected atmospheric down-going events, i.e. those events where neutrinos are accompanied by detectable muons in the same air shower originating from a cosmic ray interacting with the atmosphere. This reflects the fact that part of the atmospheric background was subtracted from the experimental signal as well, using the information on muons detected in coincidence with neutrinos. A refined procedure to compute the expected number of neutrino events in the HESE analysis from neutrino fluxes is detailed in Ref.~\cite{Palomares-Ruiz:2015mka}. However, confident that this does not change the main conclusions of our study, we followed a less sophisticated (and slightly less accurate) procedure, suggested by the IceCube Collaboration and already used in many previous papers. In particular, we took into account the flavor-dependent effective detector areas provided by the IceCube Collaboration\footnote{See the URL \texttt{https://icecube.wisc.edu/science/data/access}.}, together with an exposure time of 1347 days (4-year analysis). These effective areas were convoluted with the theoretical neutrino fluxes. Our HESE predictions are shown in figure~\ref{heseevents}, in comparison with the IceCube experimental data. The number of expected atmospheric events include both a conventional (i.e. due to the decay of light mesons) component, according to the Honda predictions~\cite{Honda:2006qj} (also used by the IceCube Collaboration), extended to higher energies and reweighted using the H3a CR primary spectrum, and a prompt component, on the basis of the GM-VFNS computation presented in this paper and the same CR primary spectrum. The uncertainty in the total number of events presented in figure~\ref{heseevents} has to be understood as a lower limit to the uncertainty, because it fully accounts for the uncertainty on the prompt component, but it neglects the one on the conventional component.\footnote{At present, precise estimates on the uncertainty on the predictions for the conventional component are still missing, but one could roughly expect that, at the high energies of relevance for the HESE study, these uncertainties could vary in the range $\sim$ 10--30\%.}
From the comparison between predictions and HESE experimental data, it seems that, at energies above 1 PeV, the data is not well compatible with an interpretation just in terms of an atmospheric-neutrino component, and this conclusion, already drawn in previous theory and experimental papers, remains true even in the present work, i.e. even when considering the large PDF uncertainties accompanying our GM-VFNS prompt predictions. However, due to low experimental statistics at high energies, definite conclusions can be premature.

On the other hand, at energies between 300 TeV and 1 PeV, the experimental data lie above the central theoretical predictions, but seem to be still compatible with an interpretation in terms of prompt neutrinos, at least when considering the large uncertainties affecting both our GM-VFNS predictions and the HESE experimental data. This is in contrast with results obtained by means of other PDFs, e.g. the {\texttt{PROSA}} PDFs, characterized by a smaller PDF uncertainty band, as shown in the bottom panel of figure~\ref{heseevents}, where the total number of HESE events predicted by the {\texttt{PROSA}} computation of Ref.~\cite{Garzelli:2016xmx} is compared to the one obtained in this paper.

Also the IceCube analysis of the diffuse flux using muon tracks from the northern hemisphere is not compatible with an interpretation of these data in terms of prompt neutrinos. This is shown again in figure~\ref{heseevents}, where the IceCube upper limit on the atmospheric neutrino flux at 90\% C.L., as obtained by the IceCube Collaboration on the basis of the analysis of Ref.~\cite{Aartsen:2013eka},\footnote{This limit was obtained by the IceCube Collaboration under plausible assumptions concerning the ratio of the fluxes of neutrinos of different flavors $\phi_{\nu_e}$: $\phi_{\nu_\mu}$ : $\phi_{\nu_\tau}$, taking into account that neutrinos of all flavors are detected in the HESE study, whereas only muon neutrinos contribute to the analysis of muon tracks from the northern hemisphere.} is also plotted. This limit indeed puts stronger constraints on the fluxes than the IceCube HESE data. The central theoretical predictions by the GM-VFNS approach in association with the {\texttt{CT14nlo}} PDFs turn out to be below this upper limit in all bins, but a large part of the uncertainty band lies beyond it. We conclude that the HESE data, considered by themselves, are not capable to put strong constraints on the uncertainty band accompanying our computation of prompt-neutrino fluxes at present. However, the analysis of the diffuse muon neutrino flux, already in its old versions, has this power. The most stringent limits from this analysis available at present are those shown in figure~\ref{fig:upperlimit}. 

\section{Conclusions}
\label{sec:conclusions}
A GM-VFNS approach, in which hard-scattering matrix elements with NLO QCD accuracy are complemented by an accurate and consistent set of FFs varying with scale according to NLO evolution equations, has been developed over the years by some of the authors of this paper. This work extends the applicability of that approach to the low-transverse-momentum bins ($p_{T,h_c}$ $<$ 3 GeV) of the $p_{T,h_c}$ differential cross sections of the inclusive hadroproduction of charmed mesons/baryons $NN$~$\rightarrow$~$h_c$~+~$X$. This goal is achieved in practice by a proper choice of the factorization and renormalization scales, also taking into account that QCD theory does not dictate an univocal recipe for choosing these scales. Our predictions are compared with experimental data on charmed-meson hadroproduction from the LHCb collider at $\sqrt{S}=$ 5, 7 and 13 TeV, showing the effectiveness of this approach. The same me\-tho\-do\-lo\-gy was successfully applied in Ref.~\cite{Kniehl:2015fla} in the case of bottom-flavored mesons. The extension described in the present paper allows for an ampler usage of the approach, not only in collider physics, but also in astroparticle physics, where the emission of particles at low $p_T$ plays a fundamental role. 

Furthermore, this work provides a relevant example of how data from astroparticle physics can be considered as a tool, complementary and independent with respect to collider measurements, to constrain hadron properties and quantities of interest for collider phenomenology. In particular, the upper limits on prompt-neutrino fluxes obtained by IceCube observations give strong indication that the {\texttt{CT14nlo}} gluon PDFs, widely used in collider phenomenology, are poorly constrained, with a too large uncertainty band which overestimates the gluon density allowed at low $x$ in a nucleon. This is already hinted at in the comparison of cross section predictions to LHCb charm production data (see Appendix~\ref{sec:appendix}). The predicted uncertainty is far larger than the deviation of the experimental results from the central predictions. The uncertainty might be reduced when including these data in the fit.
In the light of these results, we believe that a revision of these PDFs is urgent, in particular as for the sets 53--56, especially if one wants to apply them to estimates of heavy-quark hadroproduction processes with sensible PDF uncertainty bands. 

Additionally, by comparing the predictions from the GM-VFNS approach, with NLO hard-scattering matrix elements complemented by the {\texttt{KKKS08}} set of FFs with NLO evolution, to those from {\texttt{POWHEGBOX~+~PYTHIA}, making use of NLO hard-scattering matrix elements in the FFNS matched to parton shower and hadronization, this work shows that the uncertainties due to different descriptions of the evolution/transition from hard-scattering partons to hadrons are sub-dominant with respect to those due to scale and PDF variation, and that predictions for prompt fluxes in the GM-VFNS are compatible with those with hard-scattering matrix elements in the FFNS, at least when considering the present uncertainties due to scale variations affecting all predictions. 

\vspace{0.3cm}
Our lepton fluxes will be made available as numerical tables for download at {\texttt{http://}}{\texttt{www.desy.de/$\sim$lepflux}}. Further predictions  can be requested from the authors of this paper by e-mail.

\subsection*{Acknowledgments}
We thank Hubert Spiesberger, Sonny Mantry, Marco Guzzi, Thomas Hahn, Sergey Alekhin, Achim Geiser, Claudio Kopper, Christopher Wiebusch and Stefano Forte for useful discussions. We are grateful to Satyendra Thoudam for clarifications and for having provided us with tables of cosmic-ray primary fluxes according to the most recent fit by the Nijmegen group. We are grateful to Mary Hall Reno and Yu Seon Jeong for having provided us with tables of their prompt-flux predictions. This work has been partially supported by the German Research Foundation DFG through the Collaborative Research Center SFB 676 ``Particles, Strings and the Early Universe'' and by the Helmholtz Alliance for Astroparticle Physics (HAP) funded by the Initiative and Networking Fund of the Helmholtz Association. 

\appendix
\section{LHCb predictions}
\label{sec:appendix}

In figures~\ref{fig:all5}, \ref{fig:all7} and \ref{fig:all13}, we collect our GM-VFNS predictions for ($D^+$ + $D^-$) transverse momentum distributions in $pp$ collisions at $\sqrt{S}$ = 5, 7 and 13 TeV, including the combined uncertainty due to scale and PDF variations. The latter is calculated as explained in Section \ref{sec:fluxes}, taking into account all 56 PDF error sets included in the {\texttt{CT14nlo}} fit, and added in quadrature to the scale uncertainty. We note that the PDF uncertainty is dominant in the low-$p_T$ bins, where low-$x$ effects, encoded in the PDF sets 52--56, play an important role and could be reduced by including the LHCb data in the PDF fit. In case of other $D$-mesons, not shown in the following, we obtain similar trends and levels of agreement of our theoretical predictions with the LHCb experimental data. 
\begin{figure}[ht]
\centering
\includegraphics[width=73mm]{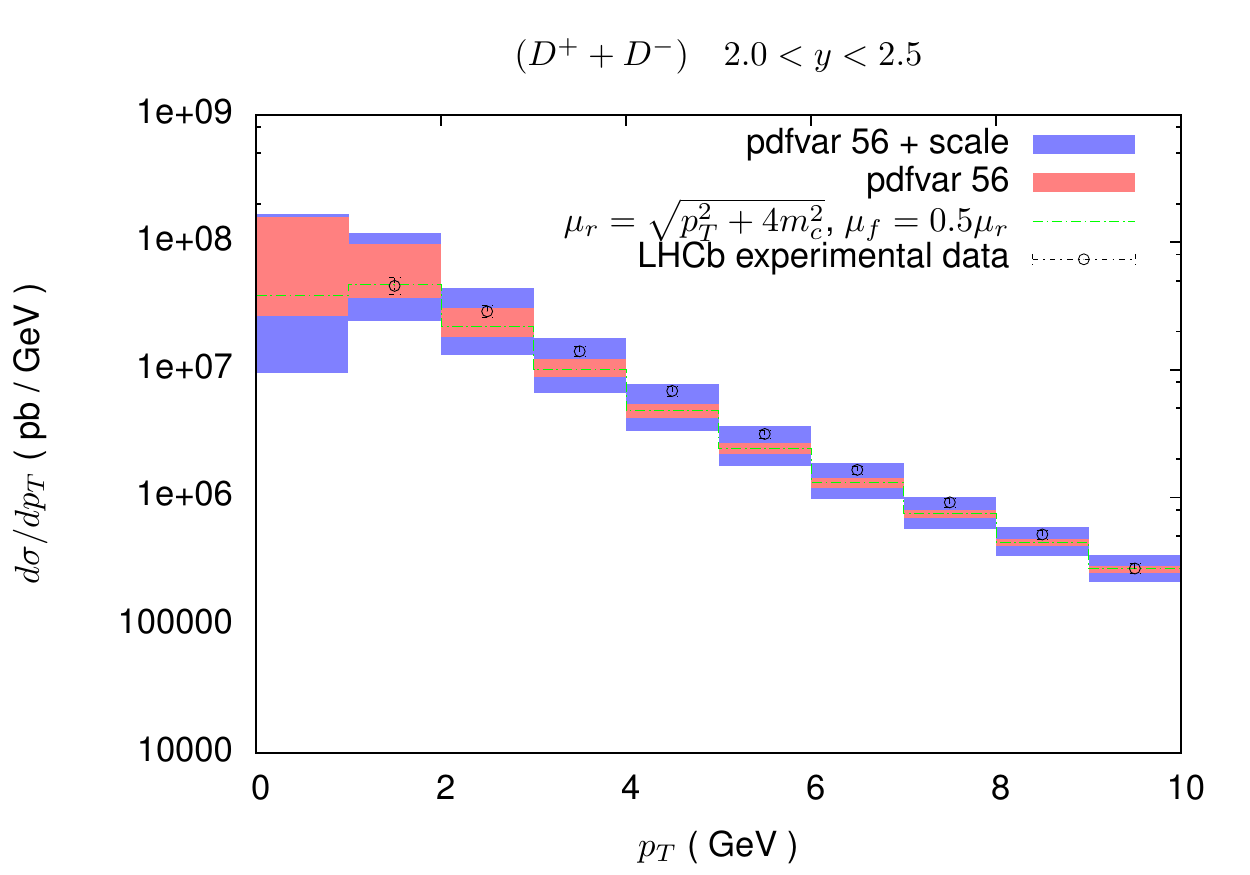}\quad\includegraphics[width=73mm]{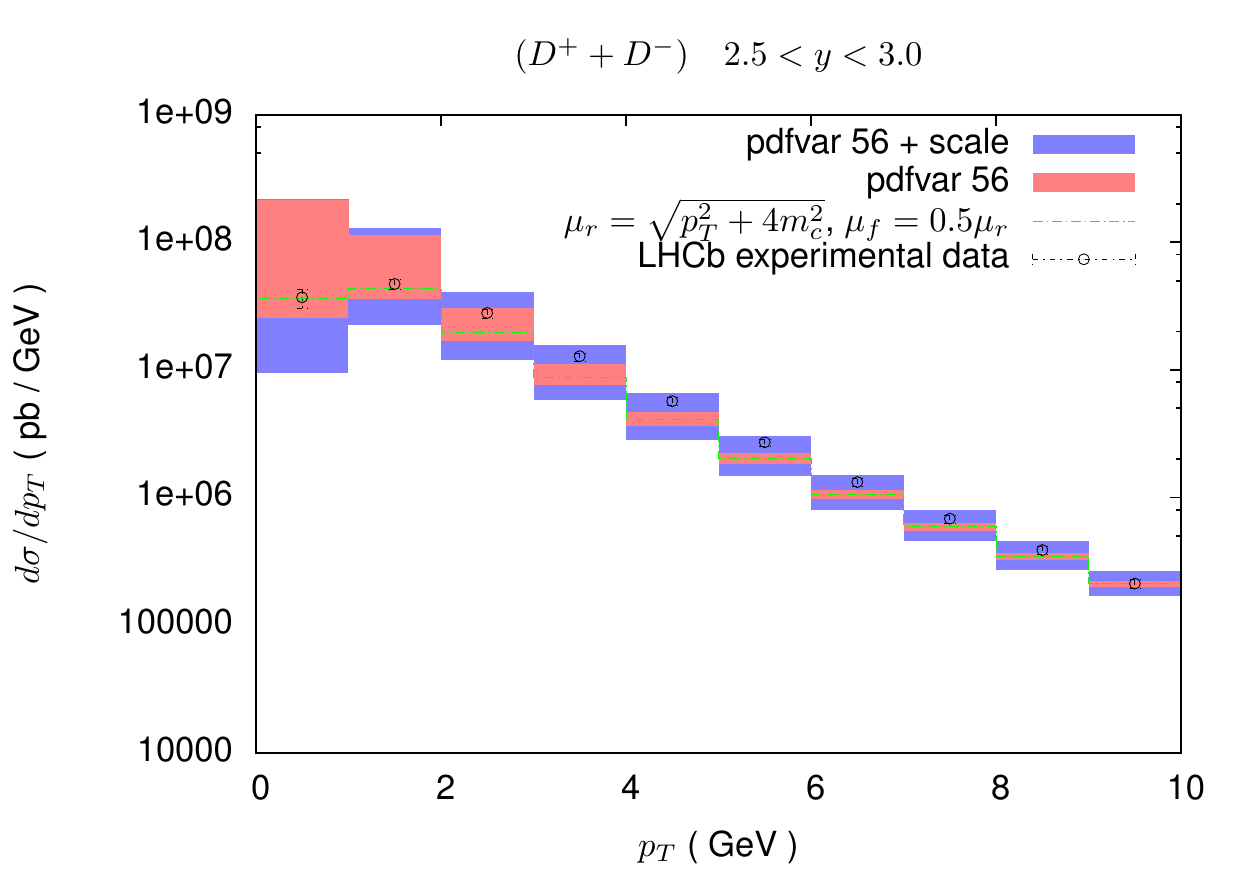}\\
\includegraphics[width=73mm]{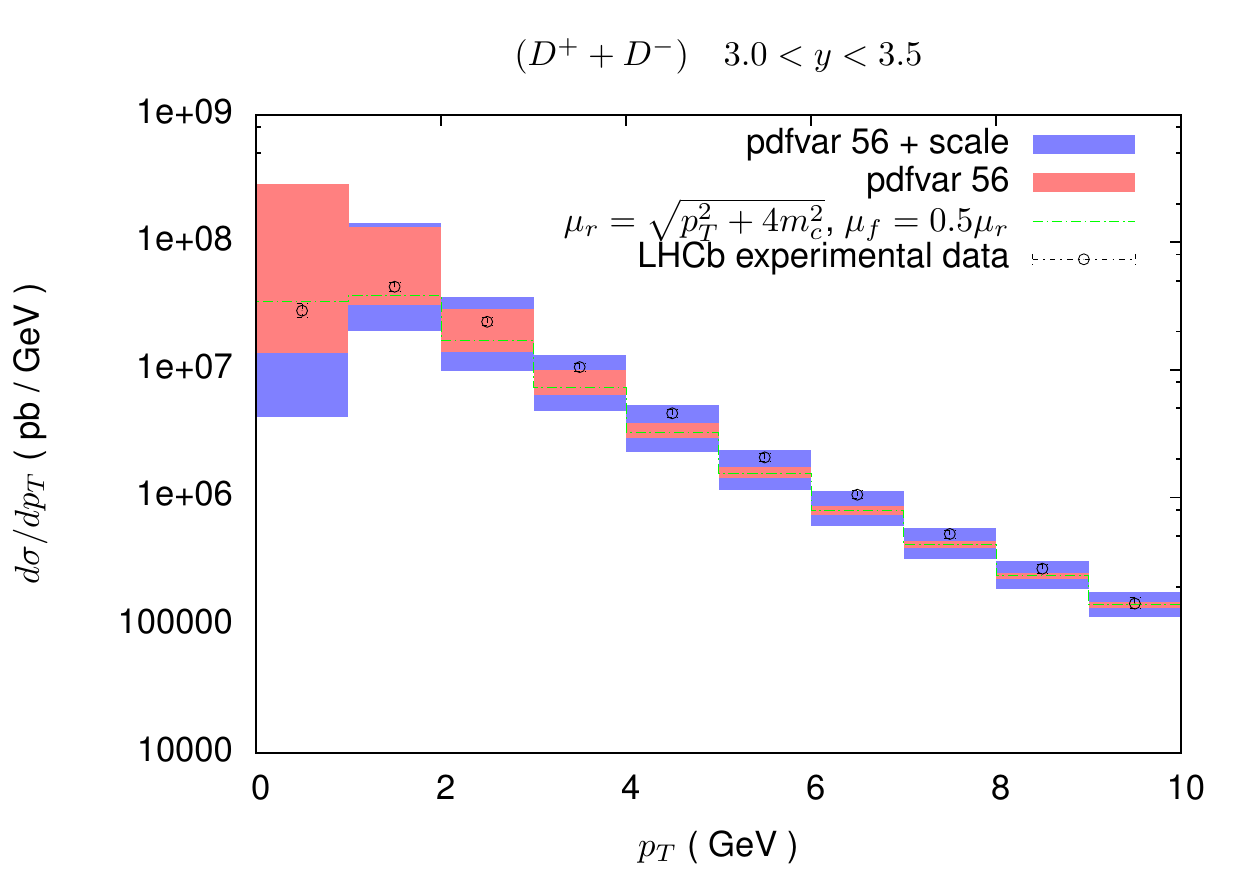}\quad\includegraphics[width=73mm]{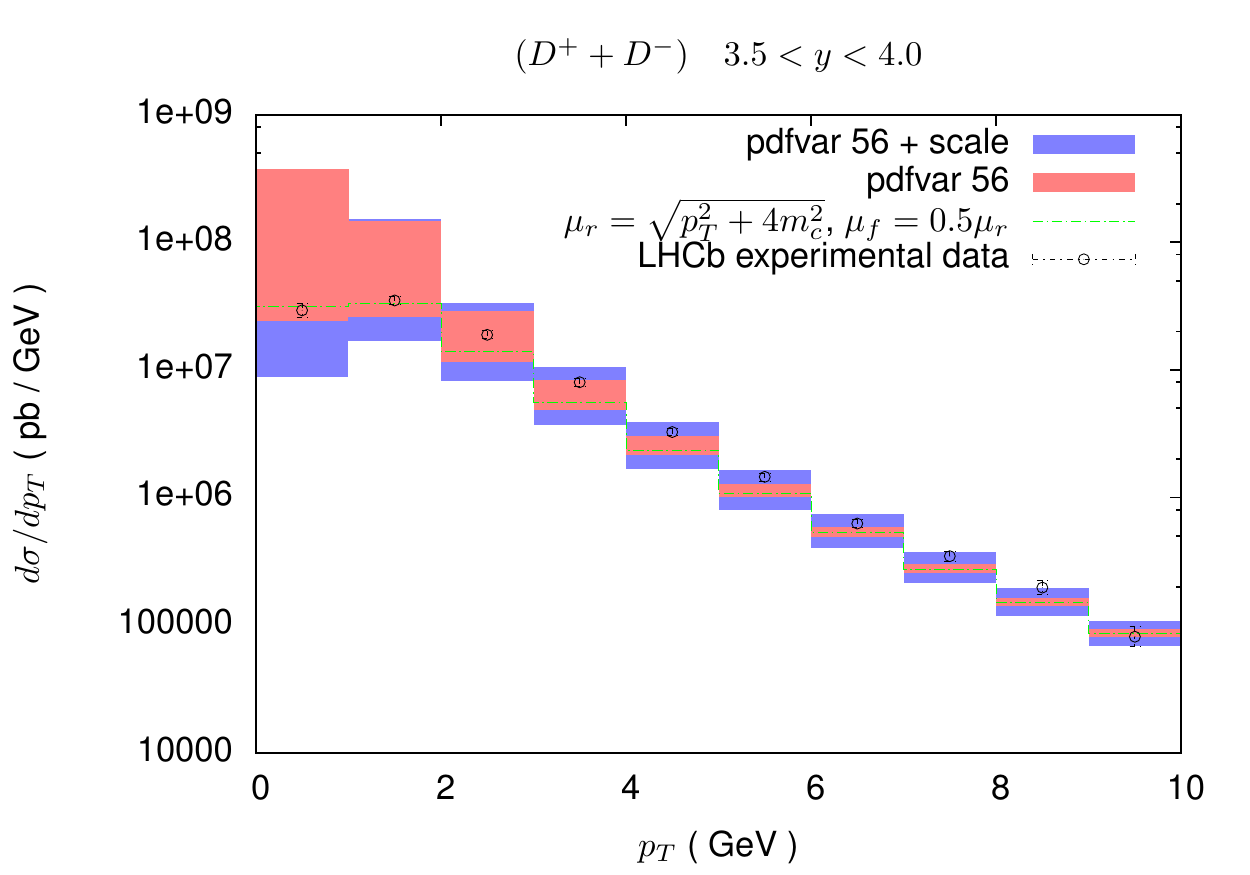}\\
\includegraphics[width=73mm]{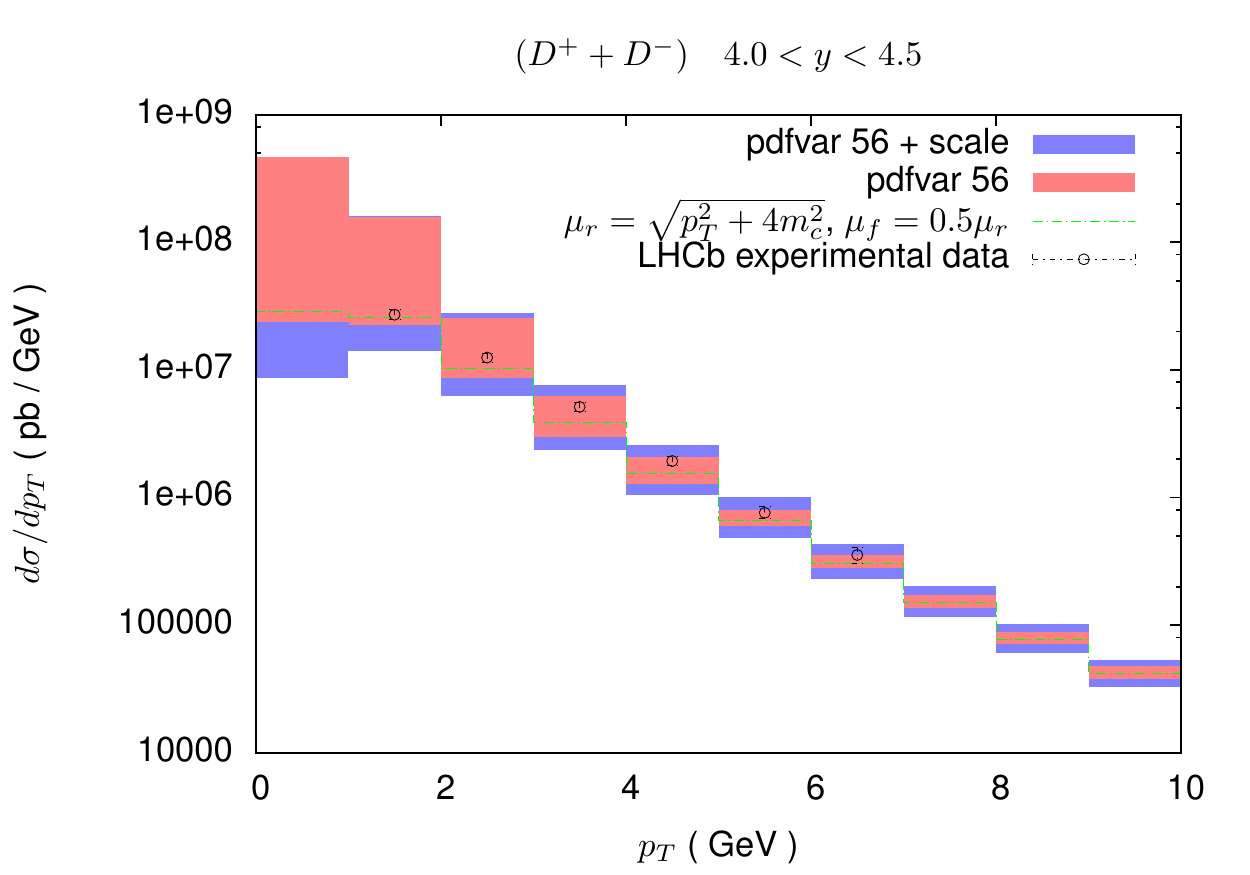}
\caption{Our GM-VFNS predictions for ($D^+$ + $D^-$) transverse-momentum distributions in $pp$ collisions at $\sqrt{S}$ = 5 TeV vs. LHCb experimental data of Ref.~\cite{Aaij:2016jht}. Each panel corresponds to a different rapidity bin in the interval 2 $<$ $y$ $<$ 4.5. The renormalization scale $\mu_r$ is chosen as $\sqrt{p_T^2+4m_c^2}$, and varied in the [0.5, 2] interval around this central value. The PDF uncertainties are computed as explained in Section \ref{sec:fluxes}, while the combined uncertainty is determined by adding the scale and PDF uncertainties in quadrature.
}
\label{fig:all5}
\end{figure}

\begin{figure}[ht]
\centering
\includegraphics[width=73mm]{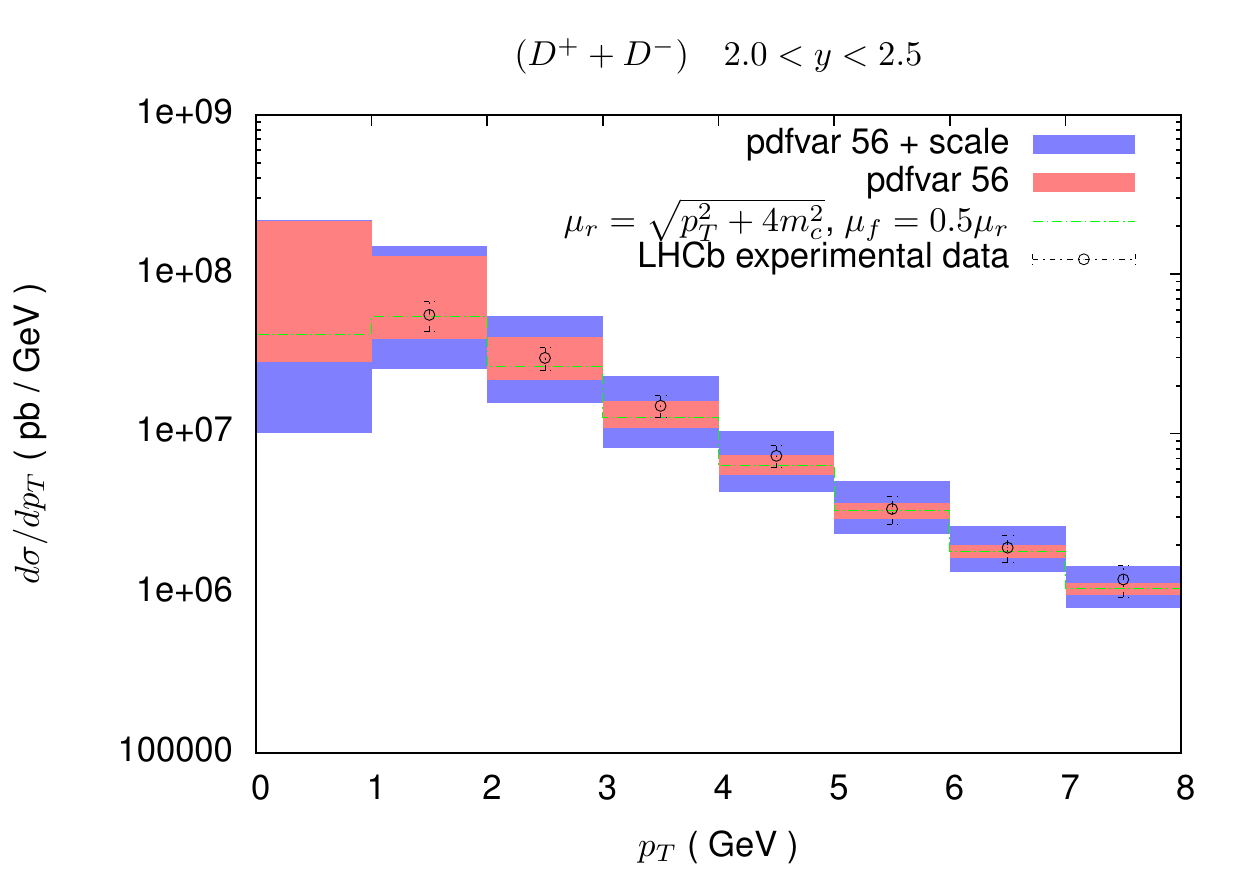}\quad\includegraphics[width=73mm]{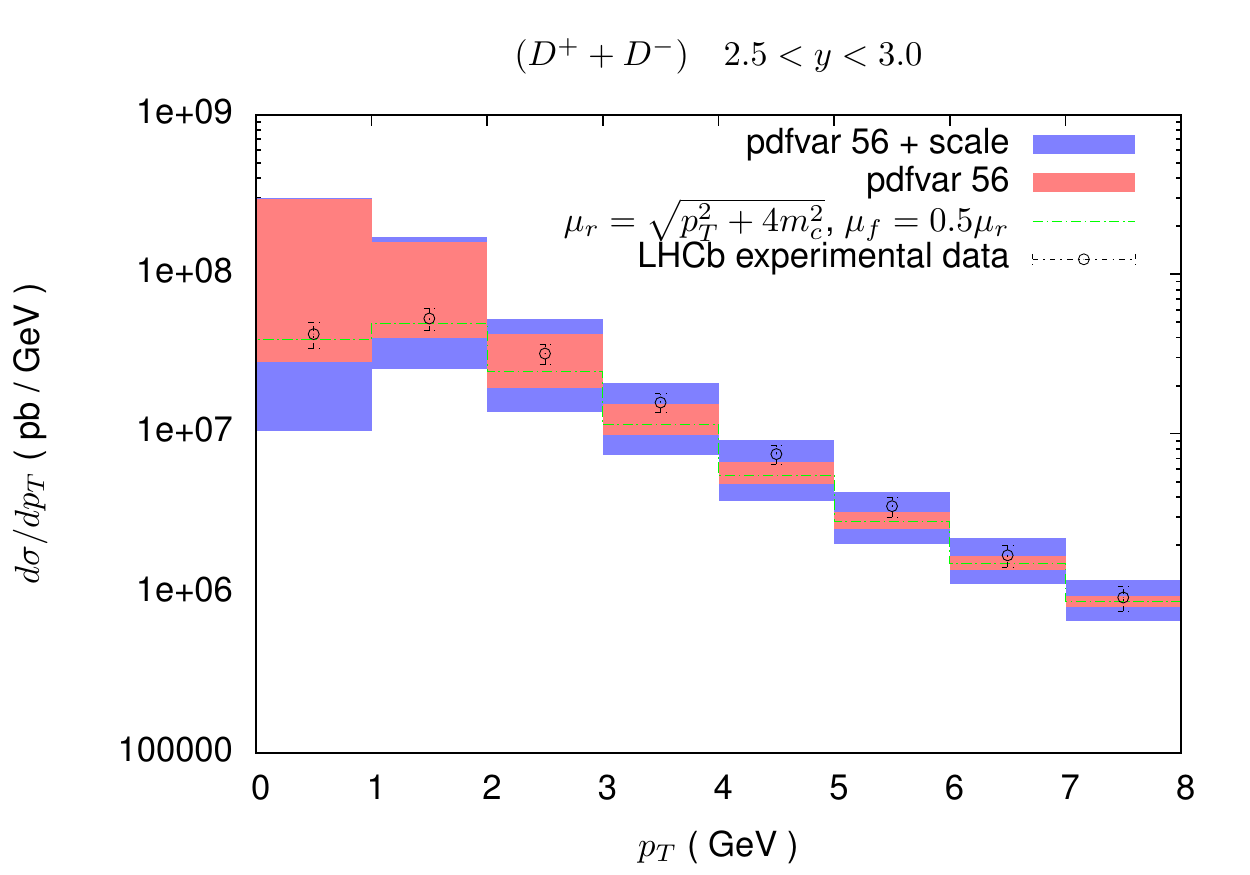}\\
\includegraphics[width=73mm]{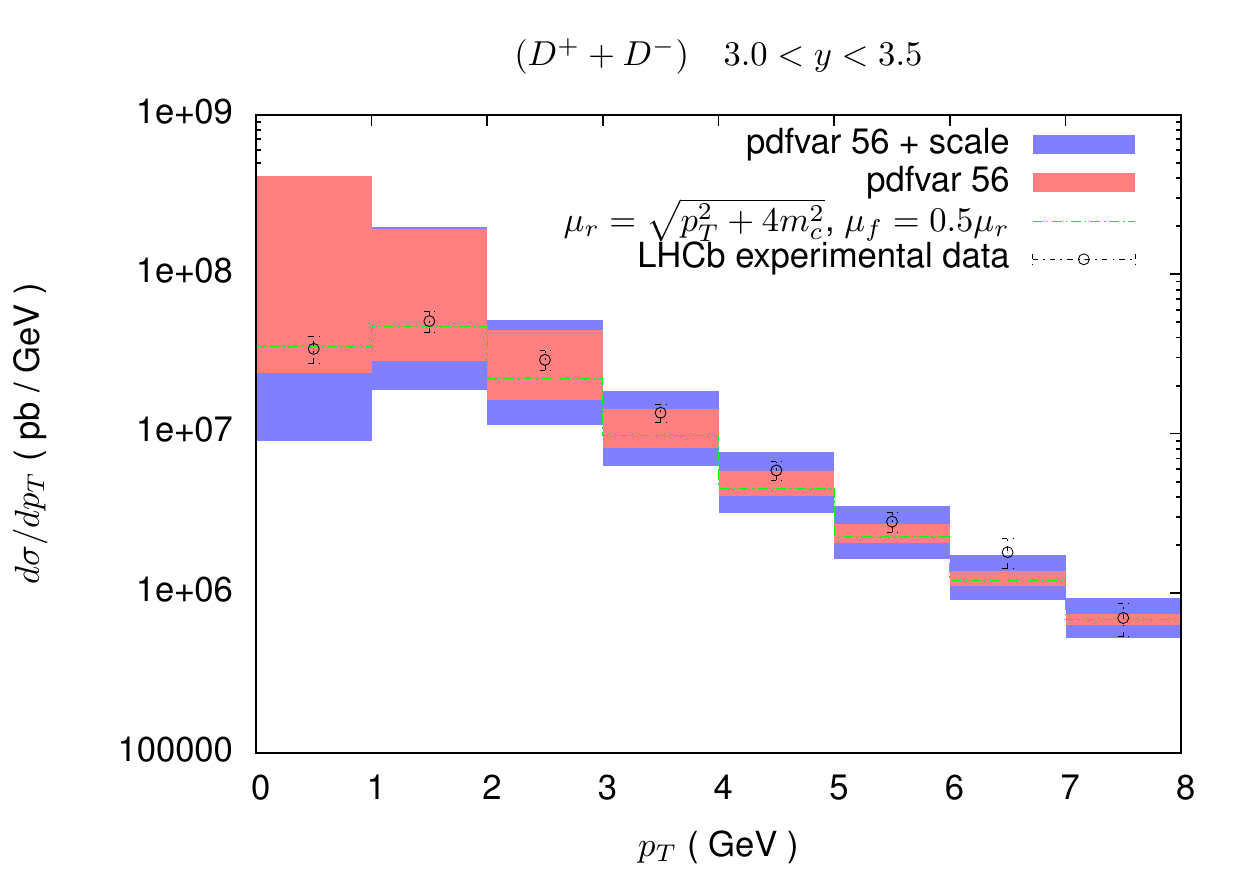}\quad\includegraphics[width=73mm]{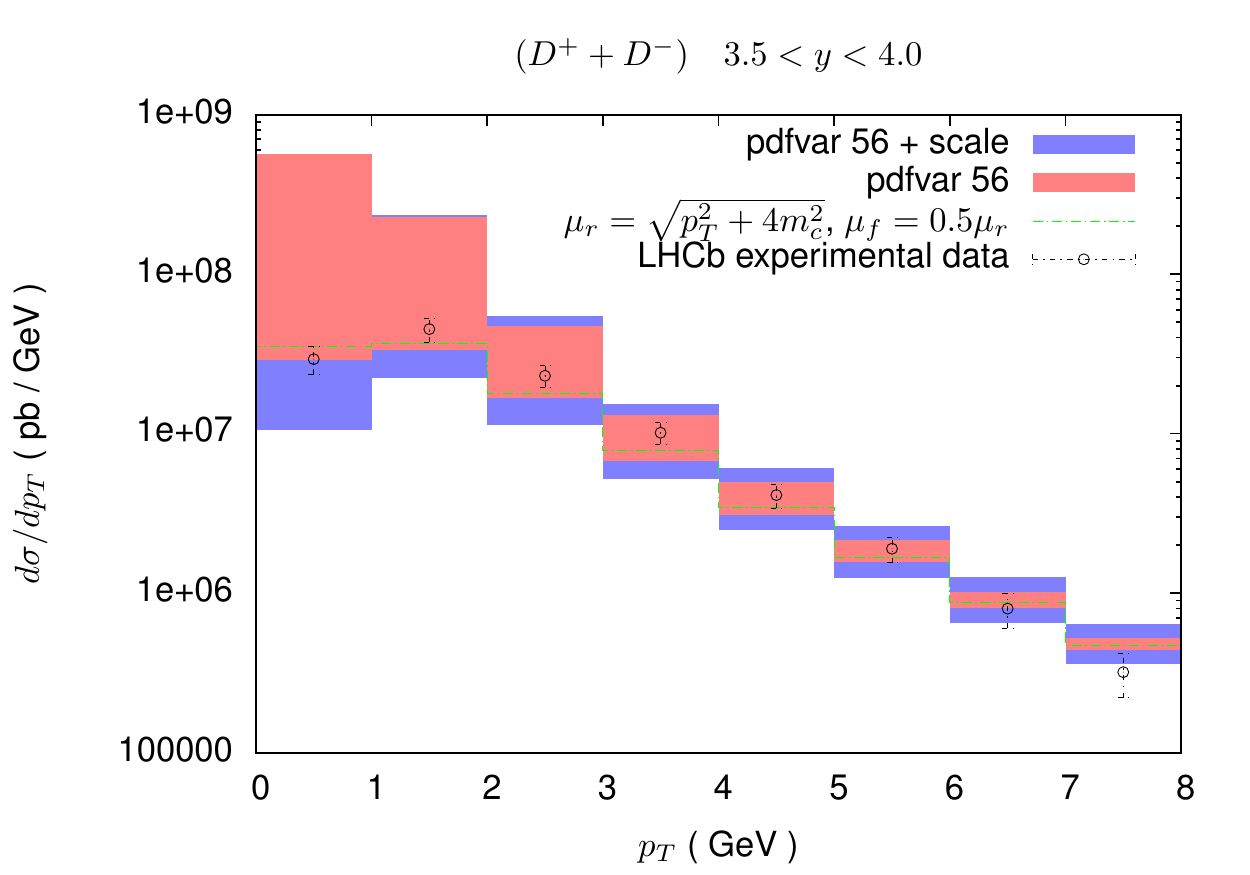}\\
\includegraphics[width=73mm]{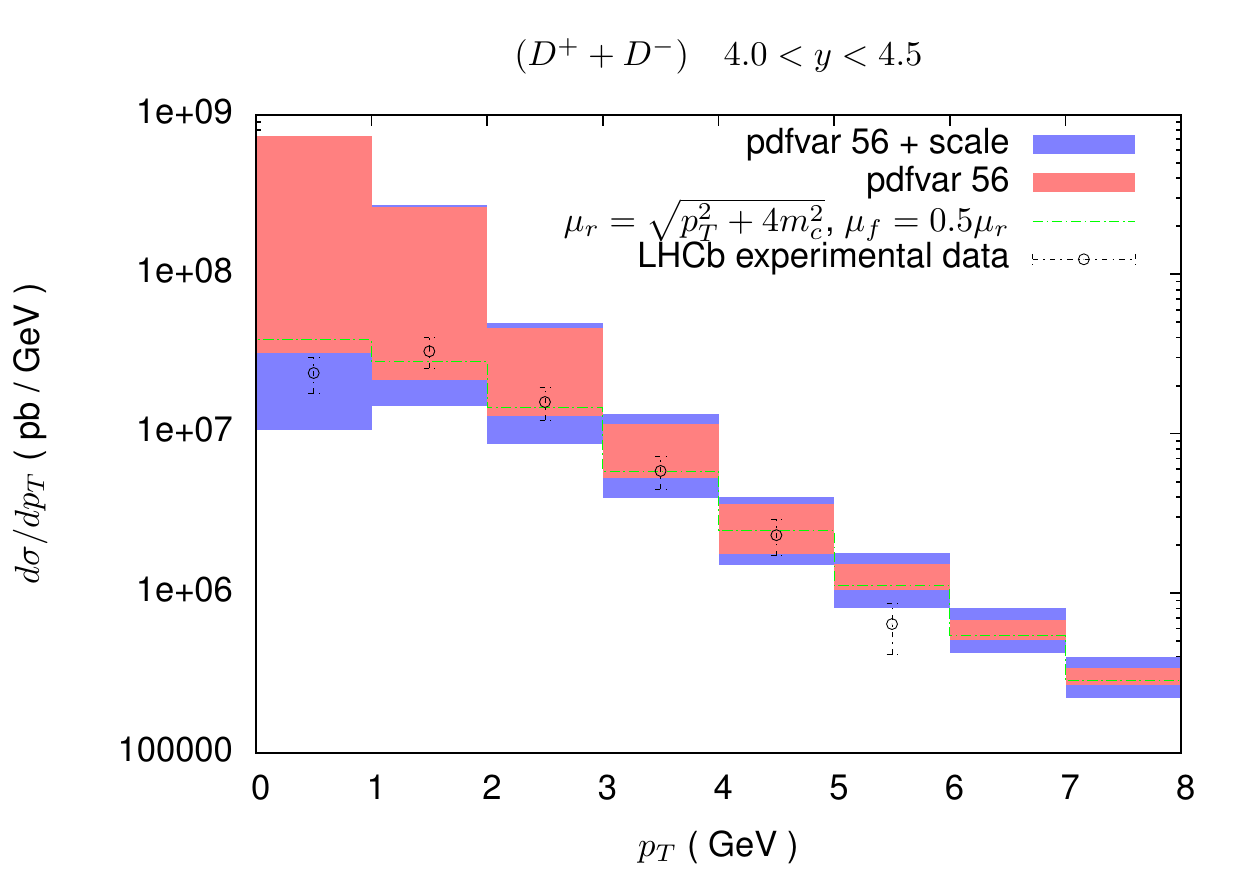}
\caption{Our GM-VFNS predictions for ($D^+$ + $D^-$) transverse-momentum distributions in $pp$ collisions at $\sqrt{S}$ = 7 TeV vs. LHCb experimental data of Ref.~\cite{Aaij:2013mga}. Each panel corresponds to a different rapidity bin in the interval 2 $<$ $y$ $<$ 4.5. The PDF uncertainties and the scale plus PDF uncertainties are computed as in figure~\ref{fig:all5}.
}
\label{fig:all7}
\end{figure}

\begin{figure}[ht]
\centering
\includegraphics[width=73mm]{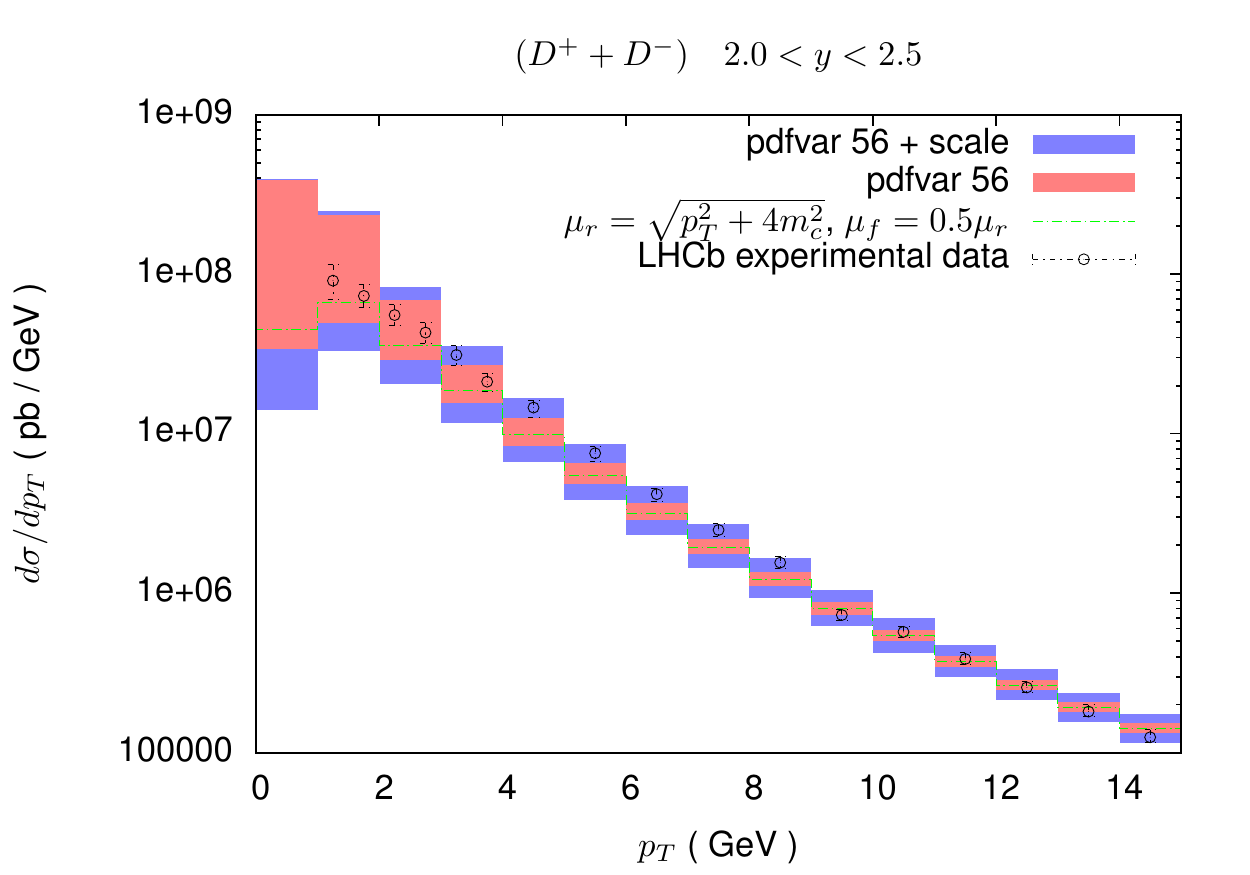}\quad\includegraphics[width=73mm]{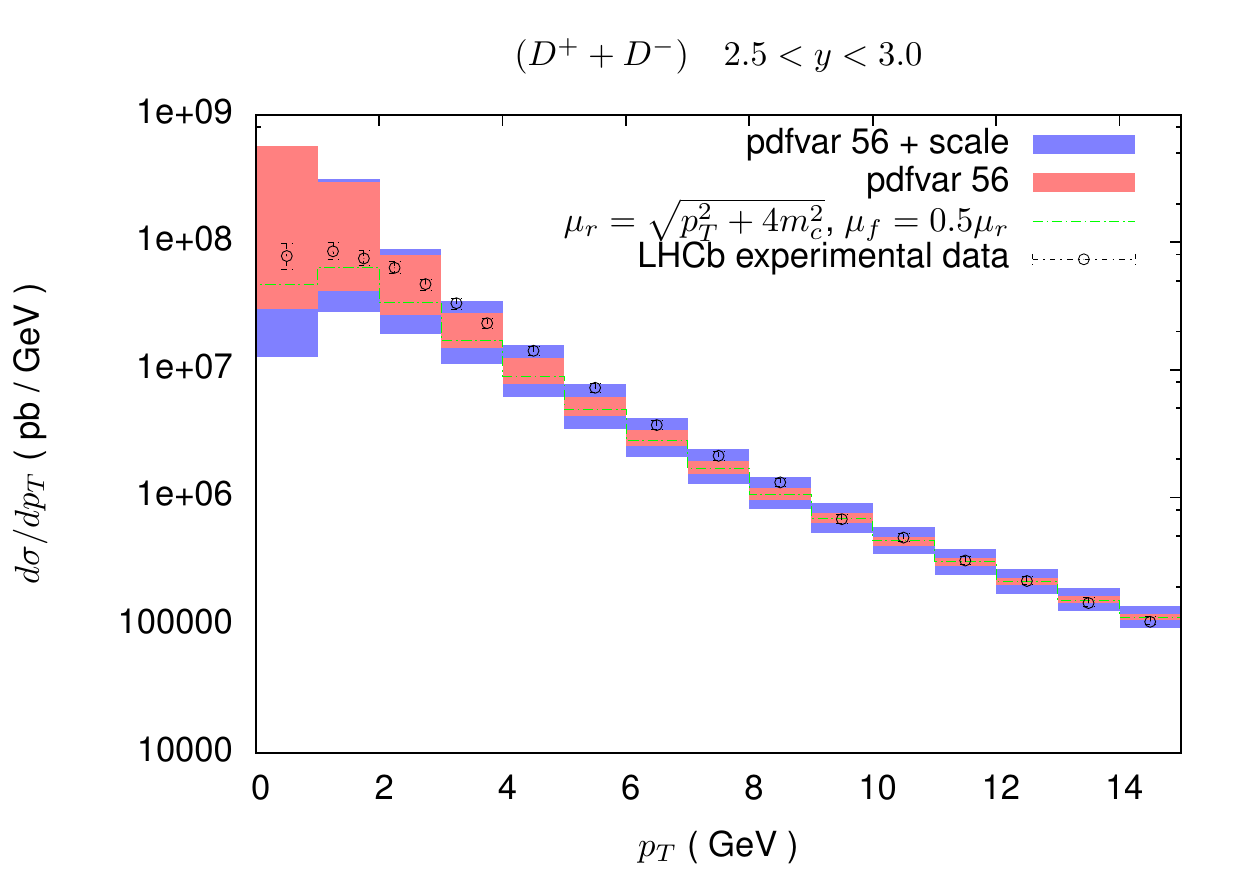}\\
\includegraphics[width=73mm]{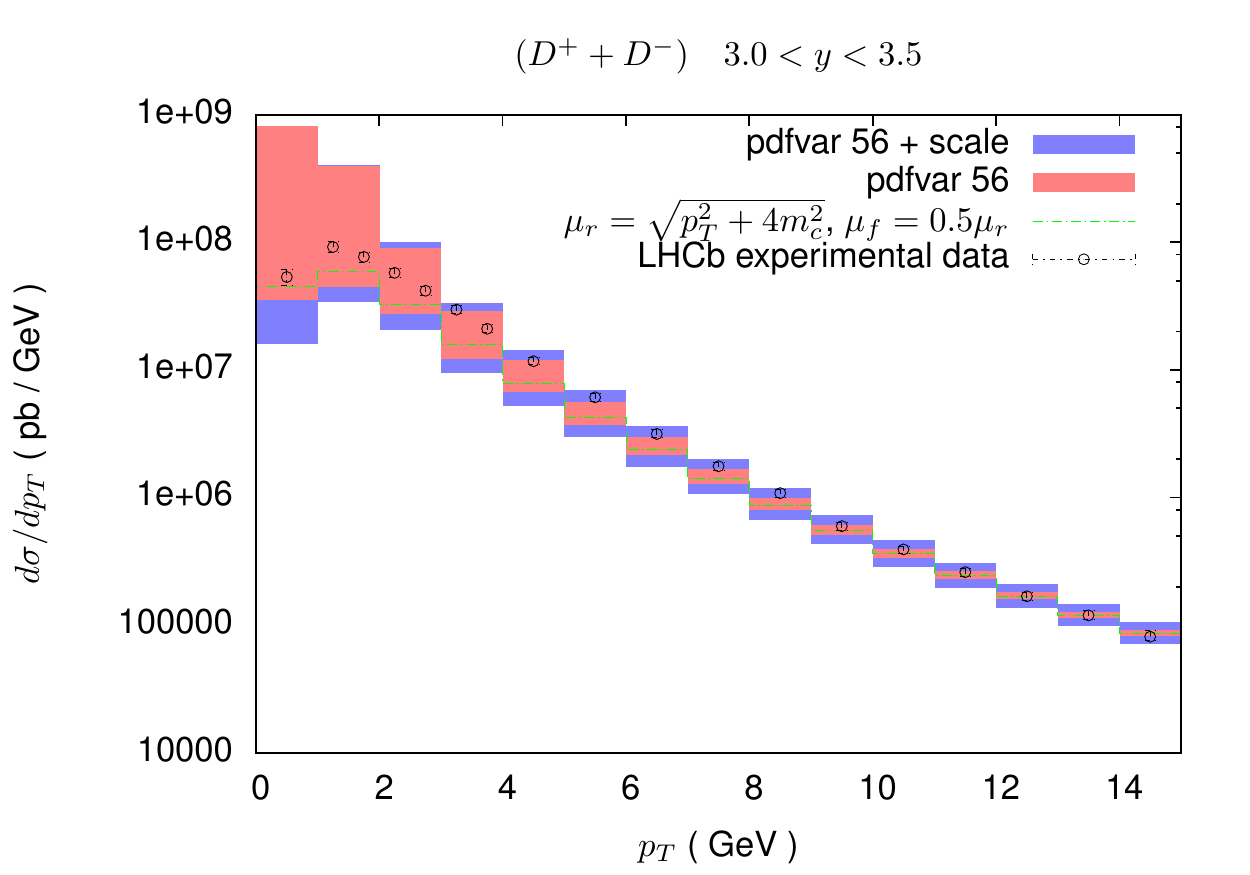}\quad\includegraphics[width=73mm]{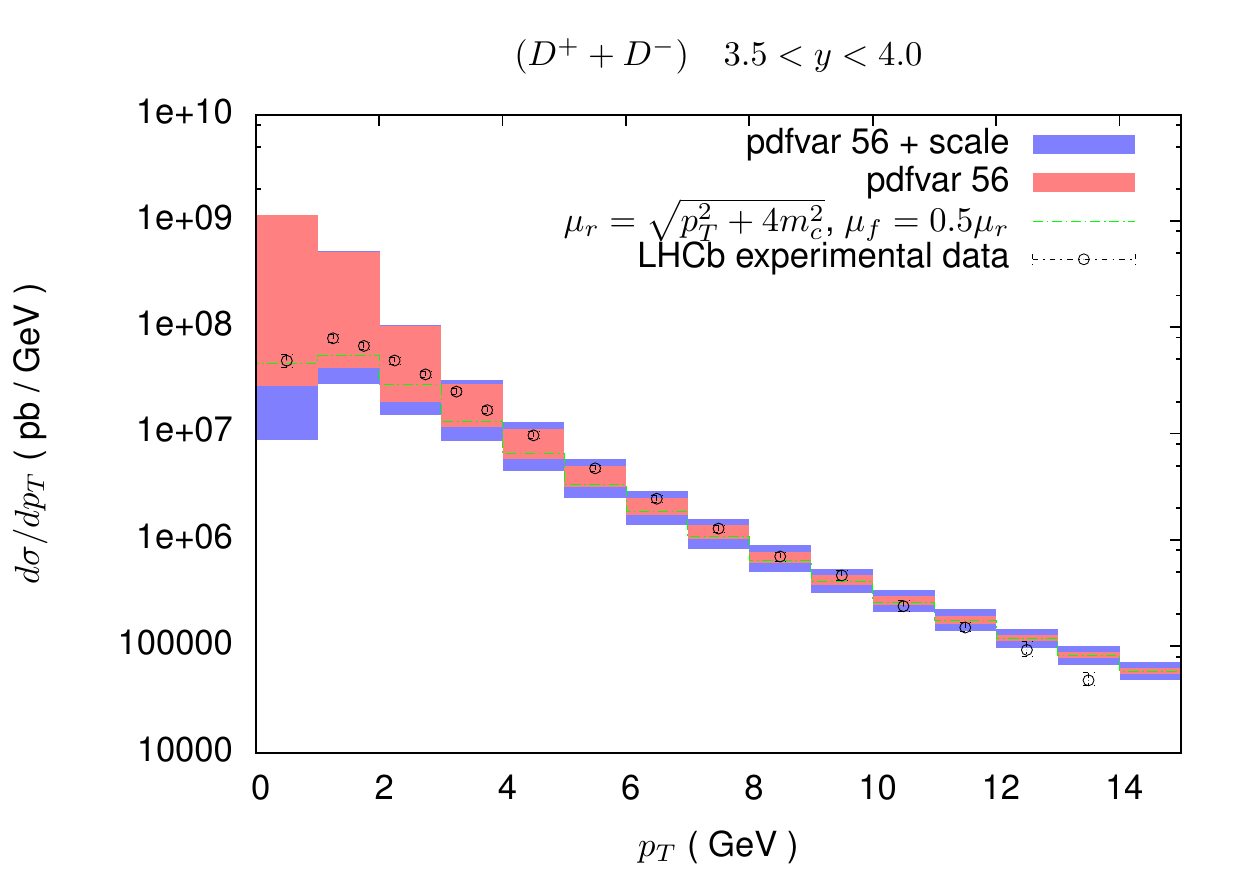}\\
\includegraphics[width=73mm]{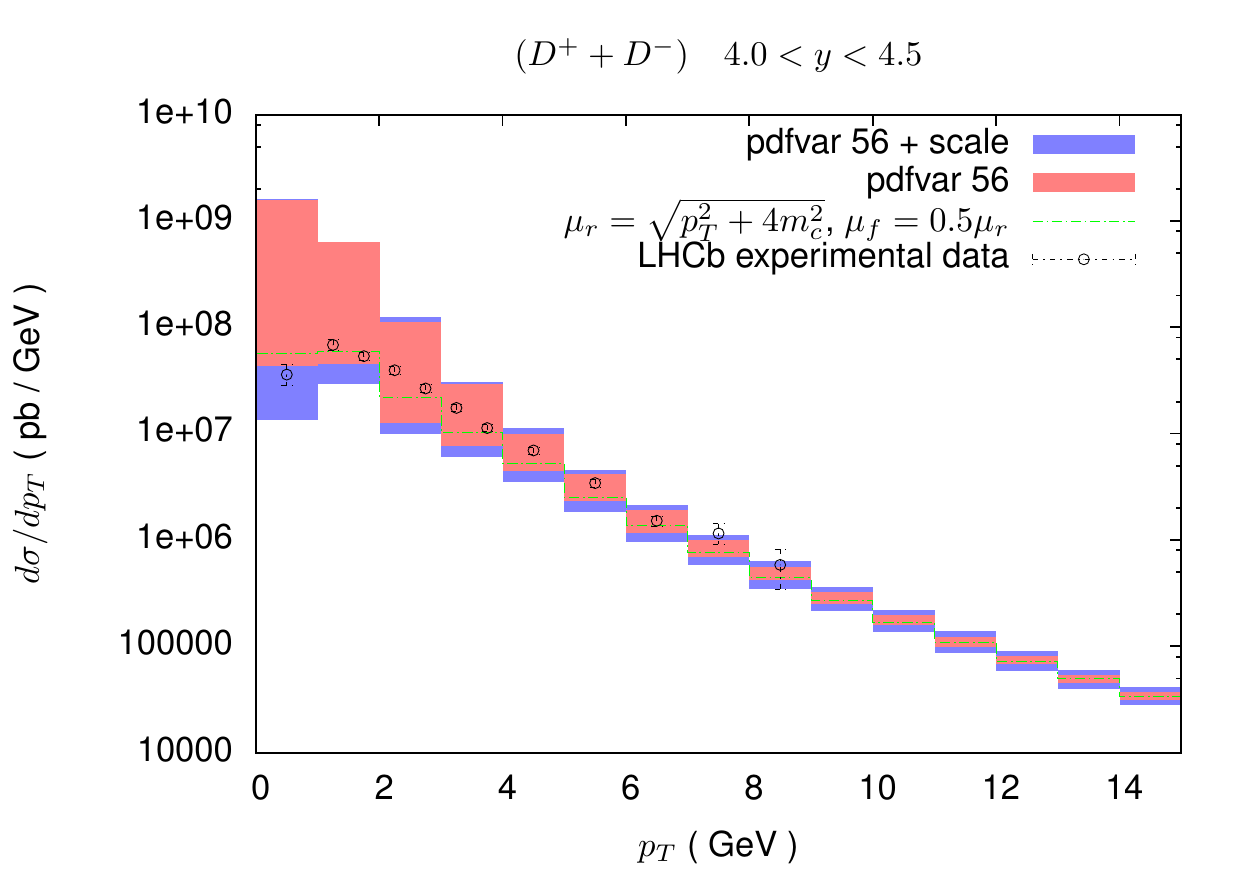}
\caption{Our GM-VFNS predictions for ($D^+$ + $D^-$) transverse-momentum distributions in $pp$ collisions at $\sqrt{S}$ = 13 TeV vs. LHCb experimental data of Ref.~\cite{Aaij:2015bpa}. Each panel corresponds to a different rapidity bin in the interval 2 $<$ $y$ $<$ 4.5. The PDF uncertainties and the scale plus PDF uncertainties are computed as in figure~\ref{fig:all5}.
}
\label{fig:all13}
\end{figure}

\section{Predictions with the NNPDF3.0+LHCb PDF set}
\label{sec:nnpdf}

Throughout this paper, we have used the \texttt{CT14nlo} PDF set~\cite{Dulat:2015mca} as the input to our computations. As we have seen, there is a large uncertainty due to the badly constrained gluon PDF at very low $x$. On the other hand, recent updates of the \texttt{NNPDF3.0}+\texttt{LHCb}~\cite{Gauld:2016kpd} PDF fit\footnote{We have used the \texttt{NNPDF3.0}+\texttt{LHCb} PDF version dating back to the end of June 2017, corresponding to the revised version v2 of Ref.~\cite{Gauld:2016kpd}. This version includes corrected LHCb data from the revised LHCb papers \cite{Aaij:2016jht,Aaij:2015bpa}, dating back to May 2017.} use the latest LHCb data on charm hadroproduction in order to reduce the uncertainty. However, as opposed to the \texttt{CT14nlo} set, they are not positive definite and turn out to have large negative values for very small $x$ and low scale $\mu_f$, as can been seen in the left panel of figure~\ref{fig:nnpdfprint}.
\begin{figure}[ht]
\centering
\includegraphics[width=73mm]{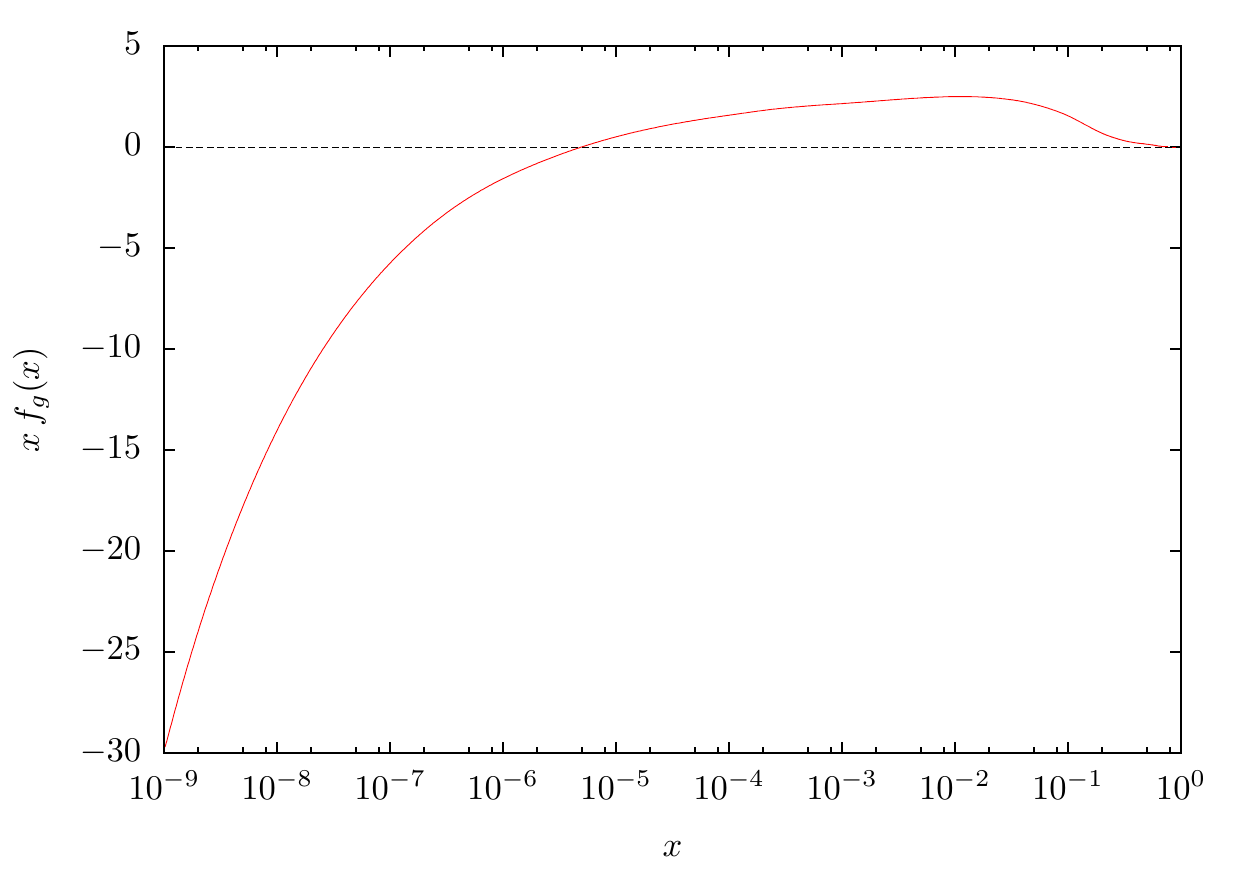}\quad\includegraphics[width=73mm]{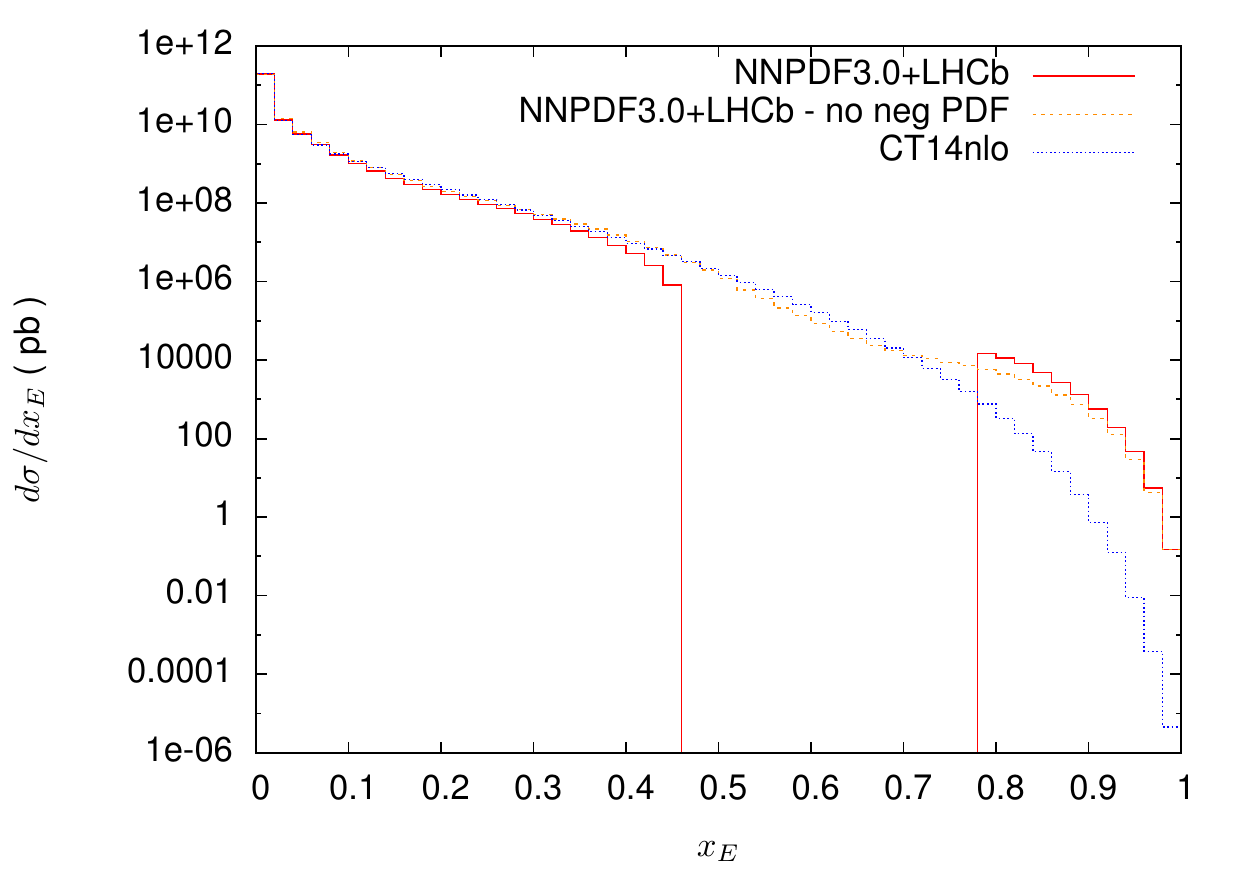}
\caption{Left: Gluon distribution according to the latest version of the \texttt{NNPDF3.0}+\texttt{LHCb} PDF set at $\mu_f=1.5\,$GeV. Right: Differential cross sections d$\sigma$/d$x_E$ of inclusive $D^0$ hadroproduction, $pp$ $\rightarrow$ $D^0 + X$, for initial-state protons with a laboratory energy $E_{p,\,\text{lab}}$~=~$10^8\,$GeV. The orange graph is generated by setting the \texttt{NNPDF3.0}+\texttt{LHCb} PDFs to zero when they are negative.}
\label{fig:nnpdfprint}
\end{figure}
This in turn yields negative cross sections in phase space regions with large $y$ and small $p_T$ of the charmed hadron. Running our computation with the latest version of the \texttt{NNPDF3.0}+\texttt{LHCb} PDFs, we can observe this behavior explicitly. In the right panel of figure~\ref{fig:nnpdfprint}, we show the $x_E$ spectrum for $D^0$ hadroproduction for $E_{p,\,\text{lab}} = 10^8\,$GeV using different PDF sets. We note that \texttt{CT14nlo} and \texttt{NNPDF3.0}+\texttt{LHCb} yield very similar results for the dominant region of small $x_E$. However, between $x_E\approx 0.46$ and $0.78$ the latter becomes negative. This issue is present at all energies, where for larger $E_{p,\,\text{lab}}$ the negative region moves towards lower $x_E$. Although NLO PDFs can be negative in principle, the physical quantities computed from them (like cross sections) must always be positive. One possible ad-hoc prescription to deal with the negative cross sections is to set all PDFs to zero, when they are negative, which gives rise to an uncontrolled uncertainty, not encoded in the PDF uncertainty bands generated by considering all members of the set. Nevertheless, we compare the effect of this prescription on our computation with respect to the standard case in the right panel of figure~\ref{fig:nnpdfprint}. Again, we find good agreement between all graphs for small $x_E$. For larger $x_E$, the positive definite version of \texttt{NNPDF3.0}+\texttt{LHCb} yields a positive cross section, while changing its curvature a few times.

In figure~\ref{fig:nnpdfflux}, we compare the central values of the ($\nu_\mu$ + $\bar{\nu_\mu}$) flux, obtained with the ad-hoc prescription explained above, with those from an alternative prescription, in which the positive and negative signs of the PDFs are retained, but the negative cross sections are set to zero when using them in the cascade equations. Our reference predictions with the {\texttt{CT14nlo}} PDF set are also shown.  
\begin{figure}[ht]
\centering
\includegraphics[width=0.75\textwidth]{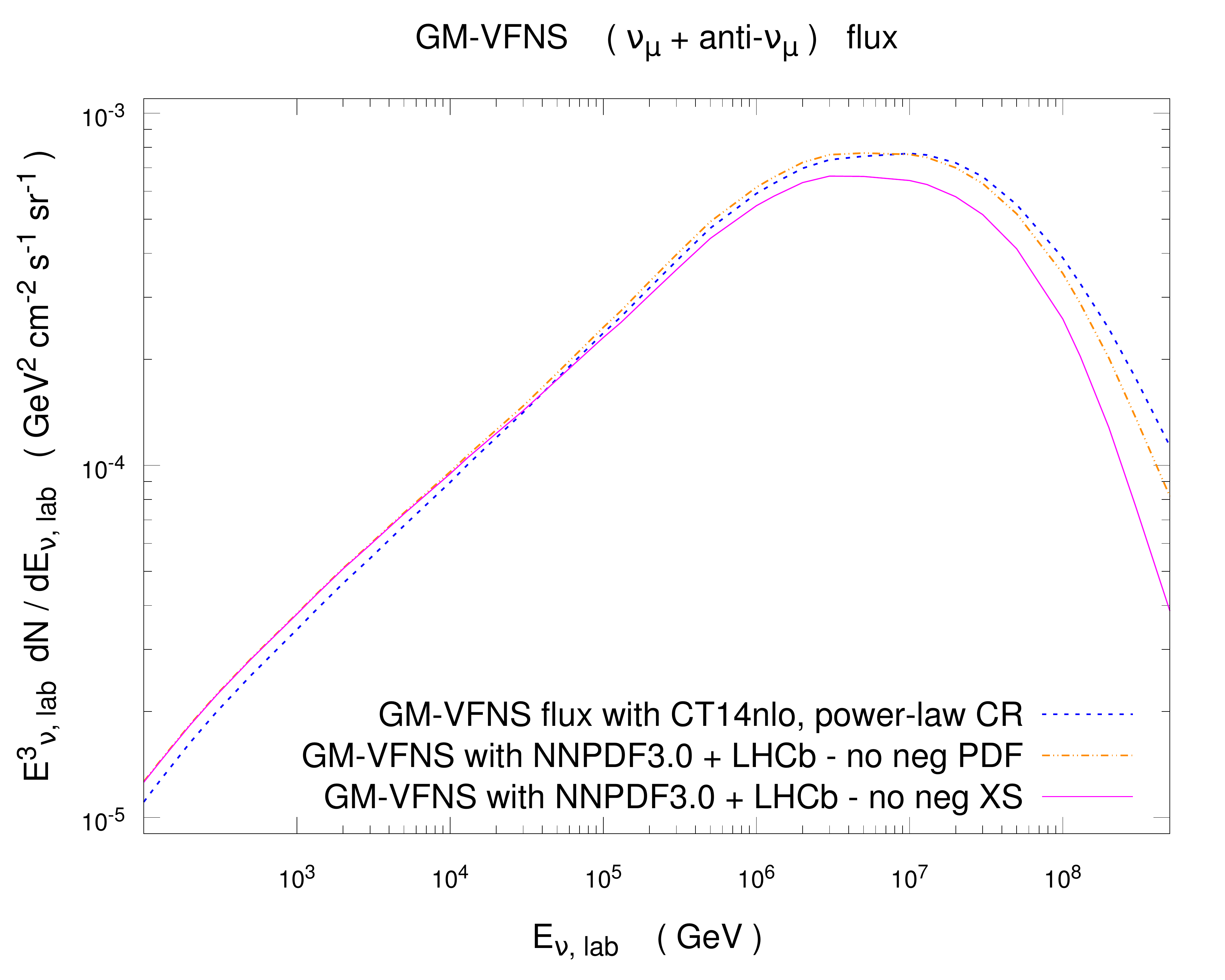}
\caption{Prompt-($\nu_\mu$ + $\bar{\nu}_\mu$) fluxes as a function of neutrino energy $E_{\nu,\,\text{lab}}$. Our standard \texttt{CT14nlo} predictions are compared with those obtained using \texttt{NNPDF3.0}+\texttt{LHCb} with two prescriptions for the treatment of negative values (no negative PDFs vs. no negative cross sections) as explained in the text.}
\label{fig:nnpdfflux}
\end{figure}
There is a good agreement among the predictions at the lower neutrino energies, as can be expected, since the positive region of small $x_E$ dominates the cross section. For increasing energies, the discrepancies grow, and indeed, we cannot exclude even bigger differences for other prescriptions dealing with the negative PDFs. 

In conclusion, we find that including the latest LHCb data in the PDF fits is an important step to reduce the large error bands of the cross section predictions. However, the large negative gluon PDFs appearing in the {\texttt{NNPDF3.0}+\texttt{LHCb}} fit at $x$ values smaller than those constrained by LHCb, introduce a new uncertainty, which needs to be better controlled. Prescriptions proposed here are only tentative a-posteriori prescriptions to deal with the issues. We leave it to the authors of the {\texttt{NNPDF3.0}+\texttt{LHCb}} PDF fit to provide appropriate solutions. 

\clearpage
\bibliographystyle{JHEP}
\bibliography{gmvfnsAstro.bib}

\end{document}